\documentclass[fleqn,usenatbib]{mnras}

\usepackage{newtxtext,newtxmath}

\usepackage[T1]{fontenc}
\usepackage{ae,aecompl}

\usepackage{graphicx}	
\usepackage{amsmath}	
\usepackage{amssymb}	

\newcommand{\mdmunits}{{\rm pc \, cm^{-3}}} 
\newcommand{\dmunits}{$\mdmunits$}
\def\dm#1{{\rm DM_{\rm #1}}}
\newcommand{\mdmacosmic}{< \! {\rm DM}_{\rm cosmic} \! >}
\newcommand{\dmacosmic}{$\mdmacosmic$}
\newcommand{\mdmcosmic}{\dm{cosmic}}
\newcommand{\dmcosmic}{$\mdmcosmic$}
\newcommand{\mdmhalos}{\dm{halos}}
\newcommand{\dmhalos}{$\mdmhalos$}
\newcommand{\dmism}{$\dm{ISM}$}
\newcommand{\dmovii}{$\dm{OVII}$}
\newcommand{\dmhvc}{$\dm{MW,cool}$}
\newcommand{\mdmhalo}{\dm{halo}}
\newcommand{\dmhalo}{$\mdmhalo$}
\newcommand{\mdmigm}{\dm{IGM}}
\newcommand{\dmhost}{$\dm{host}$}
\newcommand{\dmigm}{$\dm{IGM}$}
\newcommand{\dmfrb}{$\dm{FRB}$}
\newcommand{\dmlgm}{$\dm{LGM}$}
\newcommand{\mdmmwh}{\dm{MW,halo}}
\newcommand{\dmmwh}{$\mdmmwh$}
\newcommand{\dmlmch}{$\rm DM_{\rm LMC}^{\rm halo}$}

\newcommand{\mmmin}{M_{\rm min}}
\newcommand{\mmin}{$\mmmin$}
\newcommand{\mmhalo}{M_{\rm halo}}
\newcommand{\mhalo}{$\mmhalo$}
\newcommand{\mrvir}{r_{\rm 200}}
\newcommand{\rvir}{$\mrvir$}
\newcommand{\mrmax}{r_{\rm max}}
\newcommand{\rmax}{$\mrmax$}
\newcommand{\mfhalo}{f_{\rm halo}}
\newcommand{\fhalo}{$\mfhalo$}
\newcommand{\mmcgm}{M_{\rm CGM}^{\rm cool}}
\newcommand{\mcgm}{$\mmcgm$}
\newcommand{\mmbhalo}{M_{b,\rm halo}}
\newcommand{\mbhalo}{$\mmbhalo$}
\newcommand{\mfbb}{f_b}
\newcommand{\fbb}{$\mfbb$}
\newcommand{\mfhb}{f_{b, \rm halo}}
\newcommand{\fhb}{$\mfhb$}

\newcommand{\mfovi}{f_{\rm OVI}}
\newcommand{\fovi}{$\mfovi$}
\newcommand{\mrperp}{R_\perp}
\newcommand{\rperp}{$\mrperp$}
\newcommand{\mdmrperp}{{\rm DM}(\mrperp)}
\newcommand{\dmrperp}{$\mdmrperp$}

\newcommand{\mzfrb}{z_{\rm FRB}}
\newcommand{\zfrb}{$\mzfrb$}

\newcommand{\nhi}{$N_{\rm HI}$}

\newcommand{\mNh}{N_{\rm H}}
\newcommand{\Nh}{$\mNh$}
\def\cm#1{\, {\rm cm^{#1}}}
\def\N#1{N({\rm {#1}})}

\newcommand{\mmsun}{M_\odot}
\newcommand{\msun}{$\mmsun$}

\newcommand{\Novi}{$N_{\rm OVI}$}

\newcommand{\lya}{Ly$\alpha$}

\newcommand{\hi}{\ion{H}{i}}

\newcommand{\ovi}{\ion{O}{vi}}
\newcommand{\ovii}{\ion{O}{vii}}
\newcommand{\civ}{\ion{C}{iv}}
\newcommand{\ciii}{\ion{C}{iii}}

\newcommand{\mgii}{\ion{Mg}{ii}}

\def\rtp{\, \right  ) }
\def\ltp{\left  ( \,}
\def\intl{\int\limits}
\newcommand{\mkms}{{\rm km \, s^{-1}}}

\defcitealias{mcquinn14}{M14}
\defcitealias{mb04}{MB04}
\defcitealias{faerman17}{YF17}

\title[FRBs and Halos]{Probing Galactic Halos with Fast Radio Bursts} 

\author[Prochaska \& Zheng]{
J. Xavier Prochaska$^{1,2}$
Yong Zheng$^{3,4}$
\\
$^{1}$Department of Astronomy \& Astrophysics, UC Santa Cruz, USA \\
$^{2}${Kavli Institute for the Physics and Mathematics of the Universe (Kavli IPMU; WPI),
The University of Tokyo, Japan} \\
$^{3}$Department of Astronomy, UC Berkeley, USA\\
$^{4}$Miller Institute for Basic Research in Science, UC Berkeley, USA
}

\date{Accepted XXX. Received YYY; in original form ZZZ}

\pubyear{2018}

\begin{document}
\label{firstpage}
\pagerange{\pageref{firstpage}--\pageref{lastpage}}
\maketitle

\begin{abstract}
The precise localization ($<1''$) of multiple fast radio bursts (FRBs)
to $z>0.1$ galaxies has confirmed that the dispersion
measures (DMs) of these enigmatic sources afford a new opportunity to
probe the diffuse ionized gas around and in between galaxies.
In this manuscript, we examine the signatures of gas in dark matter
halos (aka halo gas) on DM observations in current and forthcoming
FRB surveys.  Combining constraints from observations of the high
velocity clouds, \ion{O}{VII} absorption, and the DM to the 
Large Magellanic Cloud
with hydrostatic models of halo gas, we estimate that our Galactic
halo will contribute $\mdmmwh \approx 50-80 \mdmunits$
from the Sun to 200\,kpc independent of any contribution  
from the Galactic ISM.
Extending analysis to the Local Group,
we demonstrate that M31's halo will be easily detected by 
high-sample FRB surveys (e.g. CHIME) although signatures from
a putative Local Group medium may compete.  
We then review current empirical constraints on halo gas
in distant galaxies and discuss the implications for their
DM contributions.
We further examine the DM probability distribution function of 
a population of FRBs at $z \gg 0$ 
using an updated halo mass function and new models
for the halo density profile.
Lastly, we illustrate the potential
of FRB experiments for resolving the baryonic fraction of halos
by analyzing simulated sightlines through the CASBaH survey.
All of the code and data products of our analysis are available
at https://github.com/FRBs.
\end{abstract}




\section{Introduction}

Precise measurements of the light element ratios -- Helium/Hydrogen and especially Deuterium/Hydrogen
--
coupled with Big Bang Nucleosynthesis theory have provided a tightly
constrained estimate for the cosmic baryonic mass density 
$\rho_b = \Omega_b \rho_c = 0.044 h_{70}^{-2}$ 
\citep[e.g.][]{bt96,omeara+01,steigman10,cooke18}.
These analyses have been complemented and confirmed by independent
analysis of the cosmic microwave background \citep[CMB;][]{Planck2015}, 
a triumph for
experimental and theoretical astrophysics and cosmology.

With the baryonic cosmic mean established, 
observers have sought
to perform a census of baryons throughout the universe and
across cosmic time \citep{fhp98,pt09}.
At early times ($z \sim 3$), before the growth of substantial
structure, it is generally accepted that the majority of baryons
are in a cool ($T \sim 10^4$\,K), 
diffuse ($n \sim 10^{-5} \, \rm cm^{-3}$) plasma
that fills the space between galaxies, aka 
the intergalactic medium  \citep[IGM; e.g.][]{sargent80,mco+96}.
Observationally, this plasma gives rise to the so-called \ion{H}{I}
\lya\ forest in the spectra of high-$z$ sources \citep{rau98}.
When combined with estimates
of the extragalactic UV background (EUVB),
analysis of the optical depth of the \lya\ forest indicates that 
$\gtrsim 90\%$ of the baryons reside in the IGM \citep{rau98,fpl+08}.
Indeed, researchers now invert the experiment to leverage 
\lya\ forest observation and cosmological simulation 
to assess the EUVB and other cosmological parameters
\citep[e.g.][]{flh+08,palanque+13}.

Running the clock forward, dark matter collapses into galactic halos
and into larger structures bringing baryons along with it.
As this gas streams into a halo, it is predicted to 
shock-heat to the virial temperature (e.g.\ $T \sim 10^6$\,K for 
halos with mass $\mmhalo = 10^{12} \mmsun$)
and yield a circumgalactic medium (CGM) of hot, diffuse gas.
A fraction $\sim 10\%$ of this gas cools
and drives galaxy formation near the center of the halo.
By $z \sim 0$,  the population of 
dark matter halos with $\mmhalo \ge 10^{10} \mmsun$
are predicted to contain $\sim 35$\%\ 
of the all dark matter and, potentially, a similar fraction of 
the baryonic mass.
Of principal interest to this paper, and future studies by the
authors, is to measure the mass fraction of gas within halos.

At the highest mass -- galaxy clusters -- X-ray observations
reveal a hot, virialized plasma referred to as the intracluster
medium (ICM) which has sufficient mass to nearly close 
the baryon census within the structure, i.e.\ 
$M_{\rm ICM}/\mmhalo \approx \Omega_b/\Omega_{\rm m} = 0.158$ \citep{allen02}.
Stepping down in mass to galaxy groups ($\mmhalo \sim 10^{13} \mmsun$)
and individual galaxies ($\mmhalo \sim 10^{12} \mmsun$),
the X-ray experiment becomes increasingly difficult to perform
due to both the lower masses and virial temperatures.  
Current lore based on 
these data is that such halos are deficient in baryons relative
to the cosmic mean  \citep[e.g.][]{dai10},
i.e.\ $M_b/\mmhalo < \Omega_b/\Omega_{\rm m}$
for halos with $\mmhalo < 10^{14} \mmsun$ 
where $M_b$ is the total baryonic mass of the system.
We emphasize, however,  that these X-ray 
experiments have insufficient sensitivity to precisely assess
the baryonic mass fraction
$f_b \equiv M_b/\mmhalo$ in lower mass halos \citep{anderson13,li+18}.

Nevertheless,
there are also persuasive arguments from galaxy formation theory
that lower mass halos have $\mfhb \ll \Omega_b/\Omega_{\rm m}$.
For example, attempts to place galaxies within dark matter halos
suggest that $L \ll L^*$ galaxies comprise a very small
fraction of the available baryons \citep{moster+10}.  
One possibility is that the gas streams onto
the galaxy and is then ejected from the system prior to 
forming stars. 
Such feedback is frequently invoked to match semi-analytic models
or computer simulations of galaxy formation to observed luminosity
functions \citep[e.g.][]{somerville+15,muratov15,christensen18}. The majority of these models even envision the gas escaping the halo
to pollute the surrounding IGM \citep[e.g.][]{shen+14}.
At high $z$, there is evidence for this scenario \citep{booth+12},
but empirical confirmation at low $z$ is lacking \citep{pwc+11}.

Given the insufficient sensitivity of X-ray observations to 
L* galaxies (much less lower-mass systems),
researchers have had to pursue alternate approaches to place constraints on 
$f_b$ for galactic halos.  A long-standing technique has been to apply
similar techniques used to assess the IGM, i.e.\ absorption-line 
analysis of sightlines that coincidentally intersect galactic
halos \citep[e.g.][]{lbt+95,pwc+11,tumlinson+13}.
These surveys have demonstrated that galaxies which
show a great diversity
of mass and star-formation history all exhibit a substantial, cool
($T \sim 10^4$\,K) plasma in their halos.  This plasma is manifest as 
\ion{H}{I} Lyman series lines and lower ionization transitions of
heavy elements \citep[e.g.][]{werk+13}.  
Assuming this cool CGM is photoionized by the
EUVB (or local sources), researchers have crudely estimated the
cool gas mass $M_{\rm CGM}^{\rm cool} \sim 10^{10} - 10^{11} \mmsun$
implying a significant baryon fraction from cool gas alone,
e.g.\ $\mfhb > 0.1 \, \Omega_b/\Omega_{\rm m}$
\citep{stocke13,werk+14, keeney17}.
However,
the substantial uncertainty in this estimate, driven by modeling
assumptions and the small sample sizes, has sparked substantial
debate as to whether $L*$ galaxies are missing baryons.
Furthermore, the community awaits results from ongoing surveys to
estimate \mcgm\ for sub-$L*$ and dwarf galaxies \citep[e.g.][]{bordoloi14}.

The same absorption-line datasets that probe the cool CGM
also offer measurements of high-ions, especially O$^{+5}$.
High column density measurements of O$^{+5}$ 
have revealed
a more highly-ionized plasma, interpreted by many as a 
tracer of the predicted $T \gtrsim 10^6$\,K virialized halo gas
\citep[e.g.][]{beno+16,faerman17,mp17}.
Estimating the mass of this putative, hot-phase is even more
challenging because it is difficult to assess the degree
of \ion{H}{I} absorption associated with it and one rarely
has access to neighboring ions of the same element
to constrain the ionization state of the gas.
Put simply, far-UV observations are limited in their ability
to trace a $T > 10^6$\,K plasma \citep[but see][]{savage+11b,burchett18}.  
And, while X-ray absorption-line spectroscopy of transitions like
\ion{O}{VII} ($\lambda=21$\AA) offer promise \citep{fang+15,nicastro18}, 
current technology is insufficient
for statistically meaningful conclusions beyond
our Galaxy.

A promising, and still developing, complementary technique is to
search for the Sunyaev-Zeldovich (SZ) signal from the hot gas in halos.
The all-sky Planck experiment has enabled analysis of halos with 
$\mmhalo \ge 10^{13.3} \mmsun$ at $z \sim 0$ \citep{planckXI}.
Their results suggest that these halos are `closed', i.e. they
contain approximately the cosmic fraction of baryons for their mass.
Estimates for lower mass halos (i.e.\ galactic halos) is beyond
current SZ sensitivity and are further challenged by the poor
spatial resolution and one's ability to precisely select a sample
of such halos \citep{hill18}.

Recently \citep{chatterjee17,tendulkar17},
it was confirmed that a new technique had emerged to constrain
the distribution of baryons in the universe: 
survey the dispersion measure (DM) of distant sources 
-- the fast radio bursts (FRBs).  
Provided a short and coherent burst of radiation, the plasma 
it travels through imposes an index of refraction that
retards the group velocity as a function of frequency \citep{hirata14}.
The precisely measured DM value, from observations of the photon arrival
time versus frequency, is given by

\begin{equation}
    {\rm DM} = \int \frac{n_e \, ds}{1+z} \;\;\; ,
    \label{eqn:DM}
\end{equation}
which offers an integral constraint on the electron distribution
along the path between the source and Earth.  This includes the 
IGM \citep{inoue04,zheng14,shull18}, our Galaxy \citep{ne2001a}, our Local Group,
the galaxy hosting the FRB \citep{xu15},
and the baryons residing in other galactic halos 
near the sightline \citep[][hereafter \citetalias{mcquinn14}]{mcquinn14}.
With the confirmation that FRBs are extragalactic in origin,
the door has opened for an entirely new approach
to assessing the baryonic distributions of the IGM 
and dark matter halos.


In this manuscript, we examine several aspects of using
FRB observations to constrain the nature of baryons in the CGM.
Our emphasis, in contrast to previous work (\citetalias{mcquinn14}),
is largely empirical, i.e.\ we examine the scientific potential of FRBs
in the context of modern CGM observations and analysis.  
We are also motivated, in part, to rectify a number of 
misconceptions in the literature on what is known 
(and not known!) about gas in dark matter halos.
Further, we examine our own Galaxy and its halo
in detail including gas from the Local Group.  Lastly,
we formulate several experiments motivated by upcoming FRB 
surveys and discuss follow-up strategies that may optimize
constraints.

This paper is organized as follows.  Section~\ref{sec:ism} briefly comments
on the contributions to DM from the interstellar medium (ISM) of galaxies
and is followed by section~\ref{sec:models} which introduces a set of
models for the gas distributions of halos.
In Section~\ref{sec:Galaxy}, we detail current constraints on the distribution
of ionized gas in our Galactic halo and its DM contribution to FRBs.
Section~\ref{sec:LG} extends the discussion to the Local Group
with emphasis on M31 and the Magellanic Clouds.
In Section~\ref{sec:exgal} we consider gas from halos in the distant
universe and the typical DM contribution from the most massive halos.
Lastly, Section~\ref{sec:discuss} offers a discussion of several illustrative
examples of DM distributions that may be revealed by 
ongoing and forthcoming FRB surveys. And we conclude in Section~\ref{sec:sum}.
Throughout, we use the Planck15 cosmology as encoded in {\sc astropy}.
This paper also makes extensive use of the halo mass
function code 
Aemulus\footnote{https://github.com/tmcclintock/Aemulus\_HMF} developed 
and kindly distributed by T. McClintock and J. Tinker. 

\section{A Brief Section on the ISM}
\label{sec:ism}

This paper focuses on diffuse gas that lies beyond a galaxy's interstellar
medium (ISM).  This means the gas in galactic halos (aka the CGM)
and the gas in between halos (aka the IGM).
We recognize, however, that free electrons
in the ISM of galaxies -- including our own -- will contribute 
signatures to the observations of FRBs.  On this topic,
we refer the reader to the excellent NE2001 model\footnote{
Ported from FORTAN to Python by Baror \& Prochaska; 
https://github.com/FRBs/ne2001}
developed by \cite{ne2001a,ne2001b} from observations of pulsars in
and around our Galaxy.  
Where relevant, we evaluate the Galactic ISM contribution to DM
using this model and refer to it as \dmism\footnote{There are additional
models of the Galactic ISM \citep{gaensler08,YMW17} but we recommend the
NE2001 model with the updates included in the Python distribution.}. Note that the \dmism\ values from NE2001 include free electrons associated with the warm ionized medium (WIM) which extends a few kpcs beyond the Galactic disk \citep[e.g.][]{reynolds91, sembach+00}. The velocities of the \hi\ counterpart of WIM are typically at $40 \lesssim |v_{\rm LSR}|\lesssim$ 100 km s$^{-1}$ -- the so-called intermediate velocity gas \citep{albert04, wakker04}. To avoid double counting, in \S\ \ref{sec:hvc} we refrain from this velocity range and only focus on the high-velocity gas ($|v_{\rm LSR}|>100$ km s$^{-1}$) when evaluating DM contribution of the cool Galactic halo. Regarding the ISM of distant galaxies intercepting FRB sightlines, we point the interested
reader to \cite{pn18}, who demonstrate that intervening
galaxies are unlikely to have substantial impact on
FRB observations.

\section{Halo Models}
\label{sec:models}

In this section, we consider several models for the 
distribution of halo gas in dark matter halos and describe
several general implications for DM measurements with FRBs.  
For convenience and clarity, we list in Table~\ref{tab:quantities}
quantities referred to throughout the paper.

\begin{table}
\centering
\caption{List of Quantities Referenced in the Paper}\label{tab:quantities}
\begin{tabular}{@{}cl@{}}
\hline
Quantity & Description \\
\hline
\multicolumn{2}{c}{Halo Properties} \\
\hline
\mhalo\ & Total halo mass (within \rvir) \\
\rvir & Virial radius \\
$M_b$ & Total baryonic mass in the halo \\
\mbhalo & Diffuse halo gas mass (excludes stars, ISM) \\
\fbb & Total baryonic mass fraction, $M_b/\mmhalo$ \\
\fhb & Halo gas mass fraction, $\mmbhalo/\mmhalo$ \\
\rmax & Adopted physical extent of the halo \\
\hline
\multicolumn{2}{c}{DMs} \\
\hline
\dmfrb & Total DM measurement of an FRB \\
\dmism & DM of the Galactic ISM \\
\dmhvc & DM of cool gas in our Galactic halo \\
\dmmwh & DM of all gas in our Galactic halo \\
$\dm{LMC}$ & Total DM measured to the LMC \\
$\rm DM_{\rm LMC}^{\rm halo}$ & DM estimate for Galactic halo gas to the LMC \\
$\dm{LGM}$ & DM of the Local Group medium  \\
$\dm{LG}$ & Total DM from all Local Group contributions \\
$\dm{halo}$ & DM from a galaxy halo \\
\dmhalos & DM from an ensemble of galaxy halos \\
\dmigm & DM from the IGM (gas between halos) \\
\dmcosmic & DM from all cosmic gas for an event \\
\dmacosmic & Average DM from all cosmic gas (IGM+halos) \\
\dmhost & DM from the FRB host galaxy \\
\hline
\end{tabular}
\end{table}

Since the pioneering work of \cite{nfw97}, cosmologists have established
a paradigm for the density distribution of the dark matter
in collapsed halos: the NFW profile

\begin{equation}
\rho(r) = \frac{\rho_0}{y (1+y^2)} \;\; ,
\label{eqn:NFW}
\end{equation}
where $y \equiv c (r/\mrvir)$, $c$ is the concentration
parameter, and \rvir\ is the virial radius, defined
here as the radius within which the average density is
200 times the critical density\footnote{This widely adopted
definition has the unfortunate complication of coupling
\rvir\ to the expansion of the universe, i.e.\ 
$\rho_c \propto (1+z)^3$.
Our future analyses may instead consider the so-called psuedo-\rvir\
\citep[e.g.][]{dimer14}.
},  $\rho_c \equiv 3 H^2 / 8 \pi G$.
While difficult to test observationally,
the NFW profile has withstood numerous numerical experiments,
and constraints from lensing experiments offer solid 
observational support 
\citep[at least for more massive halos; e.g.][]{slacsVII,sonnenfeld18}.  
Numerical studies, meanwhile, show a relation between the
concentration, halo mass, and redshift,

\begin{equation}
c_{200} = 4.67 (M_{200} / 10^{14} h^{-1} \mmsun)^{-0.11} \;\; ,
\label{eqn:c}
\end{equation}
which we adopt throughout.

\begin{figure}
	\includegraphics[width=\columnwidth]{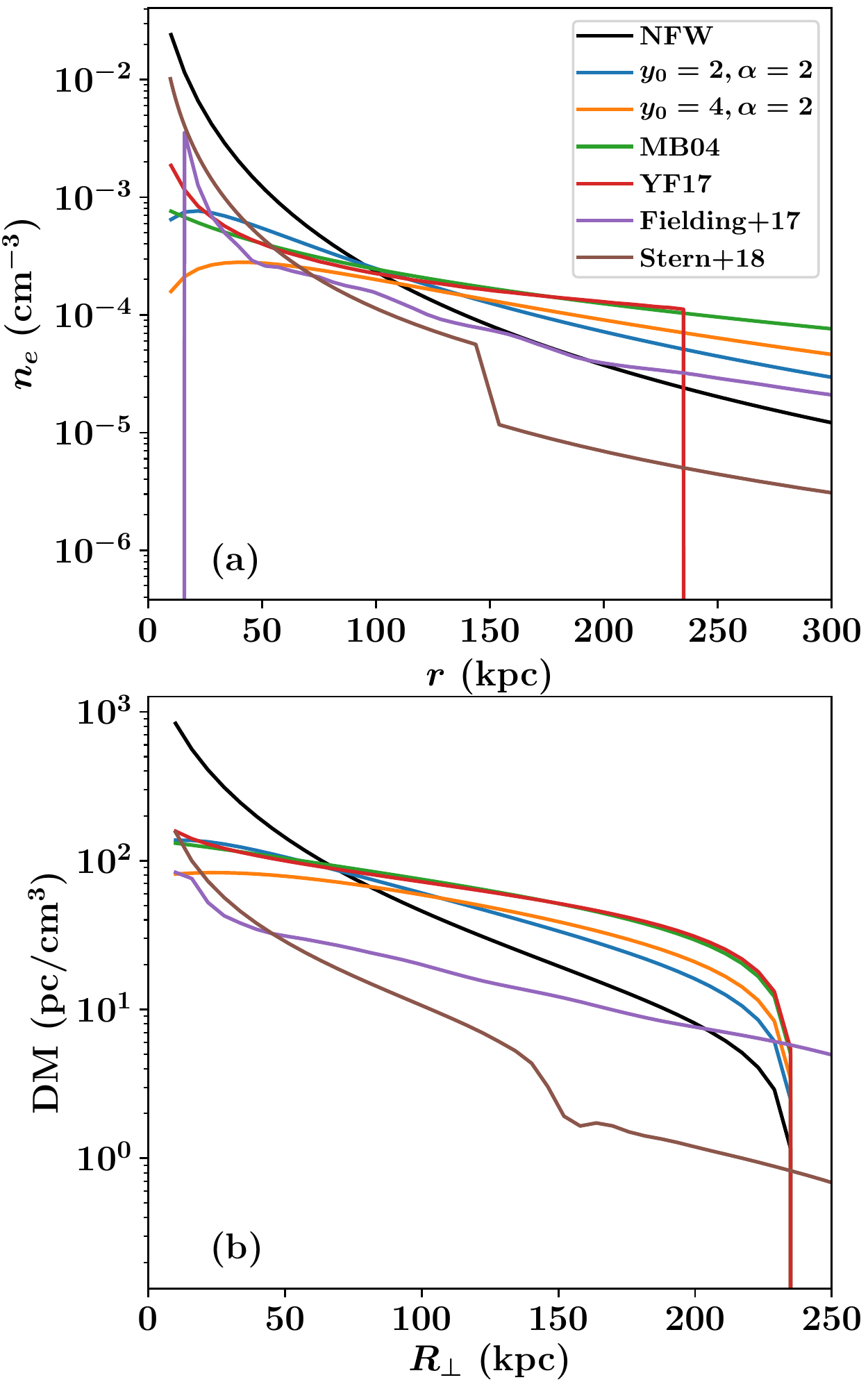}
    \caption{(a): Electron density profiles for a series of models
    which might describe the ionized, halo gas in a galaxy like
    our Milky Way.  Most of the models have been scaled to a halo mass of
    $\mmhalo = 1.5 \times 10^{12} \mmsun$ ($\mrvir = 236$\,kpc)
    and an assumed halo gas fraction $\mfhb = 0.75$.
    See \S\ \ref{sec:models} for additional details.
    (b): Integrated dispersion measures \dmrperp\ for sightlines 
    passing through the halo at a series
    of impact parameters \rperp.  The evaluation of Equation~\ref{eqn:DM_Rperp}  
    assumes $r_{\rm max} = \mrvir$.
    \label{fig:cgm_model}
    }
\end{figure}
The simplest model for baryons in the halo, therefore, is
for them to trace
the underlying dark matter distribution.  This is depicted
with the black curve 
in Figure~\ref{fig:cgm_model} for 
a dark matter halo
with $\mmhalo = 1.5 \times 10^{12} \mmsun$ and concentration
parameter $c=7.7$ which are characteristic values for our Galaxy.   
The total baryonic mass within the dark
matter halo is given by 
$M_b = \mfhb \mmhalo (\Omega_b/\Omega_{\rm m})$, where 
\mhalo\ is the integral of Equation~\ref{eqn:NFW}
to \rvir\ and \fbb\ is the fraction of cosmic baryons
in the halo.  
We further define
$\mmbhalo = \mfhb \mmhalo (\Omega_b/\Omega_m)$
as the mass of diffuse baryons in the halo (i.e.,
ignoring stars and the dense ISM) and \fhb\ 
specifies the mass fraction. 
As a default, we will assume $\mfbb=1$ and
a fiducial value of $\mfhb = 0.75$. 
This presumes that the system has
retained the cosmic mean of baryons and that
25\%\ of these baryons
are in the galaxy as stars, collapsed objects, and ISM \citep[e.g.][]{fhp98}
which contribute a negligible number of free electrons. 
Of course, these fractions may well vary with
halo properties \citep[e.g.][]{behroozi10}.
For sightlines intersecting dark matter halos, the 
quantity of greatest relevancy to FRBs is the dispersion
measure profile \dmrperp, i.e.\ the DM value
recorded as a function of impact parameter \rperp\ to the
center of the halo.
The \dmrperp\ profile 
(shown in Figure~\ref{fig:cgm_model}b) is defined as
\begin{equation}
\mdmrperp = 2 \intl_0^{\sqrt{r_{\rm max}^2 - \mrperp^2}} n_e ds \;\;\; ,
\label{eqn:DM_Rperp}
\end{equation}
where $r_{\rm max}$ is the maximum radius of integration through the
halo, typically taken to be \rvir, and 
\begin{equation}
n_e = \mu_e \frac{\rho_b}{m_p \mu_{\rm H}}
\label{eqn:ne}
\end{equation}
with $\mu_{\rm H} = 1.3$ the reduced mass (accounting for Helium)
and $\mu_e = 1.167$ accounts for fully ionized Helium and Hydrogen. 
Corrections for heavy elements are negligible.
The steeply rising slope of the
NFW density profile lends to very high DM
values at $\mrperp < 50$\,kpc.  Indeed, it is argued that
such a high gas density 
is unsustainable because the implied
cooling time is so short that the gas would rapidly condense
onto the ISM \citep[e.g.][]{mm96}.
As such, this model for the gas profile is greatly disfavored.

\citeauthor{mb04} (2004; hereafter \citetalias{mb04}) proposed a modified density profile 
for halo gas
based on the hydrodynamic relaxation of the gas 
predicted by numerical
work \citep{frenk99} and an additional treatment
of metal-line cooling.
Figure~\ref{fig:cgm_model}b shows the \dmrperp\ profile based on the
\citetalias{mb04} density distribution using the same underlying dark matter
halo and a cooling radius $R_c = 147$\,kpc (the results are largely
insensitive to this parameter).   
The obvious distinction from the NFW profile is the greatly
reduced gas density at small radii (Figure~\ref{fig:cgm_model}a) and the concomitant increase
in \dmrperp\ at $\mrperp > 100$\,kpc which conserves 
the total baryon mass.
The \citetalias{mb04} profile has the added `benefit' that it is consistent
with the electron column density estimate of our halo
towards pulsars in the LMC (see \S\ \ref{sec:Galaxy})
and also Galactic X-ray emission \citep{fang+13}.

Alternative versions of the halo gas profile have also
been derived in the context of feedback which modifies
the distribution away from the NFW profile by altering the
entropy distribution of the gas. 
\cite{mp17} introduced such a model to investigate the 
observed profile of \Novi\ absorption in present-day
$L^*$ galaxies. 
We generalize their modified NFW (mNFW) model as,
\begin{equation}
\rho_b = \frac{\rho_b^0}{y^{1-\alpha} (y_0 + y)^{2 + \alpha}} \;\; ,
\end{equation}
and show two examples in Figure~\ref{fig:cgm_model} for our
fiducial dark matter halo. One notes that the mNFW models with $\alpha=2$, $y_0=2$ and $\alpha=2$, $y_0=4$ show similar DM profiles as \citetalias{mb04}; throughout the manuscript we adopt the mNFW model with $\alpha=2$, $y_0=2$ as the default.

In a related work, \citeauthor{faerman17} (2017; hereafter \citetalias{faerman17}) synthesized all of the available
constraints on the Galactic halo and
developed a phenomenological model 
comprised of warm ($T \sim 10^6$\,K) and hot ($T \sim 10^7$\,K)
phases. Summing the density profiles of these two phases
and scaling the integrated mass to our fiducial 
($\mmhalo = 1.5 \times 10^{12} \mmsun$),
we recover the DM profile shown in Figure~\ref{fig:cgm_model}b.
Tellingly, the \citetalias{faerman17} profile also tracks the DM curves of \citetalias{mb04} 
and the mNFW models; all impose hydrostatic equilibrium
and these models also predict the gas 
in $L*$ galaxy halos is predominantly ionized by collisions
(i.e.\ $T \gg 10^4$\,K). An alternate scenario for the physical nature of halo gas 
asserts that virialization is limited to radii $r < \mrvir$
and that the gas at larger radii is cool and photoionized \citep{stern+16,stern+18}.
The density and \dmrperp\ curves, calculated for a 
$\mmhalo = 1.5 \times 10^{12} \mmsun$ halo, are shown
in Figure~\ref{fig:cgm_model}. 
One notes a break in the density profile as one
transverses the virial shock. This model shows 
a systematically lower \dmrperp\ profile than the
rest. Furthermore, it predicts a much smaller 
\fhb\ fraction than the other models.

Lastly, we present the DM profile based on the numerical 
simulations of \cite{fielding+17} by scaling their $10^{12} \mmsun$ halo (with $\eta=2$) to $\mmhalo = 1.5 \times 10^{12} \mmsun$.
This profile is steep in the inner halo, nearly tracking
the shape of the NFW profile.  It also has a low \fhb\ value,
and systematically lower DM values than the models with
hydrostatic equilibrium.
Ultimately, we wish to test whether these predictions
from modern galaxy formation theory offer a better
description than the simpler models of Figure~\ref{fig:cgm_model}.

\begin{figure}
	\includegraphics[width=\columnwidth]{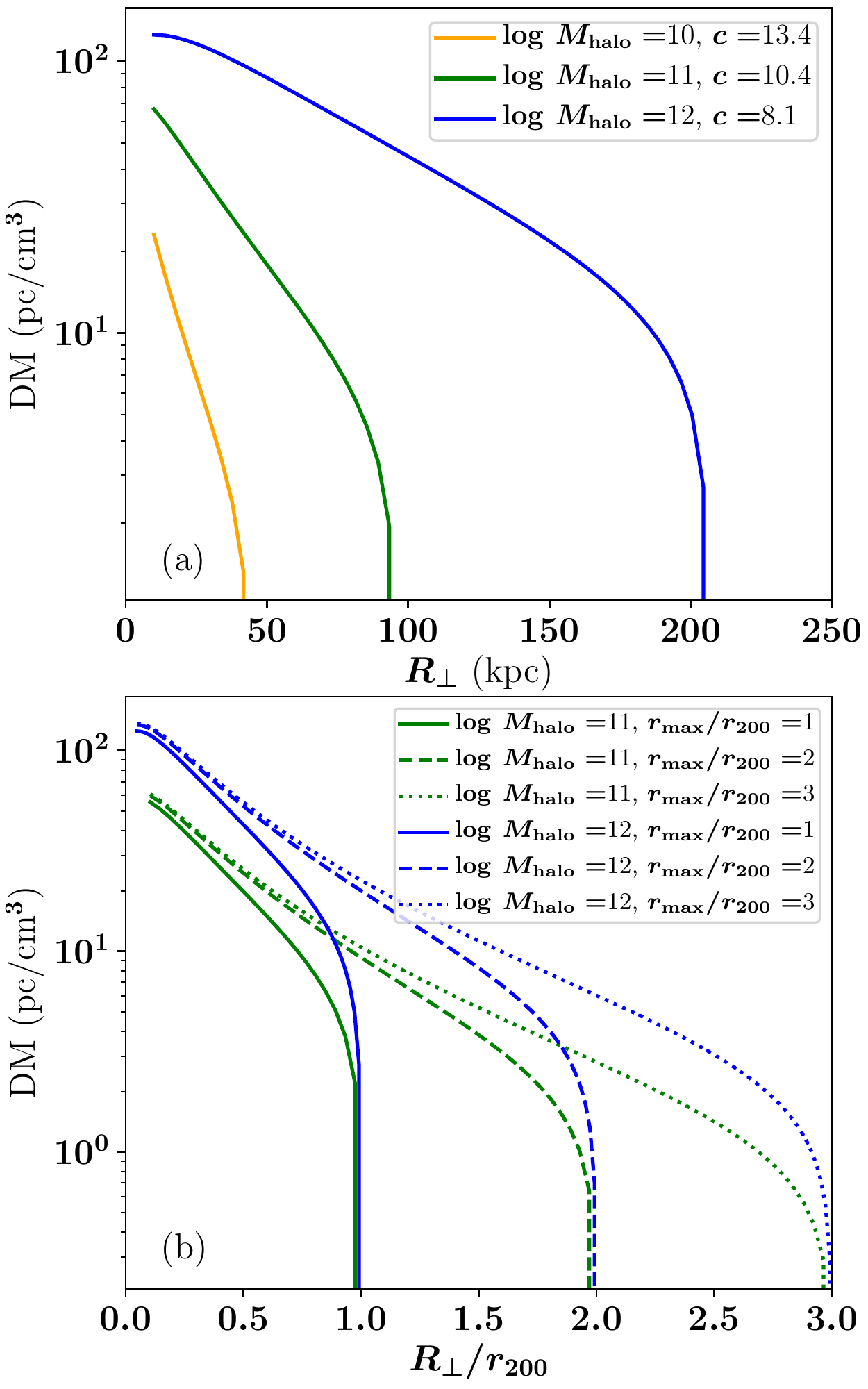}
    \caption{(a) DM profiles for halos with varying mass and concentration ($c$)
    assuming the modified NFW density profile ($y_0=2, \alpha=2$)
    and with Equation~\ref{eqn:DM_Rperp} evaluated to $r_{\rm max} = \mrvir$.
    (b) \dmrperp\ profiles for halos with $\log M_{\rm halo} = [11,12]$
    and with $n_e$ integrated to increasing values of $r_{\rm max}$.
    It is clear the outer portion of these halos can contribute tens
    of \dmunits\ and, therefore, defining the `edge' of a halo is
    critical to evaluating its contribution.
    \label{fig:vary_halos}
    }
\end{figure}

It is also illustrative to examine the \dmrperp\ 
profiles for halos with a range of properties.
In Figure~\ref{fig:vary_halos} we show results after
varying the dark matter halo mass ($\mmhalo = 10^{11} - 10^{13} \mmsun$)
and concentration using the mNFW model 
($y_0=2, \alpha=2$) and assuming
$\mfhb = 0.75$ throughout\footnote{
There are predictions
from galaxy-formation theory that \fhb\ decreases with decreasing
\mhalo \citep{hafen+18};  if desired, one can scale down most of
the results linearly with \fhb.}.
Each of the solid curves in Figure~\ref{fig:vary_halos}a
shows the integration of DM to $r_{\rm max} = \mrvir$.
Clearly, the DM `footprint' of a $10^{10} \mmsun$
halo is small,  and while such halos are predicted to
be $\approx 50$ times more common than $\approx 10^{12} \mmsun$ 
halos (per logarithmic mass bin), 
these may not dominate the integrated DM of long 
pathlengths through the universe (e.g. \citetalias{mcquinn14}). 
We return to this point later in the manuscript.

We also emphasize that the sharp cutoff in the DM 
profiles in Figure~\ref{fig:vary_halos}a, which occurs
at $\mrperp = \mrvir$, is largely artificial.
While all of the models assume that the density profile 
at large radii falls off steeply ($\rho_b \propto r^{-3}$ for
most cases),
the virial radius does not demarcate a physical edge to the halo.
Extending the integration of Equation~\ref{eqn:DM_Rperp}
from $r_{\rm max} = \mrvir$ to $2 \mrvir$
significantly increases 
DM at impact parameters $\mrperp > 0.5 \mrvir$ as
illustrated in Figure~\ref{fig:vary_halos}b.
At $\mrperp = 1.5 \mrvir$, the DM increases
from a negligible value to several tens \dmunits\ for the
$10^{12} \mmsun$ halo.
Even in galaxy formation models
that predict a majority of baryons are lost from the inner
regions of halos,
the gas is generally located within a few \rvir\ \citep{dimer14}.
Therefore, it will still contribute significantly to the DM of that halo.
Any discussion on probing 
halo gas with FRBs needs to precisely define a halo and its
`sphere of influence'.

\section{DM of the Galactic Halo}
\label{sec:Galaxy}

Here we examine the key ionized structures 
that a FRB sightline may intercept in our Galactic halo
giving \dmmwh. 
Our analysis goes as a function of distance from the Sun, starting with the nearby ionized/neutral high-velocity clouds (HVCs), and
the extended hot Galactic halo. 
We defer the analysis of the Magellanic Clouds to the following section.


\subsection{The Cool Galactic Halo} 
\label{sec:hvc}

Since the discovery of high velocity clouds (HVCs; $|v_{\rm LSR}|\gtrsim100$ km s$^{-1}$) in the 21\,cm emission-line in 1960s \citep{verschuur75, Wakker91, vanwoerden04}, the neutral hydrogen residing in the MW disk and in the inner Galactic halo has been extensively mapped. The seminal Leiden/Argentine/Bonn (LAB) survey 
\citep{LAB05} provides an all-sky \ion{H}{I} 21\,cm emission map at 
an angular resolution of $36'$, and a decade later this map is superseded by the HI4PI survey \citep{hi4pi} with an angular resolution of $16.2'$ 
and higher sensitivity. In recent years, significant attention has been given to 
the ionization state of HVCs and sightlines that pass close to
them \cite[e.g.][]{foxetal06, lh10}.
Ionized HVCs detected in a series of ionic
species (e.g., Si$^+$, Si$^{++}$) 
have been found ubiquitous over the Galactic sky 
with covering fractions typically larger than 70\%\ 
\citep{Shull09, collins09, lehner+12, richter+17}. 
While the majority of this neutral and ionized gas is now known to arise from complexes that are relatively near the Galaxy at a 
few tens kpc \citep[e.g.][]{gibson01, thom08}, 
the HVCs nevertheless contribute significantly to the column density of gas along any sightline. Here we estimate the contribution of Galactic electrons associated with neutral/ionized HVCs to the DM.



\begin{figure}
	\includegraphics[width=\columnwidth]{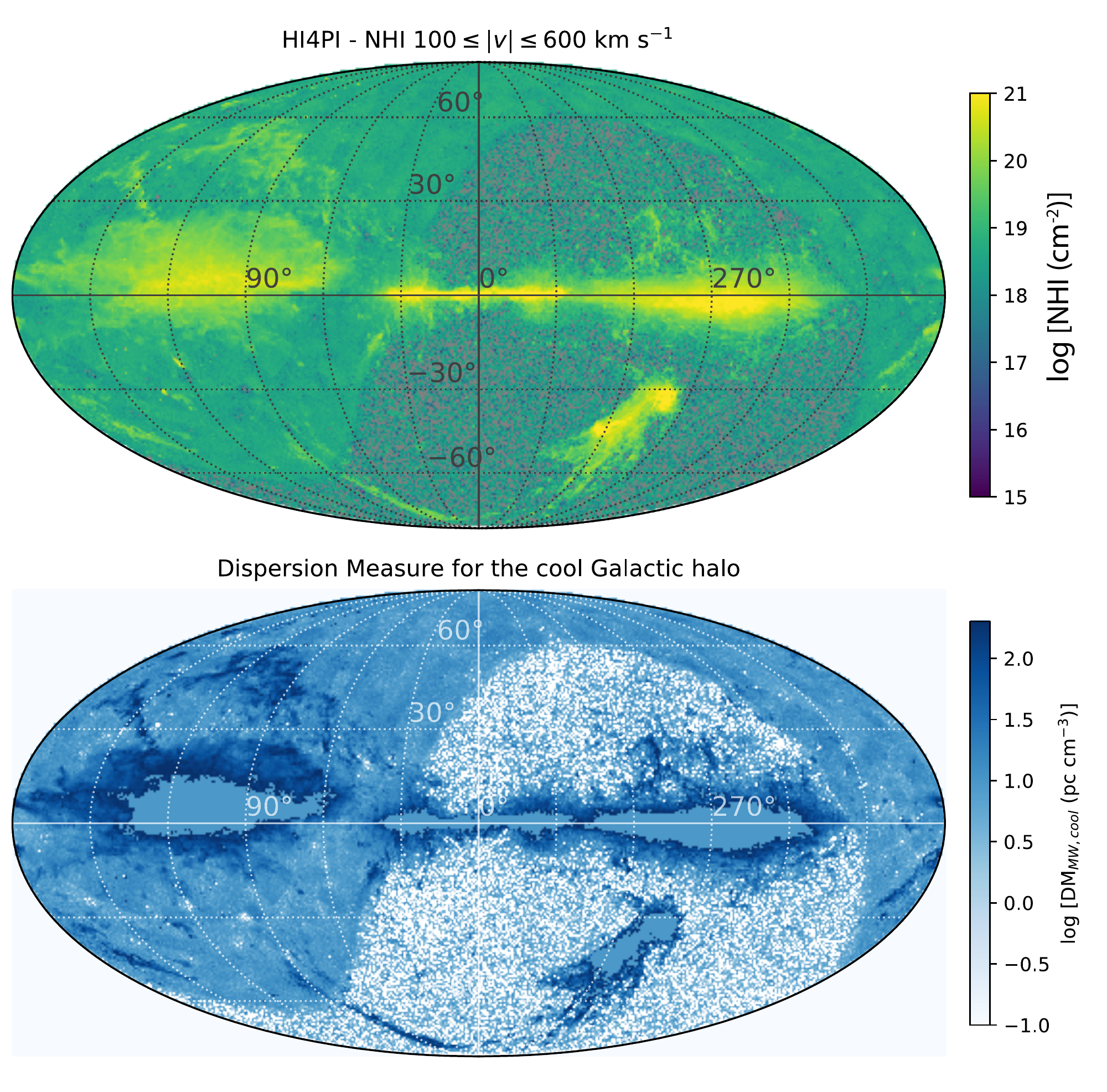}
    \caption{Top: All-sky \ion{H}{I} column density integrated over $100\leq|v_{\rm LSR}|\leq600$ km s$^{-1}$ in Mollweide projection with Galactic coordinates. The map is generated with the HI4PI dataset \citep{hi4pi}; it is the same as figure 2 in \citep{hi4pi}, with the Galactic emission ($|v_{\rm LSR}|<100$ km s$^{-1}$) removed. Since HI4PI combines the EBHIS survey in the north (DEC $>-5$ deg; \citealt{ebhis}) and the GASS survey in the south (DEC $<1$ deg; \citealt{gass}), the overlap region of the two surveys results in a clear boundary near the Equatorial plane. The difference in the two hemispheres is due to different noise characteristics and scan strategies: since the EBHIS survey was convolved from 10.8' to 16.2' to be consistent with GASS, the equatorial northern sky appears to be smoother. The $1\sigma$ value for the adopted velocity range is $\sigma_{\rm N_{\rm HI}}=3\times10^{18}$ cm$^{-2}$. 
    Bottom: Estimated DM for the MW cool halo (\dmhvc), as converted from the $N_{\rm HI}$ map in the top panel using Equation \ref{eqn:nhi_to_dm}. We mask the emission near the Galactic plane and 
    that related to LMC/SMC/M31/M33, and replace the pixels with the median value ($\sim4.3\times10^{18}$ cm$^{-2}$) from the rest of sky. Over the high-velocity sky, the mean \dmhvc\ value is $\sim$20 \dmunits. See \S\ \ref{sec:hvc} for further detail.
    \label{fig:hvc}
    }
\end{figure}

The external (i.e., beyond the disk) \ion{H}{I} can be isolated from Galactic emission using velocity cuts. 
For our purposes, we adopt $100\leq|v_{\rm LSR}|\leq600$ km s$^{-1}$ to define the external, high-velocity \ion{H}{I} that contributes to the line-of-sight DM. We make use of the all-sky HI4PI \citep{hi4pi} dataset, and produce a high-velocity \ion{H}{I} column density map as shown in Figure \ref{fig:hvc}. We note that this velocity definition of HVCs only coarsely separate the Galactic and halo \hi. The emission at $|b|\lesssim15$ deg is Galactic \ion{H}{I} 
shifted into the high-velocity regime due to differential 
rotation of the Galaxy \citep{Wakker91}. 
Other than this, we find the sky at $|b|\gtrsim15$ degree is well
covered by diffuse neutral hydrogen 
with $N_{\rm HI}\sim10^{19}$ cm$^{-2}$. 
The Magellanic System (i.e., LMC/SMC and the \ion{H}{I} stream; 
\citealt{putman03, nidever08, nidever10}) can be seen with $N_{\rm HI}\gtrsim10^{20}$ cm$^{-2}$, together with some bright HVC structures such as Complex C ($\ell\sim90, b\sim45$), K ($\ell\sim60, b\sim30$), A ($\ell\sim150, b\sim30$), and WD ($\ell\sim270, b\sim20$) \citep{wakker01, wakker04, vanwoerden04}.

For the assessment of \dmhvc\ we must consider 
the ionized gas associated with HVCs. We adopt an ionization fraction of $x_{\rm HI} = 0.3$ based on mass estimates for H\,I and ionized HVCs in the literature. \cite{putman12} find a total mass of $M_{\rm HI, HVC}=2.6\times10^7 \mmsun$ excluding the Magellanic System, while estimates for ionized HVCs range from $M_{\rm HII, HVC}=4.3\times10^7 \mmsun$ \citep{richter+17} to $1.1\times10^8 \mmsun$ \citep{lh11}. Therefore, we find $x_{\rm HI}\equiv M_{\rm HI}/(M_{\rm HI}+M_{\rm HII})\approx0.2-0.4$ and adopt the mean. \cite{fox+14} provide mass estimates for the H\,I and ionized components of the Magellanic Stream with $M_{\rm HI, MS}=4.9\times10^8\mmsun$ and $M_{\rm HII, MS}=1.5\times10^9 \mmsun$, which indicates that the $x_{\rm HI}$ for the Magellanic Stream is $\sim0.25$, similar to the non-Magellanic HVC estimate. We note that the $x_{\rm HI}$ value may vary significantly from sightline to sightline (e.g., \citealt{fox+05, howk06}), the mean value adopted here provides a bulk estimate of the contribution of the Galactic ionized hydrogen to DM.

With $x_{\rm HI}=0.3$, we may then convert the $N_{\rm HI}$ map into an electron column density map with 
\begin{equation}
N_{\rm e, cool} \equiv N_{\rm H^+, cool} = 
  \mu_e N_{\rm HI, HVC} \ltp \frac{1-x_{\rm HI}}{x_{\rm HI}} \rtp 
  \ltp \frac{1}{f_{\rm HVC}} \rtp \;\;. 
\label{eqn:nhi_to_dm}
\end{equation}
In this equation, we include $f_{\rm HVC}$ to correct 
for the cool halo gas at low velocities that is not included in the HVC $N_{\rm HI}$ map in the top panel. We adopt $f_{\rm HVC}=0.4$ based on a synthetic observation of the halo gas velocity field from a simulated MW galaxy \citep{zheng+15}. 
In the bottom panel of Figure~\ref{fig:hvc}, we show the 
\dmhvc\ map of HVCs converted from the $N_{\rm HI}$ map. As noted above, the $N_{\rm HI}$ HVC map shows prominent emission from the Galactic plane, Magellanic Clouds, and M31/M33; we mask the cores of these features that have $N_{\rm HI}>10^{20}$ cm$^{-2}$, and replace the masked pixels with the median value ($N_{\rm HI, med}=4.3\times10^{18}$ cm$^{-2}$) from the rest of sky. Overall, we find that only $\sim20$\% of the sky (by area) has \dmhvc$\gtrsim20$ \dmunits, another $\sim20$\% with \dmhvc$\sim10-20$ \dmunits, and the rest of the sky with \dmhvc$\lesssim10$ \dmunits. Taking the mean of the DM map in the bottom panel of Figure \ref{fig:hvc} finds \dmhvc$\approx20$ \dmunits. 
Note that the DM values become significant near regions with dense \ion{H}{I} structures; 
this is due to the uniform $x_{\rm HI}$ factor we assume in Equation \ref{eqn:nhi_to_dm}, 
although one may expect denser \ion{H}{I} structures to be more neutral.
Nevertheless,
we conclude that the diffuse halo gas at cool ionized phases with high surface density ($T\sim10^4$ K) does not contribute greatly to the 
line-of-sight electron density. 

\begin{figure}
	\includegraphics[width=\columnwidth]{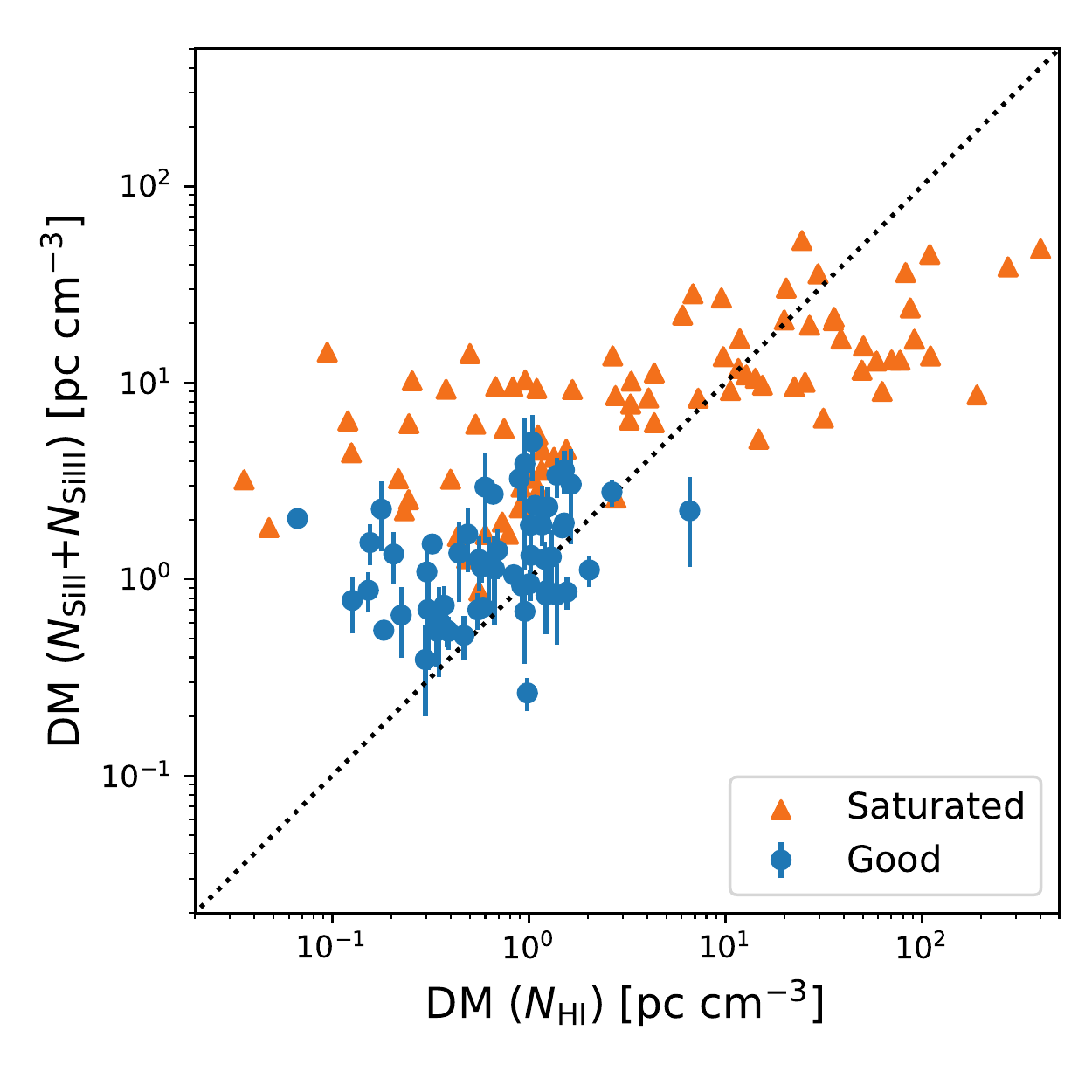}
    \caption{X axis: DM calculated from $N_{\rm HI}$, which is integrated over the same velocity ranges as the high-velocity \ion{Si}{ii} and \ion{Si}{iii} absorption lines \citep{richter+17}. The \ion{H}{i} spectra are from the HI4PI survey \citep{hi4pi}. Y axis: DM as converted from $N_{\rm SiII}$+$N_{\rm SiIII}$, assuming 0.1 $Z_\odot$ for HVCs and \ion{Si}{ii} and \ion{Si}{iii} are the dominant phases of silicon. 
    Blue dots are for data with clean detection of \ion{Si}{ii} and \ion{Si}{iii}, and orange triangles are 
    for data with saturated \ion{Si}{ii} and/or \ion{Si}{iii} lines as defined in Richter et al.\ (2017). 
    We find that the DM values from 
    $N_{\rm HI}$ and $N_{\rm SiII}+N_{\rm SiIII}$ follow 
    a rough 1:1 relation (black dotted line). Although a large scatter is present, this plot provides 
    a proof of concept for
    Equation \ref{eqn:nhi_to_dm} which generates all-sky \dmhvc\ values from an \hi\ HVC map. See \S~\ref{sec:hvc} for more detail.  
    }
    \label{fig:dm_si}
\end{figure}

Figure \ref{fig:dm_si} describes a proof of concept 
of our method for generating the \dmhvc\ map from all-sky $N_{\rm HI}$(HVC). 
We adopt the column density measurements of \ion{Si}{II} and \ion{Si}{III} from the ionized HVC survey by \cite{richter+17} and assume that \ion{Si}{II} and \ion{Si}{III} are the dominant ions of silicon at $T\sim10^4$ K. The DM related to such gas is DM$\equiv(N_{\rm SiII+SiIII})/Z_{\rm HVC}/({\rm Si/H})_\odot$, 
where $Z_{\rm HVC}=0.1 Z_\odot$ for the HVCs \citep[e.g.][]{kunth94, wakker01}
and ${\rm (Si/H)}_\odot=10^{-4.49}$ \citep{asplund09}. Meanwhile, we calculate $N_{\rm HI}$ by integrating the HI4PI spectra over the same velocity ranges as the \ion{Si}{II} and \ion{Si}{III} absorption lines, and covert the values to \dmhvc\ 
using Equation \ref{eqn:nhi_to_dm} without applying $f_{\rm HVC}$. Figure~\ref{fig:dm_si} shows a rough
1:1 correlation between the DM values from $N_{\rm HI}$ and from $N_{\rm SiII}+N_{\rm SiIII}$, the ions typical of $T\sim10^{4}$ gas.

\subsection{The Hot Galactic Halo}
\label{sec:hot_halo}

X-ray absorption-line spectra of AGN have revealed the nearly ubiquitous detection of highly ionized oxygen ions (e.g., \ion{O}{VII} at $\lambda=21.602$\AA) in our Galactic halo
\citep[e.g.][]{fang+15}. This gas must contribute a significant and separate column of electrons from the lower ionization state gas traced by HVCs and low-ion metal transitions.
Complementing these X-ray observations are far-UV surveys for \ion{O}{VI} absorption at high velocities \citep{sws+03}, i.e.\ gas distinct from the Galactic disk 
\citep{bjt+08}.   

\begin{figure}
	\includegraphics[width=\columnwidth]{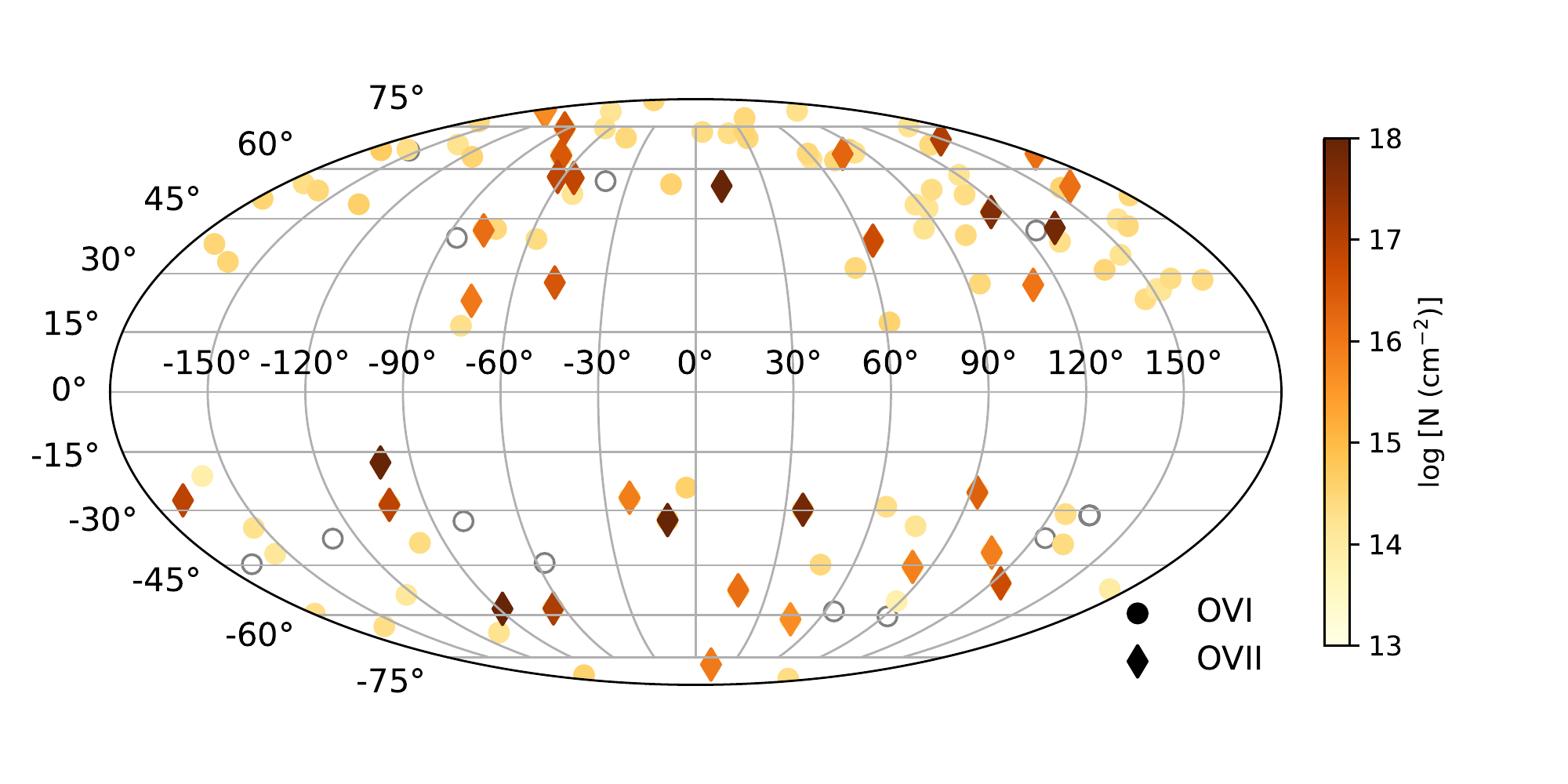}
    \caption{All-sky map (Galactic coordinates)
    of \ion{O}{VI} and \ion{O}{VII} column densities
    associated with the Galactic halo, as measured in spectra of distant
    AGN/quasars \citep{sws+03,fang+15}.  $N_{\rm OVI}$ values are only for the high-velocity halo gas with $|v_{\rm LSR}|\gtrsim100$ km s$^{-1}$ \citep{sws+03}. The $N_{\rm OVII}$ values are from the \ion{O}{VII} absorption line ($\lambda=21.602$\AA) as observed in X-ray; this line is unresolved by the low resolution of 
    X-ray spectrographs (FWHM$\sim300$ km s$^{-1}$). We find that the $N_{\rm OVII}$ values are systematically larger and exhibit much greater scatter than the smooth distribution of $N_{\rm OVI}$ measurements.
    This scatter likely reflects the greater uncertainty in 
    these measurements \citep[e.g.\ due to variations in the true
    kinematics of the gas;][]{hk2016}, although it 
    could also  indicate intrinsic variations.
    \label{fig:OVI_OVII}
    }
\end{figure}

Figure~\ref{fig:OVI_OVII} summarizes the distribution of \ion{O}{VI} and \ion{O}{VII} measurements on the sky, color-coded by the estimated column density. 
This figure attempts to isolate gas located within the Galactic halo.
For \ion{O}{VI}, we show the distribution of high-velocity, highly 
ionized gas ($|v_{\rm LSR}|\gtrsim100$ km s$^{-1}$; 
\citealt{sws+03}) beyond the Galactic disk over the sky; 
these measurements exhibit typical column densities of 
log$N_{\rm OVI} \approx 14.3$ with standard deviation of $\approx 0.2$\,dex. 
We emphasize there is also hot halo gas with $|v_{\rm LSR}|<100$ km s$^{-1}$ overlapping the Galactic disk \citep{ssw+03, wakkeretal03, wakker+12}. 
Therefore, the halo $N_{\rm OVI}$ value over the full velocity range is predicted
to be a factor of two higher \citep{zheng+15} and one should view
the values in Figure~\ref{fig:OVI_OVII} with this in mind.
As regard \ion{O}{VII}, one lacks the spectral resolution to cut 
on velocity and instead the full integration is shown.
While this may include gas from within the ISM, we argue below
that any such contribution should be minimal.


\begin{figure}
	\includegraphics[width=\columnwidth]{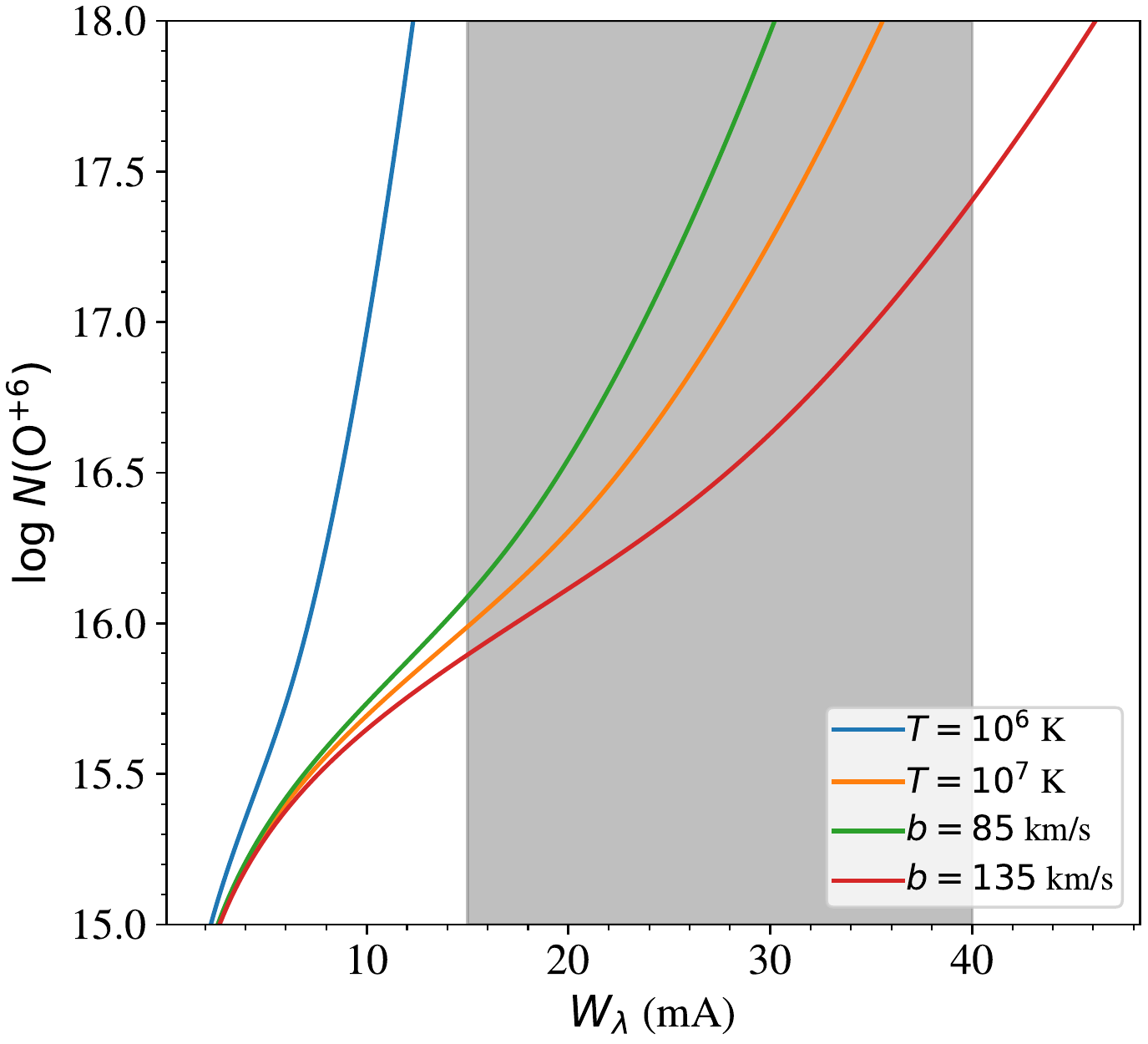}
    \caption{Curve of growth relationship between \ion{O}{VII}~21
    equivalent width $W_\lambda$ and the column density for a 
    single absorption line with varying $b$-value (colored curves).
    For a thermally broadened gas, $b = \sqrt{2 k T / m_A}$. 
    The shaded region indicates the range of $W_\lambda$ values
    observed through the Galactic halo.
    It is evident that the corresponding uncertainty in 
    $\N{O^{+6}}$ can be an order of magnitude.
    \label{fig:cog}
    }
\end{figure}

In addition to the wide-spread detection of highly-ionized
halo gas across the sky,
Figure~\ref{fig:OVI_OVII} reveals that the 
$N_{\rm OVII}$ values are several orders of magnitude higher 
than $N_{\rm OVI}$ and exhibit a much larger dispersion.  The latter 
reflects, we believe,
the greater uncertainty in the measurements, e.g.\ 
due to our assumption of a single $b$-value for all
of the absorption.
The X-ray spectra that provide the $N_{\rm OVII}$ estimates have a spectral resolution $R \sim 400$ (FWHM $\sim 700 \, \mkms$) and any absorption detected is unresolved. Furthermore, the measured equivalent widths of the \ion{O}{VII}~21\AA\ transition ($W_\lambda \sim 15-40$\,m\AA) 
place them firmly on the saturated portion of the curve-of-growth. As Figure~\ref{fig:cog} illustrates,  the inferred column densities
depend entirely on the assumed $b$-values 
(for absorption dominated by a single component).
More complex models would generally lead to lower column densities 
\citep[but see][]{pro06}.
These large uncertainties aside, Figure~\ref{fig:OVI_OVII} indicates that the dominant ionization state of oxygen along sightlines through the Galactic halo is \ion{O}{VII}. 
While the location of this \ion{O}{VII} is debated 
\citep{wangetal05,hk2016,faerman17},
the gas must contribute a large column density of electrons
for any FRB event.

\begin{figure}
	\includegraphics[width=\columnwidth]{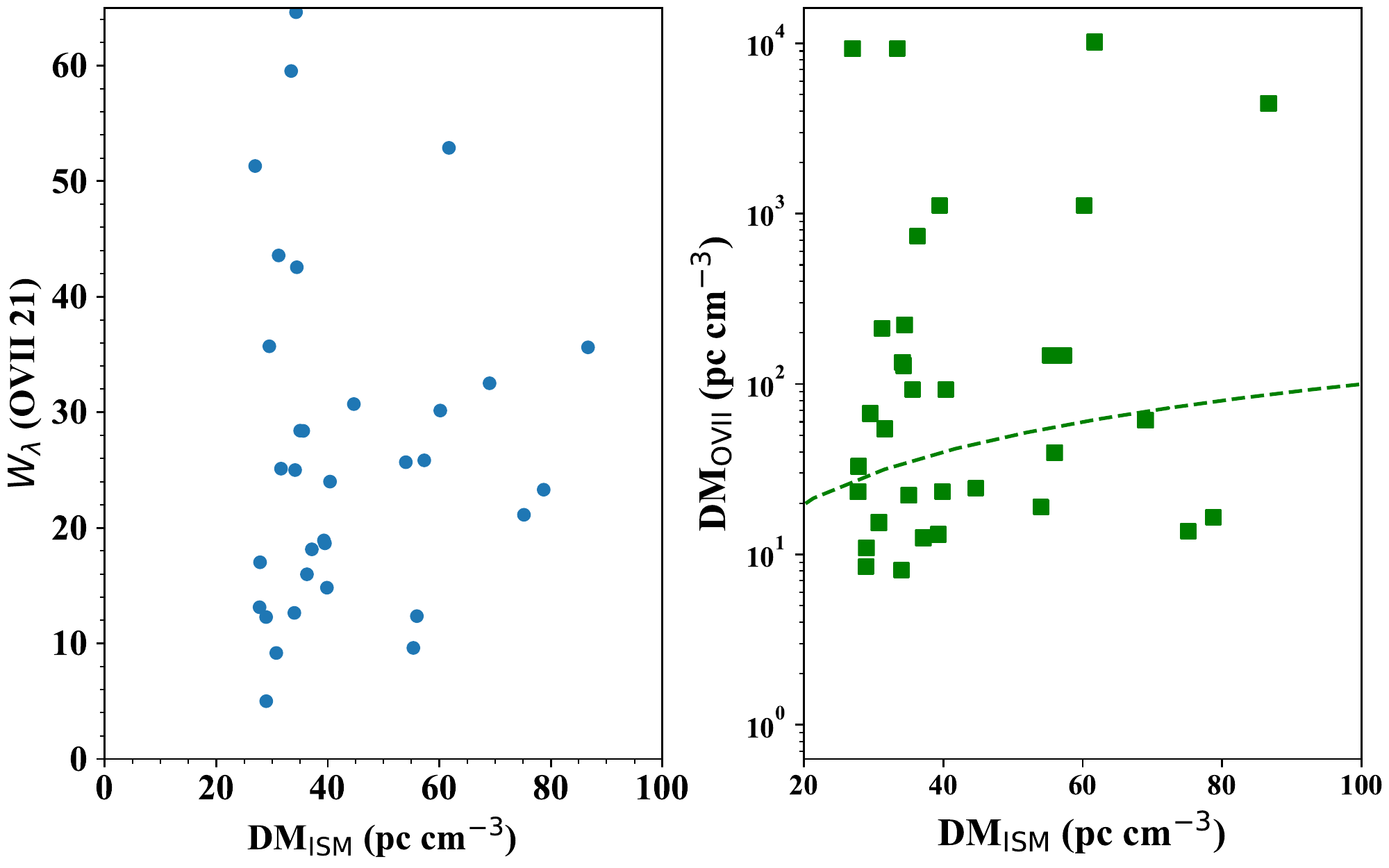}
    \caption{Left: Equivalent width $W_\lambda$ of the \ion{O}{VII}~21\AA\ transition
    against the estimated ISM dispersion measure DM$_{\rm ISM}$
    for the positive detections along sightlines to distant 
    AGN \citep{fang+15}.  There is no apparent correlation  
    (Pearson correlation coefficient of 0.04 and  $p$-value of 0.82)
    and we conclude that the \ion{O}{VII} gas lies beyond the
    Galactic ISM.
    Right: The same data but with $W_\lambda$ converted to
    $\dm{OVII}$ using Equation~\ref{eqn:DM_OVII}.  The dotted
    line shows a one-to-one relation and we note that many of the
    inferred $\dm{OVII}$ values greatly exceed the Galactic ISM estimates.
    The large scatter, however, likely reflects measurement uncertainty
    (e.g.\ variations in the assumed Doppler parameter) and possibly 
    intrinsic variations.  Higher spectral resolution X-ray observations are
    greatly desired to more precisely estimate $\dm{OVII}$.
    \label{fig:DM_vs_EW}
    }
\end{figure}

For completeness, we first consider whether the \ion{O}{VII} gas could arise in the ionized regions of the Galactic ISM. If so, we expect a tight 
correlation between the DM from the ISM and the equivalent width of \ion{O}{VII} gas. To test this hypothesis, we plot the 
DM$_{\rm ISM}$ estimated along the \ion{O}{VII} sightlines from the NE2001
model against the measured $W_\lambda$(\ion{O}{VII}~21) values
(Figure~\ref{fig:DM_vs_EW}). There is no statistically significant correlation;
we recover a Pearson correlation coefficient of 0.04 and $p$-value of 0.82. 
Furthermore, the implied $N_{\rm OVII}$ column densities imply DM values from 
the corresponding ionized hydrogen
(using Equation~\ref{eqn:DM_OVII} below)
that greatly exceed the DM$_{\rm ISM}$ values.
We consider this firm evidence that 
the gas lies beyond the Galactic ISM but
await spectra with higher resolution to confirm the
inferred column densities of $N$(\ion{O}{VII}).

It is possible, with a few conservative assumptions, to use the measured \ion{O}{VII} column densities to infer the dispersion measure related to this component, DM$_{\rm OVII}$
\citep[see also][]{shull18}.
First, we assume that the majority of highly ionized oxygen is in the \ion{O}{VII} state.  This is supported by the lower estimates of $N_{\rm OVI}$ and $N_{\rm OVII}$ and also by the predicted virial temperature of our Milky Way halo:
$T \sim 10^6$\,K, implying 
$f_{\rm OVII} \equiv n_{\rm OVII}  / n_{\rm O} \approx 1$
for a collisionally ionized gas.
Second, we assume that the gas has solar metallicity or less
$Z \le Z_\odot$, i.e. O/H $\le 10^{12-8.67}$ by number.  
With these two assumptions, we recover: 

\begin{equation}
\dm{OVII} \approx 80 \, \mdmunits \; f_{\rm OVII} 
  \ltp \frac{Z}{Z_\odot} \rtp^{-1} 
  \ltp \frac{N_{\rm OVII}}{10^{17} \cm{-2}} \rtp
 \label{eqn:DM_OVII}
\end{equation}
For $N_{\rm OVII} = 10^{16.5} \cm{-2}$, a 1/3 solar metallicity, and $f_{\rm OVII} = 1$, we estimate 
DM$_{\rm OVII} \approx 50 \, \mdmunits$
(Shull \& Danforth 2018 adopt a lower value for $N_{\rm OVII}$
and report DM$_{\rm OVII} \approx 30 \, \mdmunits$).
This value exceeds DM$_{\rm ISM}$ for essentially any extragalactic (high latitude) sightline and also the typical value assumed for the Galactic halo in current FRB literature \citep{dolag15}. We compare our estimated \dmovii\ value with the prediction (DM$_{\rm Xray}$) from hot gas halo models based on extragalactic X-ray studies \citep{sharma12}. By taking their entropy-core electron density profile (see their figure 1) and scaling it to 
$M_{\rm halo} = 1.5 \times 10^{12} \mmsun$, we find DM$_{\rm Xray}\sim$ 30 \dmunits\ within $r_{\rm max}=2r_{\rm 200}$. Note that the DM$_{\rm Xray}$ value may vary by a few given the uncertainties of the halo gas models; in general, 
we find our \dmovii\ to be in good agreement with extragalactic X-ray studies.

\begin{figure}
	\includegraphics[width=\columnwidth]{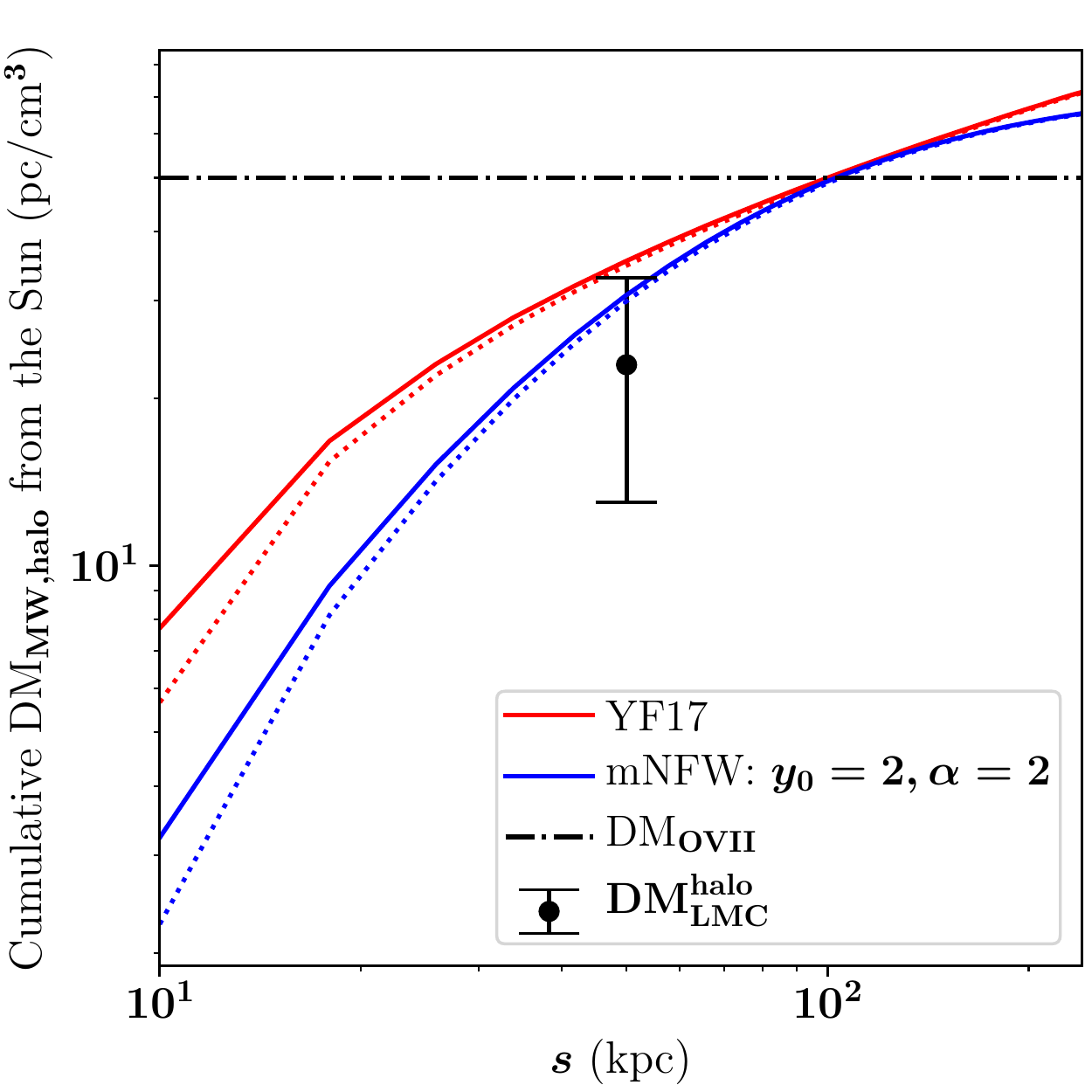}
    \caption{
    The color curves show the cumulative \dmmwh\ values
    through our Galactic
    halo for two models of the baryon distribution.
    The calculations originate at the Sun and travel a distance
    $s$ in two directions through the halo: 
    (solid) $\ell = 0, b=90$; 
    (dotted) $\ell = 280.5, b=-32.9$ corresponding to the direction
    towards the LMC.  The inner 10\,kpc of the halo has had $n_e$ set to zero
    so as to ignore any ISM contribution.
    The dash-dot horizontal line shows an estimate for 
    $\dm{OVII}$ based on the observed equivalent \ion{O}{VII}
    widths (Figure~\ref{fig:DM_vs_EW},
    Equation~\ref{eqn:DM_OVII}).  The black point shows an estimate of 
    Galactic halo contribution to DM \dmlmch\ for the sightline to
    the LMC, plotted at 
    the distance to the LMC (as described in the text).
    \label{fig:DM_from_Sun}
    }
\end{figure}

We now consider whether the halo models introduced in Section~\ref{sec:models} may account for the inferred DM$_{\rm OVII}$ values. Figure~\ref{fig:DM_from_Sun} presents the cumulative dispersion measure from the Galactic
halo \dmmwh\ for sightlines originating at the
Sun.  This estimate ignores any contribution from electrons
within 10\,kpc of the Galactic center including
the ISM.  We assume
spherical symmetry with $M_{\rm halo} = 1.5 \times 10^{12} \mmsun$
and show two different paths ($\ell,b$)
through the halo: straight up from the Sun and towards the LMC.  
We derive small variations with angle given the Sun's proximity
to the halo center.
If the scatter in $\dm{OVII}$ shown in Figure~\ref{fig:DM_vs_EW} is
real, it would imply substantial `patchiness' in our halo.

Overlaid on the model curves are
(i) the estimated $\rm DM_{\rm LMC}^{\rm halo}$ to the 
closest pulsar in the LMC 
ignoring the Galactic ISM contribution in that direction, and
(ii) a fiducial estimate of DM$_{\rm OVII} = 50 \mdmunits$
for the hot gas component discussed above\footnote{This
can well exceed \dmlmch\ because it has no
constraint on distance in the halo.}.
The two models reproduce the fiducial \dmovii\ at 
a distance $s = 100$\,kpc from the Sun
and well exceed the value by \rvir.
Any excess relative to DM$_{\rm OVII}$ is acceptable as
there may be substantial mass at large radii with a
very low oxygen abundance.
For the mass of the Milky Way adopted here, however, the
models in Figure~\ref{fig:DM_from_Sun} predict a higher
density at 50\,kpc than the constraint published by
\cite{salem+15} based on ram-pressure stripping 
modeling of the LMC ($n \approx 1 \times 10^{-4} \cm{-3})$.  
For our Galaxy, therefore, one might adopt a shallower density
profile in the inner halo than the ones depicted.
We further emphasize that these models imply significantly
greater DM contributions ($2-3\times$)
from the Galaxy halo than those typically
adopted in the FRB literature \citep{dolag15}, where
a lower halo mass and steeper density profile were
been assumed.  
We recommend $\mdmmwh=50-80 \mdmunits$
for an integration to \rvir, and even larger values
if one extends beyond.

Models of the Galactic halo in the literature
have been heavily influenced by the LMC
constraint.  And one notes that 
the models in Figure~\ref{fig:DM_from_Sun} exceed
the central value of the ISM-corrected estimate 
by $\sim 5-10~\mdmunits$.
Given its importance,
let us reexamine the origin of this DM 
constraint on the Galactic halo towards the LMC.
To be precise, \cite{manchester06} reported DM measurements
for 14 pulsars discovered towards the Magellanic clouds,
and argued 12 were located within them (although
none have confirmed distances).  Standard treatment has been
to assume that the total DM from Earth to the
LMC $\rm DM_{\rm LMC} = 70 \mdmunits$ based on several of the
pulsars showing approximately this value \citep[e.g.][]{anderson10}.
To estimate the contribution from the Galactic halo,
one then requires an estimate
to $\rm DM_{\rm LMC}$ from the Galactic ISM.
The latter is dominated by the thick disk component and the
NE2001 model estimates 
is $\dm{ISM} = 52 \mdmunits$ (49\dmunits\ using YMW17).
Empirically, {\it none} of the pulsars within 10\,deg of the LMC
have a parallax measurement, i.e.\ there is no test of the 
thick disk model in this direction.
From the \cite{gaensler08} analysis, one derives an even lower
$\dm{ISM} = 47 \mdmunits$ value but, unfortunately, 
there are few  $|b| \approx 30$\,deg pulsars to test it.
Given the lack of empirical constraints on the distance 
of any of the pulsars
and the uncertainty of the thick disk model, 
we adopt a $1\sigma$ systematic uncertainty of 
10\dmunits\ for $\dm{ISM}$ towards the LMC.
Last, we estimate the DM of the Galactic halo alone
towards the LMC from the difference:
$\rm DM_{\rm LMC}^{\rm halo} = 23 \pm 10 \mdmunits$, 
as illustrated  in Figure~\ref{fig:DM_from_Sun}.  
Clearly, both models used in Figure~\ref{fig:DM_from_Sun} 
are consistent with this estimate.
We encourage the community to further refine estimates
of $\dm{LMC}$ and \dmlmch\ through
new constraints on the distances to pulsars
towards the LMC.

\section{The Local Group}
\label{sec:LG}

The Milky Way is not an isolated galaxy.
Its own halo contains the Magellanic
clouds and tens -- if not hundreds -- of 
satellite galaxies.  Furthermore, at $\sim 1$\,Mpc
distance lies M31 and its own system of satellites \citep{McConnachie12}.
Together with M33, these massive spiral galaxies 
form the Local Group.  It is possible, if not
probable, that this Local Group also contains a distinct
intragroup plasma that will contribute to the
DM of FRB events.  At the very least we expect
contributions from the halo gas of our Galaxy's
neighbors.  In this section, we provide estimates
for these Local Group contributions focusing
on hot gas with $T \gg 10^4$\,K.

\subsection{Magellanic Clouds}

The Magellanic Clouds are also believed to reside within their
own dark matter halos and may contain an ionized phase of halo
gas.  At the least, there is evidence for galactic-scale outflows
that may be polluting the regions around them
\citep{hoopes+02,lh07,barger16}.
One also observes that the
Magellanic stream spans hundreds of deg across the sky,
whose contribution
component is already captured in our HVC
analysis (\S~\ref{sec:hvc}; Figure~\ref{fig:hvc}).
Here, we focus on putative halo components localized
to the satellites themselves.

\begin{figure}
	\includegraphics[width=\columnwidth]{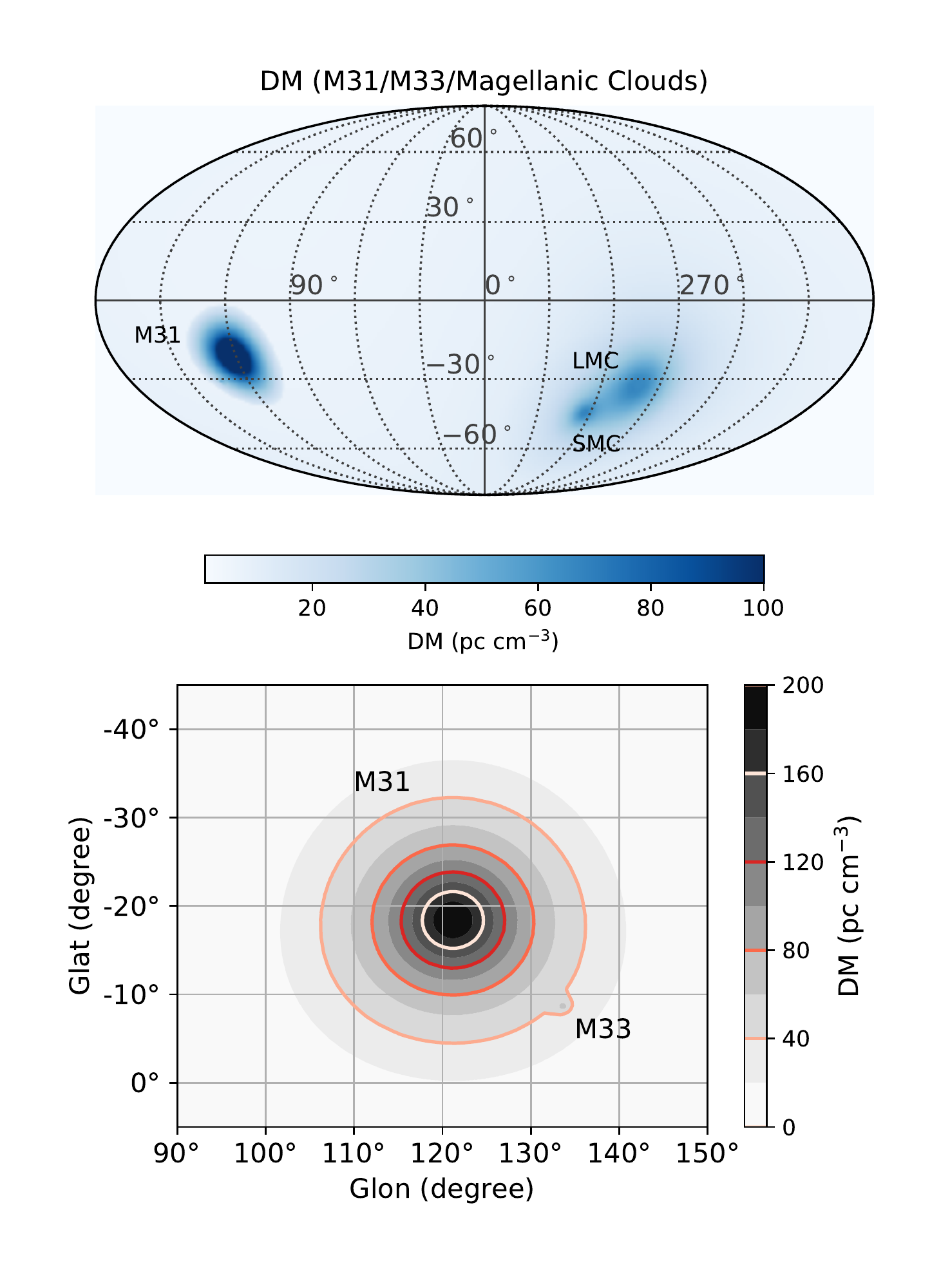}
    \caption{Top: 
    All-sky Mollweide projection (Galactic coordinates) of the DM contribution of M31's halo ($\ell=121$\,deg, $b=-21.6$\,deg), M33's ($\ell=133.6$\,deg, $b=-31.3$\,deg), the LMC's ($\ell=280.5$\,deg, $b=-32.9$\,deg), and the SMC's ($\ell=302.8$\,deg, $b=-44.3$\,deg). See the text for details on their halo properties. Note that M33 halo is overwhelmed by M31 in this panel. Bottom: Zoom-in on M31 and M33's halo contribution. M33 can be seen in the lower right corner.
    \label{fig:M31}
    }
\end{figure}

Figure~\ref{fig:M31} illustrates the potential
contributions of the Clouds
to DM measurements using the following
assumptions of mass, concentration, and
\fhb\ \citep{df16}:  
$M^{\rm LMC}_{\rm halo} = 1.7 \times 10^{10} \mmsun$,
$c^{\rm LMC} = 12.1$,
$\mfhb^{\rm LMC} = 0.75$;
$M^{\rm SMC}_{\rm halo} = 2.4 \times 10^{9} \mmsun$,
$c^{\rm SMC} = 15.0$,
$\mfhb^{\rm SMC} = 0.75$.
We have adopted the mNFW model ($\alpha=2, y_0=2$)
and distances of $d_{\rm LMC} = 50$\,kpc and
$d_{\rm SMC} = 61$\,kpc from the Sun. 
While the average DMs for the Clouds is considerably
less than M31 (see below; Figure~\ref{fig:M31}), 
it is striking how large of an area
that they span across the sky.  This follows, of
course, from their close proximity. 
It is possible, therefore, that a large, Southern FRB
experiment (e.g. SKA) could search for 
diffuse halo gas associated with the Clouds.

\subsection{M31}

Analysis of the kinematics of M31's stellar disk
and satellite system indicates it has a dark matter
halo mass comparable to our Galaxy, i.e.
$\mmhalo^{\rm M31} \approx 1.5 \times 10^{12} \mmsun$
\citep{vdm2012a}.
At a distance of only $\approx 750$\,kpc \citep{riess12},
M31 subtends a large angular diameter on the sky.
Indeed, its halo spans a sufficient size that one may identify
several tens of luminous quasars behind it and probe its halo
gas in absorption \citep{lehner+15}.
These data have revealed an enriched, cool CGM surrounding
M31 with column densities comparable to but
generally less than those
observed for other L* galaxies at low-$z$ \citep{lehner+15,howk+17}.
The implication is that M31 possesses a gaseous halo of
cool material embedded, presumably, within a hot medium.



These far-UV observations, however, do not offer direct
constraints of hot halo gas for M31.
Instead, the proximity of M31 affords a unique opportunity to study
its CGM with FRBs.  
Consider the following illustrative examples.
Adopting the mNFW halo model with $y_0=2$, $\alpha=2$, 
and $\mfhb = 0.75$,
we have estimated the
halo DM contribution\footnote{Here, we ignore the ISM of M31
and have taken DM$_{\rm M31} (\mrperp<10{\rm kpc}) = 
{\rm DM}_{\rm M31} (\mrperp=10{\rm kpc})$.
}
of M31 for sightlines passing within $2 \times \mrvir$ of its
center (Figure~\ref{fig:M31}).
The proximity of M31 implies its disk alone comprises 
many sq~deg.\ 
on the sky and its halo subtends $\approx 30$\,deg. 
A Northern-sky survey for FRBs
that yields $\sim 10,000$ events (e.g. CHIME; \citealt{CHIME}),
will randomly intersect M31's halo hundreds of times.
For completeness, we also include a model for M33
taking $\mmhalo = 5 \times 10^{11} \mmsun$
and $c=8.36$.

To crudely assess constraints from such a survey,
we performed the following exercise.  We generated a 
random sample of 10,000 FRBs with redshifts uniformly
drawn from $z_{\rm FRB}=[0,0.5]$ and declination $\delta > 0^\circ$.
For each FRB, we assume the Galaxy contributes
$\mdmmwh\ = 80 \, \mdmunits$ with a $\pm 15 \, \mdmunits$
uncertainty (uniform deviate) to account for intrinsic
dispersion and systematic uncertainty in the NE2001 model.
We further assume that the FRB host galaxy contributes
$\dm{host} = 40 \pm 20 \mdmunits$ (normal distribution).
Lastly, we assume the cosmic DM model\footnote{See https://github.com/FRBs/FRB}  
of Simha \& Prochaska (2019; in prep., hereafter SP19) with 
a $\pm 20$\%\ uniform deviate.
All of these contributions are required to have  zero or greater value.
Lastly, we adopt the halo model for M31 illustrated in Figure~\ref{fig:M31}.

\begin{figure}
	\includegraphics[width=\columnwidth]{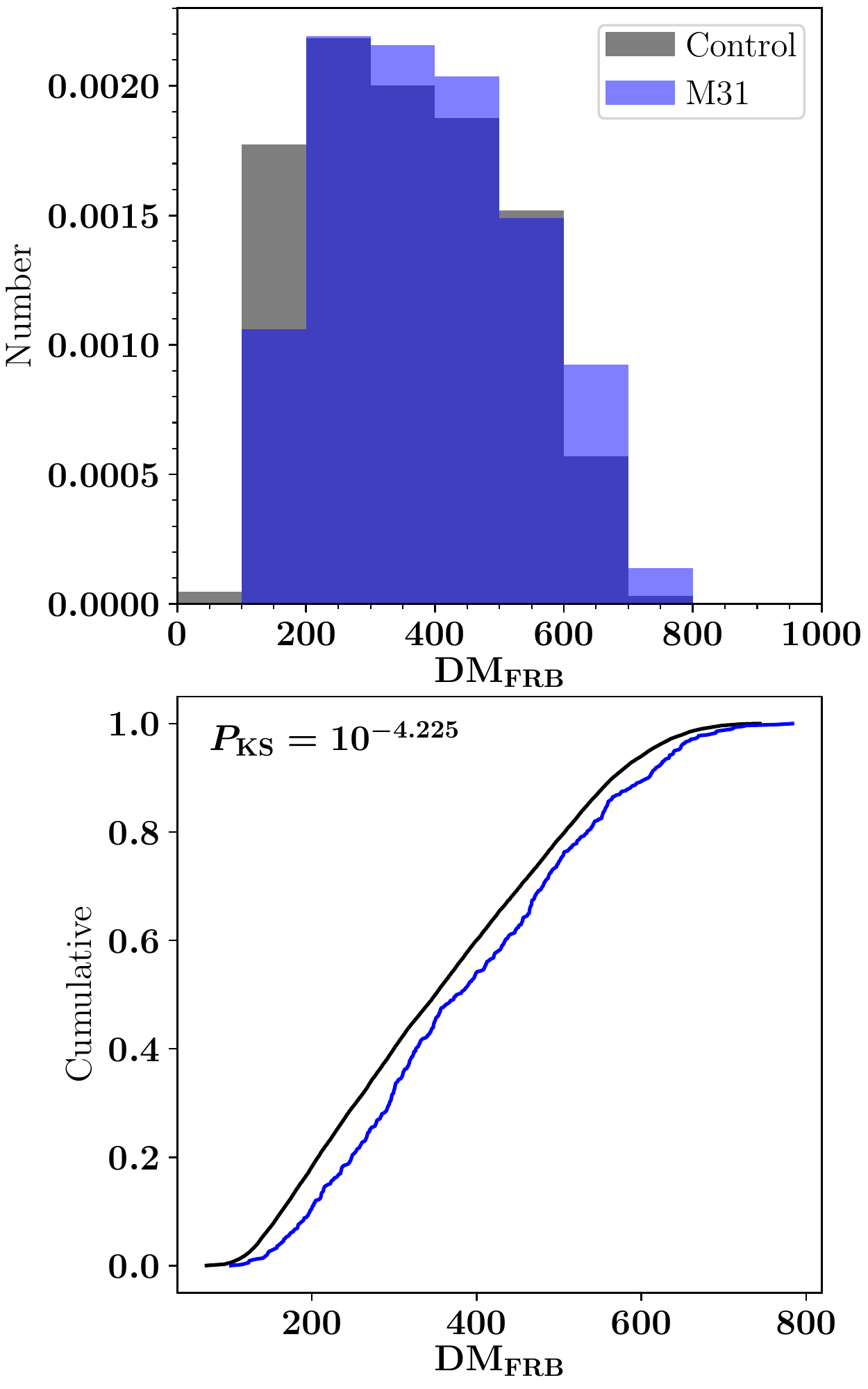}
    \caption{Top: Normalized distribution of the total \dmfrb\ measurements
    for 10,000 random FRBs distributed on the Northern sky
    (see text for details).  Blue indicates sightlines penetrating
    within $\mrvir = 237$\,kpc of M31 and black is the control
    sample away from M31. Bottom: Cumulative distributions for the two samples.
    For any random draw, we recover a very small 
    $P_{\rm KS}$ value indicating the two samples are drawn
    from distinct parent populations.
    \label{fig:random_M31}
    }
\end{figure}

The distribution of $\dm{total}$ measurements for the sample of events
away from M31 ($\mrperp > \mrvir$) versus
those intersecting its halo ($\mrperp \le \mrvir$) is shown
in Figure~\ref{fig:random_M31}.
One notes a shift in the M31 distribution of $\approx 75 \, \mdmunits$
the halo model adopted here.  A two-sided Komolgorov-Smirnov test on
the distributions yields a very small probability; this
rules out the null hypothesis that the two were drawn from the
same parent population.  The results are, of course,
sensitive to the assumed FRB redshift distribution
and the other assumptions.
Furthermore, a proper analysis would
forward model the observed distribution to perform a
maximum likelihood of M31's halo gas.
In any event, we conclude that a project like
CHIME \citep{CHIME}
will provide a very powerful constraint on M31's halo gas.

\subsection{The Local Group Medium (LGM)}
\label{sec:LGM}

Numerical studies of the Local Group have been designed to trace back
the history of our Galaxy and M31, and to predict their futures
\citep[e.g.][]{vandermar12b},
including the impending merger of the Galaxy and M31.
\cite{APOSTLE} have generated a series of hydrodynamical
simulations (named APOSTLE) 
designed to mimic our Local Group with scientific focus on its
satellite systems.
The total mass of their simulated groups are typically 
$\sim 10^{13} \mmsun$ to $r \approx 3$~Mpc, i.e.\ $\sim 5\times$ greater
than the combined masses of M31 and the Galaxy.
These simulated halos contain a highly ionized medium which
we refer to as the Local Group Medium (LGM).

Over the past decades there have been claims of an observed 
LGM in both cool gas \citep[HVCs;][]{blitz99} and an enriched hot 
phase \citep[OVII;][]{nme+05}.  These inferences, however, have been
challenged and alternatively explained by gas local to the Galactic halo
\citep[][see also the previous section]{wakker01,sws+03,fang+13}.
Nevertheless, there is strong theoretical motivation 
to expect an LGM and we
briefly explore the potential
implications while defering an extensive analysis
to a future paper (Fattahi et al., in prep.).

\begin{figure}
	\includegraphics[width=\columnwidth]{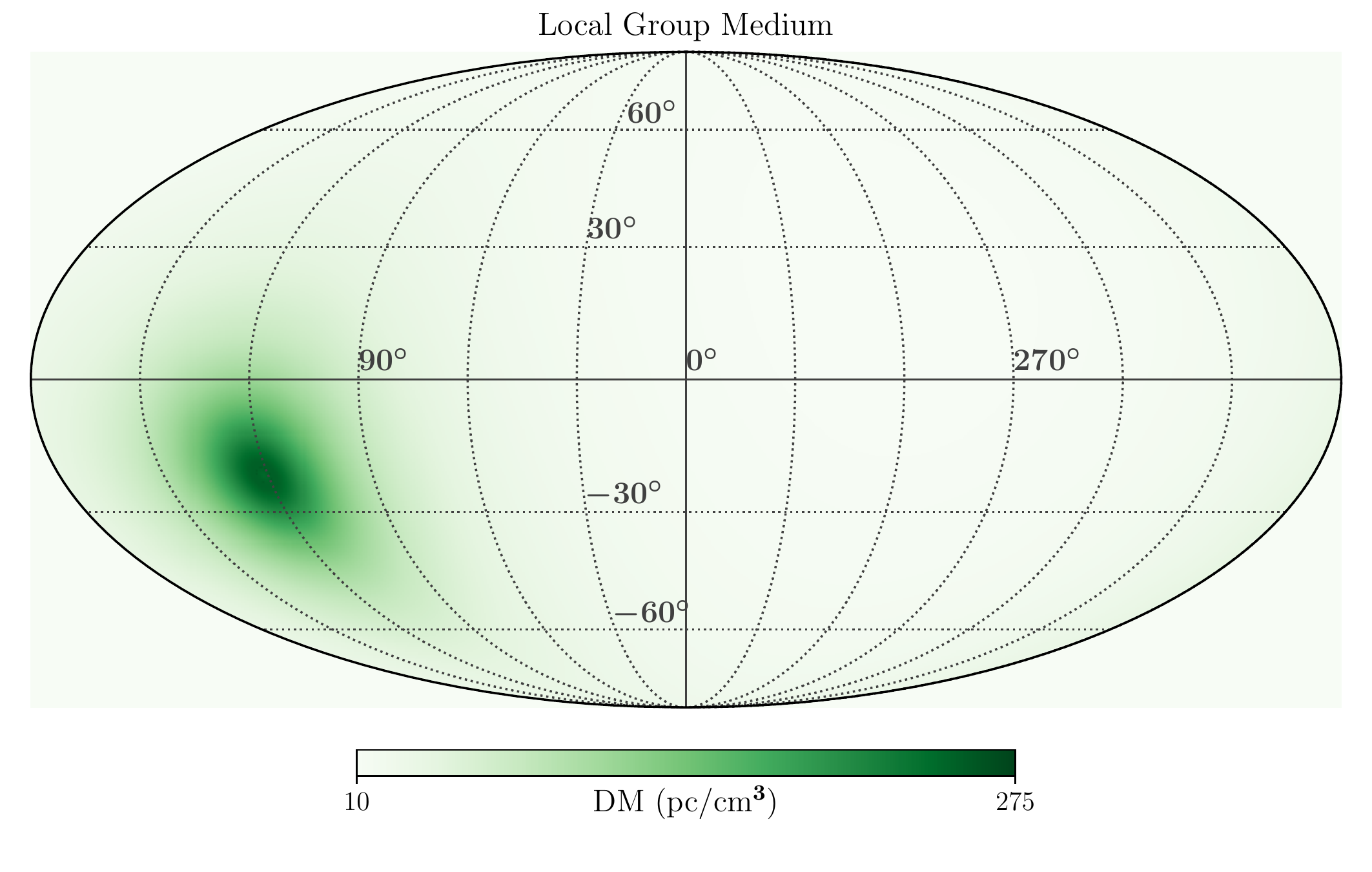}
    \caption{
    All-sky Mollweide projection (Galactic coordinates) of the
    DM contribution from the Local Group medium, adopting a halo mass of $10^{13}$ \msun\ out to 3 Mpc. The baryonic center of the Local Group is assumed to lie directly between
    the Galaxy and M31. See \S\ \ref{sec:LGM} for details. 
    \label{fig:LGM}
    }
\end{figure}

We consider a halo model centered on the barycenter of the Galaxy/M31
system with an integrated mass to 3~Mpc of $10^{13} \mmsun$.
With these simple constraints, we have generated the 
\dmlgm\ map in Figure~\ref{fig:LGM}.
As an example, we adopt the mNFW model with 
$\mmhalo = 10^{12.5} \mmsun$ to \rvir\ and $f_b = 0.8$.
Figure~\ref{fig:LGM} shows an all-sky projection of 
\dmlgm\ in the Galactic coordinate system.  The results are
similar to those we derived for M31 in Figure \ref{fig:M31} except the halo mass is larger and its center is twice closer. Clearly, there will
be a degeneracy between the two components but the same 
arguments made for resolving M31's halo otherwise apply
and are stronger for the LGM.
Future work (Fattahi et al.) will examine the combined
contributions between the Galaxy, M31, and LGM
self-consistently.

\section{Halos of Intervening Galaxies, Groups, and Clusters}
\label{sec:exgal}

In this section we review current constraints 
to describe the distribution of baryons in distant
dark matter halos.
Our goals are to describe our current (limited) knowledge and
illustrate the opportunities (and challenges) that FRB observations
may provide.  We refer to the contribution from a single
halo as \dmhalo\ and from a population of halos along a given
sightline as \dmhalos.

\begin{figure}
	\includegraphics[width=\columnwidth]{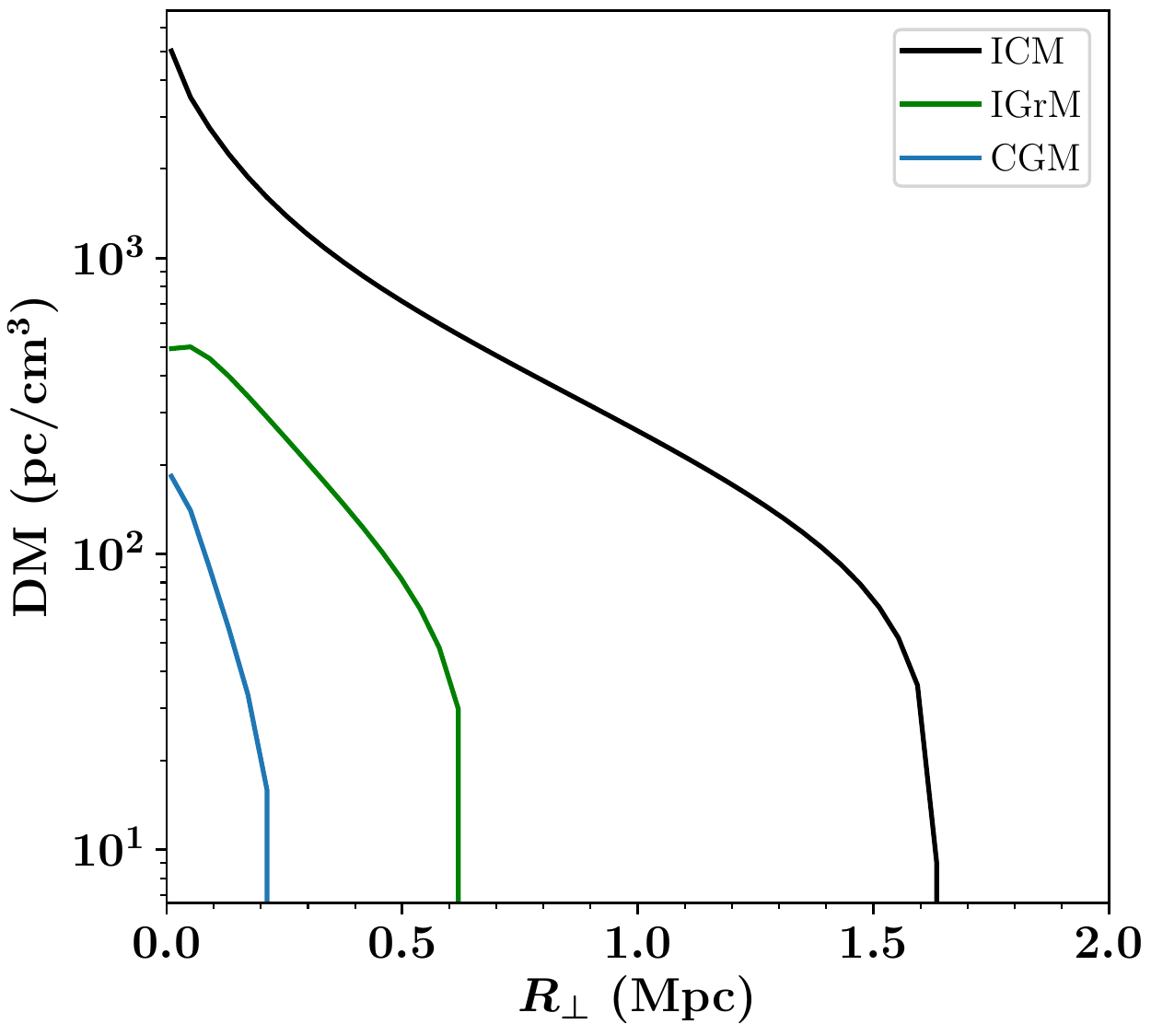}
    \caption{DM profiles for a fiducial intracluster medium (ICM)
    of a $\mmhalo = 5 \times 10^{14} \mmsun$ cluster (black),
    the intragroup medium (IGrM) of a 
    $\mmhalo = 10^{13.5} \mmsun$ halo (green),
    and the CGM of an L* galaxy (blue; mNFW model).  Clearly, the DM values
    associated with a cluster greatly exceed any contribution
    from L* galaxy or even galaxy group to beyond 1\,Mpc.
    \label{fig:ICM}
    }
\end{figure}

\subsection{Groups and Clusters}
\label{sec:clusters}

In the highest-mass dark matter halos (galaxy clusters;
$\mmhalo > 10^{14} \mmsun$),
their halo gas is virialized to $T > 10^7$\,K and
emits brightly at X-rays.
Observations of this intracluster medium (ICM)
yield model estimates for the density of this plasma
\citep[e.g.][]{vik06} with a typical parameterization:
\begin{equation}
n_p n_e = n_0^2 \frac{(r/r_c)^{-\alpha}}{(1+r^2/r_c^2)^{3\beta-\alpha/2}}
 \frac{1}{(1+r^\gamma/r_s^\gamma)^{\epsilon/\gamma}}
 \, + \,
 \frac{n^2_{02}}{(1+r^2/r_{c2}^2)^{3\beta_2}}
 \label{eqn:clusters}
\end{equation}
where all parameters are free except $\gamma=3$.
The ICM baryonic mass fraction inferred is 
$\sim 70\%$ estimated
by adopting the total dynamical mass inferred
from the gas temperature. 
Therefore, we adopt $\mfhb = 0.7$ in what follows.
For any sightline that intersects one of these rare halos, 
\dmhalo\ is substantial.  Figure~\ref{fig:ICM} shows the \dmrperp\ 
profile for a fiducial model adopting the parameter
values derived by \cite{vik06} for the cluster A907
($M_{500} = 10^{14.7} \mmsun $).
This is compared with the \dmrperp\ profile for an L* halo
($\mmhalo = 10^{12.5} \mmsun$) and the halo for a group
with mass $\mmhalo = 10^{13.5} \mmsun$.
The differences are striking and we further note that 
\dmhalo\ from a single galaxy cluster may exceed 
the total from all other contributions 
along the sightline.

While \dmhalo\ for a galaxy cluster is very large, these
halos are rare and therefore may not contribute 
significantly to \dmhalos.
We estimate the probability of intersecting a
single cluster as follows.
First, using the Aemulus halo mass function 
we calculate the average co-moving number density of clusters
from $z=0-1$ as: 
$n_c(\mmhalo \ge h_{68}^{-1} 10^{14} \mmsun) = 
8.7  \times 10^{-6} h_{68}^3 \rm \, Mpc^{-3}$
Second, the average line-density in the
cosmology-normalized Jacobian \citep{bp69}
is $\ell(X) = (c/H_0) n_c A_p$ where $A_p$ is the physical size
and $dX = (1+z)^2 H_0/H(z) dz$.
Adopting $A_p = \pi (1 \, \rm Mpc)^2$, we estimate
$\ell(X) = 0.12$.  For an FRB at $z=1$, the likelihood to
intersect a cluster is then $p \approx \ell(z) \Delta z$
with $\ell(z) = \ell(x) dX/dz \approx 0.2$. 
With $\Delta z = 1$ for a single FRB, we estimate 
$p(\mmhalo \ge 10^{14} \mmsun; \mzfrb=1) \approx 0.2$;
i.e., we expect to intersect one cluster for 5~FRBs at $\mzfrb \sim 1$.
This drives a significant fraction of the scatter predicted
for \dmfrb\ (see also \citetalias{mcquinn14}).

The contribution of galaxy groups ($\mmhalo = 10^{13} - 10^{14} \mmsun$)
to \dmcosmic\ should be as substantial
and likely greater than clusters given that their 
comoving number density is more than an order of magnitude higher. 
Groups are still sufficiently rare, however, 
that Poisson statistics will describe their effects
on FRB DM measurements.  For
example, using the same methodology as above,
we estimate 
$p(\mmhalo=10^{13.5} - 10^{14} h_{68}^{-1} \mmsun; z=1) = 0.58$
for an intersection within \rvir\ of the halo.

Existing and forthcoming experiments to measure the distribution
of massive halos in our Universe offer the opportunity to use FRBs
to probe the halo gas within them.
One viable experiment will be to cross-correlate FRB observations
with the massive halos 
`tagged' by luminous red galaxies (LRGs).  
The auto-correlation function of LRGs at $z \sim 0.6$ 
indicates these galaxies reside in dark matter
halos with $\mmhalo \approx 10^{13.5} \mmsun$ \citep{zhai17}.
Furthermore, researchers have scoured all-sky images for 
concetrations of LRGs and other red galaxies to identify groups
and clusters to $z \sim 1$ \citep{rykoff16}.
Imaging surveys
on both sides of the planet provide millions of LRGs across
the sky and several million will have have spectroscopically
measured redshifts by 2020 \citep{DESI}.
For precisely localized FRBs with redshifts, a sample of 
$\sim 50$ events at $z \sim 1$ (i.e.\ $\Delta z = 50$)
should intersect 10~clusters and 30 massive groups.
Combining \dmfrb\ with the incidence of massive halos 
intersected would provide terrific new insight into the
nature of the ICM/IGrM at large radii.
Furthermore, we can envision experiments that cross-correlate
more poorly localized FRBs with LRGs to generate
results similar to Figure~\ref{fig:random_M31}.
Of course, one can also include the large and
growing samples of clusters detected through
X-ray and SZ observations.

\subsection{L* Galaxies}
\label{sec:lstar}

In contrast to clusters,
the halo plasma temperature 
and luminosity of galactic halos ($\mmhalo < 10^{13} \mmsun$)
are sufficiently low that X-ray detections from individual
halos are scarce and generally precluded \citep[e.g.][]{li+18}.
The conventional wisdom from the X-ray community, however, is 
that such 
galactic halos have $\mfhb \ll 1$, i.e.\ they
are `missing' a non-negligible fraction of baryonic mass.
This assertion, however, is based on modeling these 
challenging X-ray observations of the inner halo 
with unrealistically steep density profiles for
the gas \citep[e.g. NFW;][]{anderson10,fang+13}.
Indeed, analysis of the SZ signal from 'stacked'
dark matter halos show no apparent decline in \fhb\
down to current sensitivity limits 
\citep[ $\mmhalo \sim 10^{13} \mmsun$][]{planckXI}.
In short, we have limited constraints on
the distribution and mass of $T \gtrsim 10^6$\,K
baryons in distant, L* halos \citep[see][for first results]{burchett18}.

An alternate approach for tracing baryons in
galactic halos is through absorption-line analysis.
Spectroscopy of background sources whose sightlines penetrate
the halo yield precise estimates on the column densities
of atoms and ions, including \hi, \ovi, \civ, \ciii, and \mgii.
The species with lower ionization potentials ($h \nu \sim 1$\,Ryd)
trace the cool ($T \sim 10^4$K), less-ionized plasma within
the halo, while ions like \civ\ and \ovi\ probe more highly ionized
and (possibly) warm-hot gas with $T \sim 10^5 - 10^6$K.
Together, these diagnostics reveal the incidence and surface
density of gas with $T \le 10^6$\,K around galaxies
that have a diverse range of properties and environments.

Focus first on the cooler, less-ionized gas traced by 
\hi\ and lower ion stages.  
Several decades of research, accelerated in recent years by the
highly sensitive COS spectrometer on {\it HST}, have provided surveys
of the CGM in a diverse set of galaxy populations at $z < 1$
\citep[e.g.][]{cwt+10,pwc+11,stocke13,tumlinson+13,werk+13}.
These data yield direct 
measurements on the projected surface density
of neutral hydrogen (\nhi) as a function of impact parameter
from targeted galaxy populations.
The generic result is a high incidence of
strong, coincident \hi\ absorption 
demanding a substantial, cool CGM \citep{pwc+11,thom12,werk+14}.  
On this point, there is community-wide
agreement.  To convert the measured \hi\ column densities 
\nhi\ into total hydrogen column density \Nh,
however, one must adopt an ionization model for the predominantly
ionized medium.  The standard approach is to assume that 
photoionization dominates, primarily driven by the EUVB.
Furthermore, one assumes equilibrium
conditions apply and one may then use observations of 
metal-line absorption coincident with the \hi\ Lyman series
to constrain the ionization model and thereby infer \Nh.


\begin{figure}
	\includegraphics[width=\columnwidth]{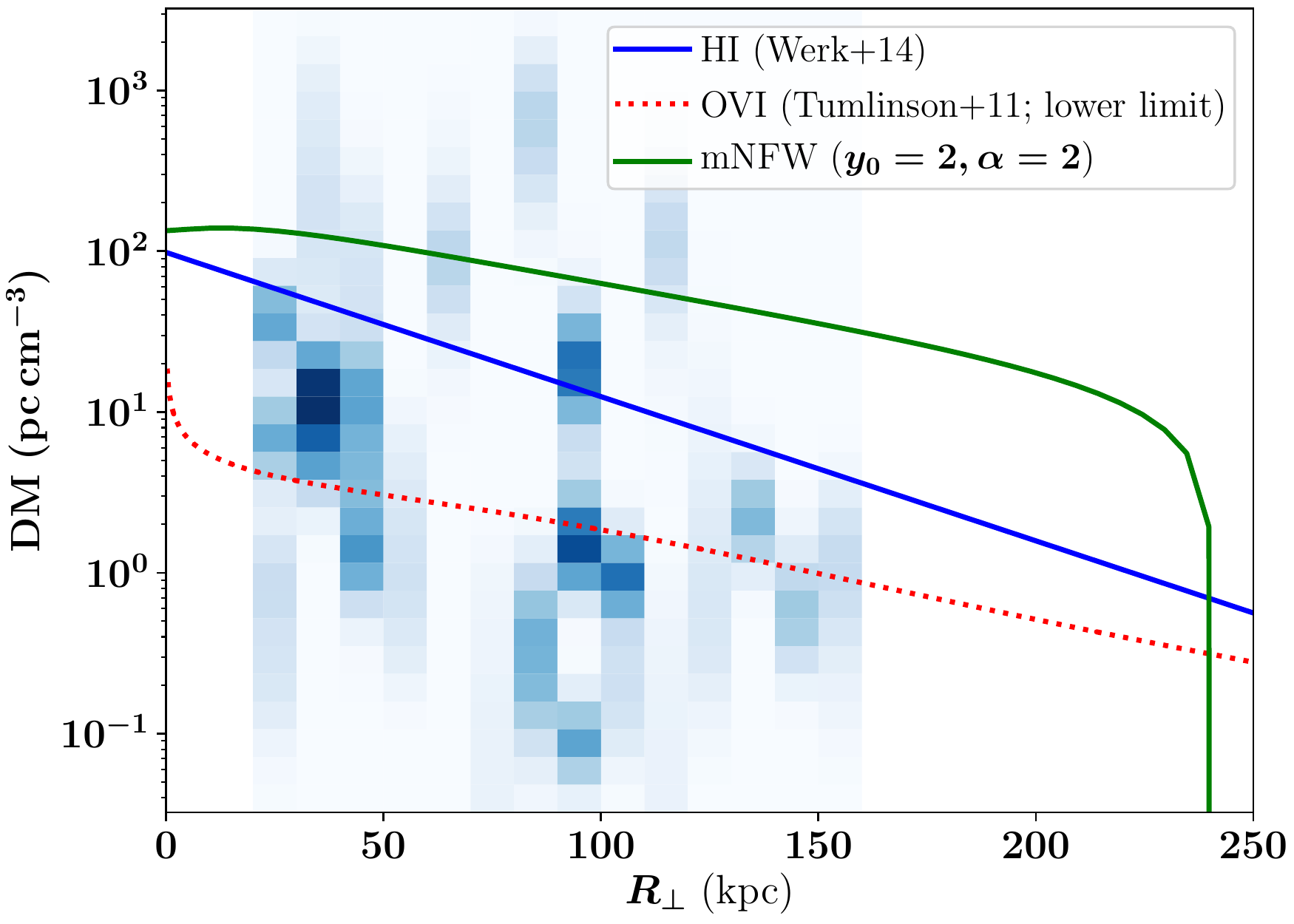}
    \caption{The blue 2D-histogram visualizes the PDF of DM$_{\rm CGM}^{\rm cool}$
    as a function of impact parameter. 
    Darker color indicates a radial
    bin with more individual measurements and less scatter within those
    measurements.
    The results were derived from ionic column density observations and
    photoionization models (yielding $N_{\rm H}$ estimates) of quasar sightlines 
    penetrating $\mrperp \le 150$\,kpc within L* galaxy halos at $z \sim 0.2$
    \citep{prochaska17}.  One notes substantial scatter in 
    $DM_{\rm CGM}^{\rm cool}$ at all radii which reflects dispersion
    in the observed column densities and, likely, systematic
    uncertainty in ionization modeling.  The curves on the plot are: 
    (red, dotted) a lower-limit estimate to DM of highly ionized
    halo gas based on observations of \ovi\ \citep{ttw+11};
    (blue) estimates of $DM_{\rm CGM}^{\rm cool}$ derived from the
    underlying PDF \citep{werk+14};
    (green) the DM profile for our fiducial Milky Way halo.
    \label{fig:cgm_empirical}
    }
\end{figure}

Figure~\ref{fig:cgm_empirical} shows a reproduction of the \Nh\ probability
distribution function (PDF) -- as a function of radial bins --
derived from the COS-Halos
survey by \cite{prochaska17}, expressed as DM values 
assuming the $\mu_e = 1.167$ correction for Helium.  
Darker color indicates a radial
bin with more individual measurements and less scatter within those 
measurements.
The PDF was estimated from
the CGM analysis of 31 $L*$ galaxies at $z \sim 0.2$
using an MCMC algorithm
that minimizes the difference between the ionization
model and observed ionic column densities.
Two results are evident:
  (i) there is a large dispersion in the inferred
  DM values of this cool phase;
  (ii) the largest values exceed even 100\,\dmunits.
On the latter point, we caution that those sightlines
with the highest inferred DM values have the largest
ionization corrections and even these uncertainties may be underestimated.
Nevertheless, one recovers integrated \dmhalo\ values of several
tens \dmunits\ to $\mrperp = 150$\,kpc and likely beyond\footnote{
The experiment was only designed to examine gas at $\mrperp \le 150$\,kpc.}.
Overplotted on Figure~\ref{fig:cgm_empirical} is an estimate of the
\dmrperp\ profile (blue curve) 
based on the fit of \cite{werk+14} to the cool CGM measurements
and the \dmrperp\ profile for our fiducial mNFW halo (green curve).
We find that the cool CGM may contain a substantial fraction
of a halo's baryons although likely less than the full 
cosmic fraction.  In any case, the cool gas in 
L* halos will contribute to the integrated DM of any distant FRBs.

Nearly every star-forming galaxy in the COS-Halos sample
also exhibits strong \ovi\ absorption.  This requires a
second, more highly ionized phase of gas within the halo 
\citep{ttw+11}.  One may estimate a lower limit to
the $\dm{OVI}$ associated with this gas by assuming conservative
estimates for the ionization fraction of \ovi\ 
(\fovi) and the gas metallicity $Z$ (analogous to 
our Equation~\ref{eqn:DM_OVII} for \ovii).  
Taking $\mfovi = 0.2$ and a one-third solar metallicity
($Z = Z_\odot/3$), we generate the curve in Figure~\ref{fig:cgm_empirical}
using the parameterization of \Novi\ by \cite{mp17}.
While this curve falls well below the DM profile for even the low
ionization phase,  we re-emphasize that the assumed 
\fovi\ and $Z$ values are conservative and order-of-magnitude
higher values are possible.
It is possible if not plausible that \ovi\ traces a warm-hot component
(where \ovii\ dominates) that 
comprises the remainder of baryons in the halos.

Designing experiments to assess the distribution of ionized
baryons in halos of L* galaxies faces distinct advantages and
challenges compared to higher mass halos.  On the positive side,
L* galaxies are far more common, e.g.\ one estimates
$p(\mmhalo=10^{11.5} - 10^{12.5} \mmsun; z=1) = 4.8$ for
$\mrperp \le 200$\,kpc.
On the other hand, the typical DM value from a single
L* halo may be only several tens \dmunits\ which is
comparable to the scatter from the signatures associated with
the Galaxy and the FRB host galaxy.  Nevertheless, a set of
well-localized low-$z$ FRBs events \citep[e.g.][]{mahony18,kb19}
where one maps the location of all $L > 0.1L*$ galaxies will offer
unique insight into galactic halo gas.



\begin{figure}
	\includegraphics[width=\columnwidth]{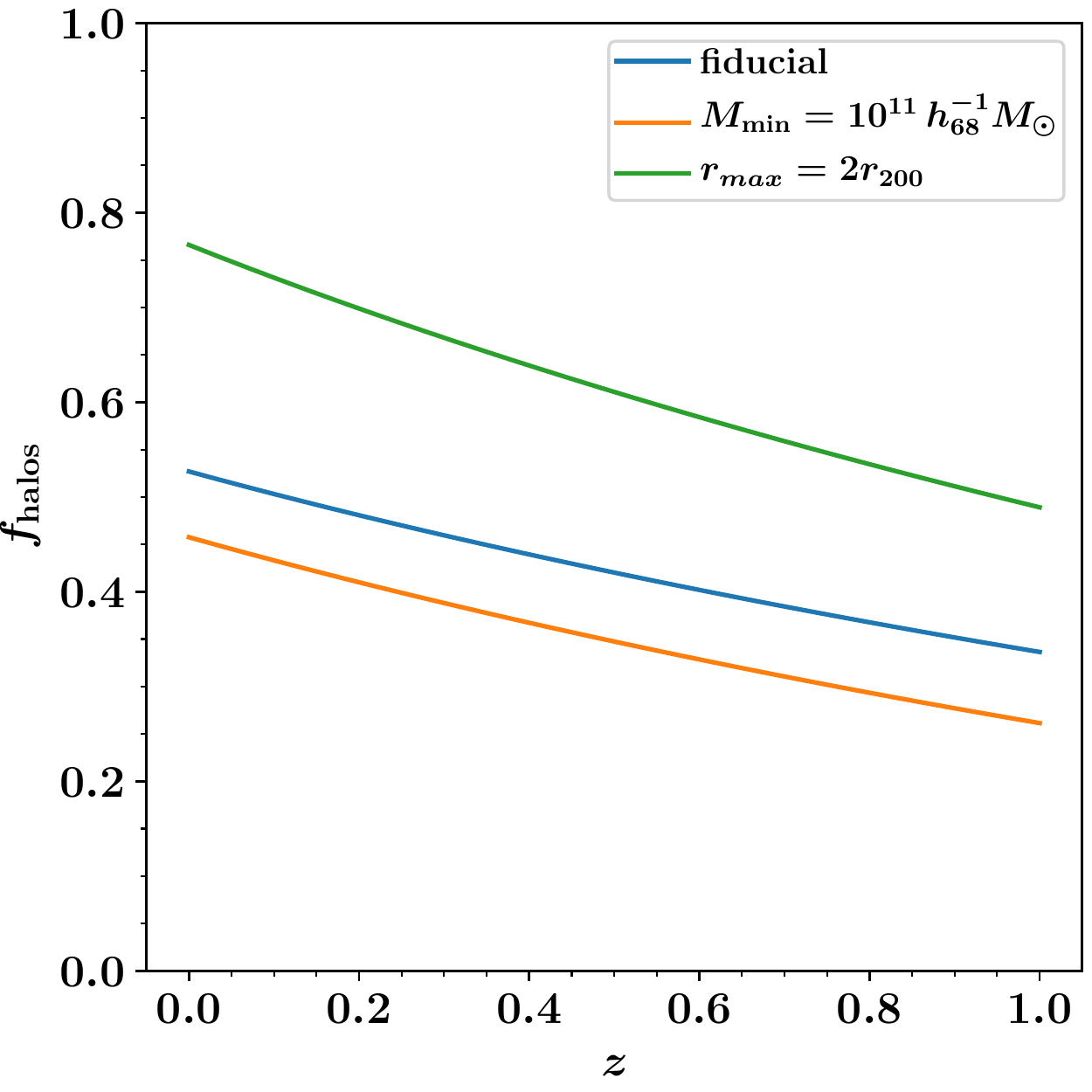}
    \caption{Estimates for the cosmic fraction of 
    dark matter in halos relative to the total dark matter
    mass density as a function of redshift.
    The fiducial model (blue) integrates the mass function down
    to $\mmmin = 10^{10} \mmsun$ and includes only matter within
    the virial radius ($\mrmax = \mrvir$).
    The other two curves vary \mmin\ (orange) and \rmax\ (green).
    At $z \sim 0$, approximately $50\%$ of the mass in the universe 
    lies within one virial radius of a dark matter halo.
    This function declines from to $z=0-1$
    indicating greater sensitivity to halo gas at
    lower redshifts.
    \label{fig:fhalo}
    }
\end{figure}

\section{Discussion}
\label{sec:discuss}

We now perform a few exercises 
motivated by the previous sections.  These are designed
to further illustrate the scientific potential and observational
challenges associated with probing halo gas using FRB
observations.

The discussion focuses on the cosmic dispersion measure 
which includes
all diffuse, ionized gas in the universe
beyond the Local Group and not including 
any contribution from the FRB host. 
The average \dmacosmic\ may be derived from the universe's
average baryonic mass density,

\begin{equation}
\mdmacosmic = \int ds \, \bar n_e / (1+z)  \;\; ,
\end{equation}
with
$\bar n_e = f_d \rho_b(z) \mu_e / \mu_m m_p$ and where
$f_d(z)$ is the fraction of cosmic baryons in diffuse ionized gas,
$\rho_b \equiv \Omega_b \rho_c$, 
and
$\mu_m = 1.3$ and $\mu_e=1.1667$ account for the mass and electrons of Helium.
This is the mean DM signal one predicts from extragalactic gas
for an ensemble of FRB events.
Any individual FRB, meanwhile, will record a specific
\dmcosmic\ value dependent on the precise distribution of 
matter foreground to it.
In the following,
we separate \dmcosmic\ into two components:
\dmhalos\ and \dmigm\, the diffuse gas within dark
matter halos and the lower density medium between them.
For an empirical discussion of \dmigm\ based on observations
of the \lya\ forest, we refer
the reader to \cite{shull18}.

\subsection{Integrated Models}

\citetalias{mcquinn14} studied the variance in the dispersion
measure to distant FRBs in the context of several halo
models.  In this sub-section, we expand on his work by adopting the mNFW halo models from Section~\ref{sec:models}
and varying several additional parameters.
Similar to \citetalias{mcquinn14} we consider models
with a minimum halo mass \mmin\ capable of
retaining the majority of its baryons.  
For $\mmhalo < \mmmin$, we set $\mfhb = 0$ and
let $\mfhb = 0.75$ otherwise.   
The other halo parameter explored is \rmax, 
which defines the radial extent of the halos.

As defined above,
the cosmic contribution (\dmcosmic) to \dmfrb\ combines
the gas within halos \dmhalos\ with the lower
density medium between them \dmigm. 
Given the build-up
of cosmic structures in time, the relative contributions
of \dmigm\ and \dmhalos\ varies with redshift,
with decreasing \dmhalos\ at higher $z$.
In Figure~\ref{fig:fhalo} we plot
$\mfhalo \equiv \rho_{\rm halos}/\rho_m$
as a function
of redshift for several assumptions on \mmin\ and \rmax.
Over the past $\sim 10$\,Gyr, the fraction of matter within
halos has risen from a few tens percent to $\sim 50\%$ today.
It is also evident that \fhalo\ has a strong dependence on
\rmax\ and only a modest dependency on \mmin.  
Another conclusion is that the contribution
of halos to \dmcosmic\ will have a stochastic nature 
that declines (in a given redshift interval)
with increasing redshift until one is left
primarily with the variance in $\dm{IGM}$.

\begin{figure}
	\includegraphics[width=\columnwidth]{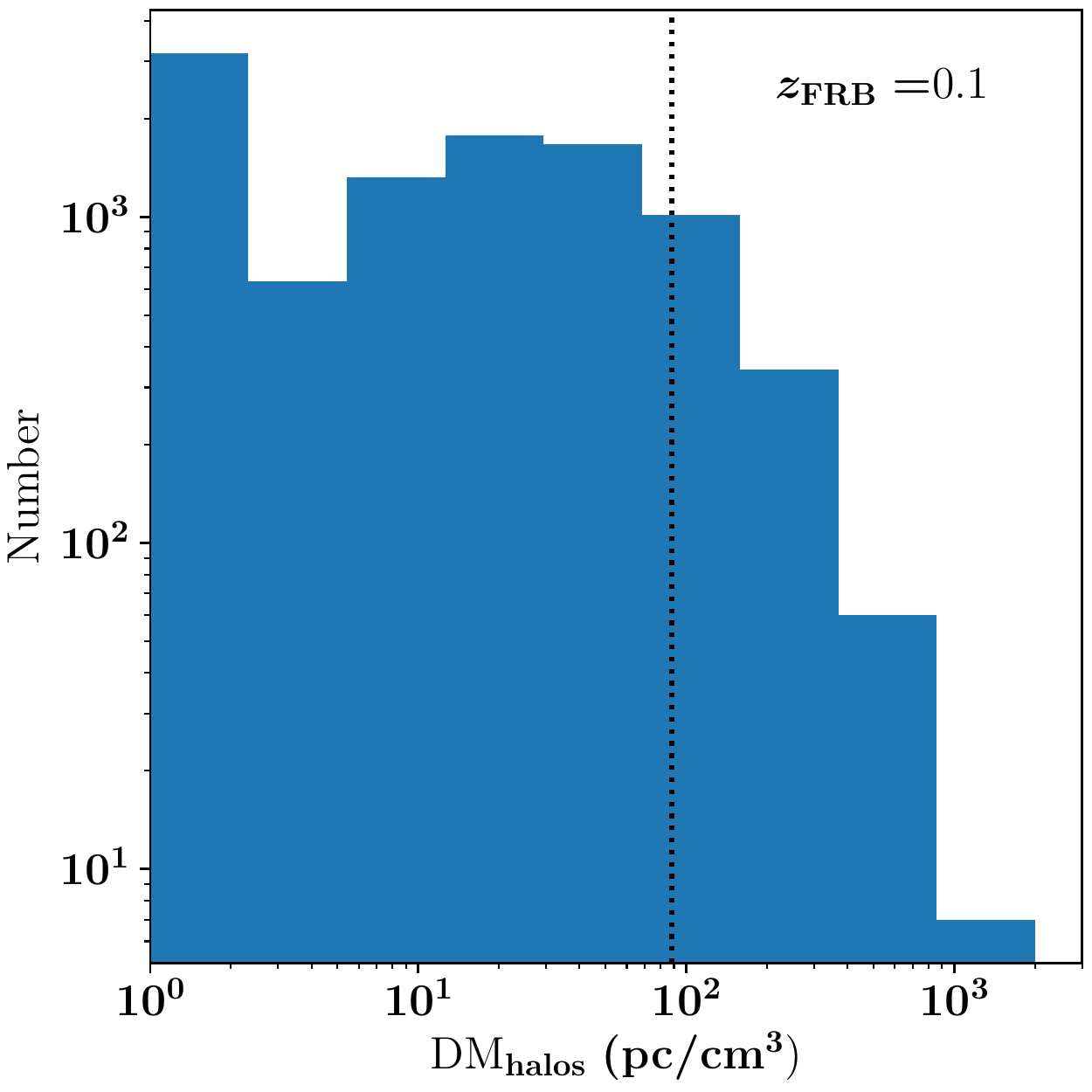}
    \caption{Probability distribution function (unnormalized)
    of $\dm{halos}$ for a population of FRBs at $\mzfrb=0.1$
    that considers only gas in foreground halos.  The halos 
    adopt the fiducial mNFW model with $\mfhb = 0.75$, 
    $\mrmax = \mrvir$, and
    $\mmmin = 10^{10} \mmsun$ and no clustering (i.e.\ the 
    halos are randomly placed; clustering would further 
    broaden the PDF).  The 95\%\ interval spans from 
    $\dm{halos} = [0, 203] \mdmunits$.  For reference, the
    dotted line shows the average \dmacosmic\ evaluated to $z=0.1$
    (halos + IGM).
    \label{fig:sigmaDM}
    }
\end{figure}

To explore this stochastic nature, we performed the following
exercise.  Using the Aemulus halo mass function, 
we generated a `random box Universe' with $l \approx 200 \, \rm cMpc$
on each side of its base and extending to $z=1$ in a series of $\delta z = 0.1$
layers.  In each layer, we add a random draw of $N_{\rm halos}$
assuming Gaussian statistics with mean 
$\bar N =  {\bar n}_c(M \ge \mmmin; z) V_{\rm layer}$ 
where ${\bar n}_c(z)$ is the average comoving number density
of halos with $M \ge \mmmin$ and 
$V_{\rm layer}$ is the co-moving volume of the layer.
We then draw halo masses \mhalo\ from the mass function at 
the mean redshift of the layer and place them randomly within
it.  Halos with $\mmhalo < 10^{14} \mmsun$ have gas profiles following the 
fiducial mNFW model with $\mfhb = 0.75$ and $\mrmax = \mrvir$.
Higher mass halos adopt the profile described
by Equation~\ref{eqn:clusters} with radial parameters
scaled by the virial radius.
Note that we have ignored the clustering of galaxies and 
large-scale structures both within each layer and between
layers, an effect we defer to future work (see also \citetalias{mcquinn14}).
Our 10,000 FRB sightlines are then drawn at regular 
intervals on a grid with 2\,cMpc spacing. 

Figure~\ref{fig:sigmaDM} shows the variation in 
$\dm{halos}$ for the 10,000 sightlines through the
first layer of this random universe.  
In this sample, there is over
three orders of magnitude variation.  The lowest
$\approx 10\%$ do not intersect a single halo and 
give zero value (here set to 1\,\dmunits).
The highest $\dm{halos}$ values pass within
a galaxy cluster halo with $\mmhalo > 10^{14} \mmsun$.
For this fiducial model, we find that halos in 
logarithmic mass intervals from 
$\mmhalo = 10^{11} - 10^{15} \mmsun$ contribute
nearly equally to \dmhalos\ on average.
The scatter, therefore, is dominated by the high 
mass halos which are Poisson distributed. 
Returning to Figure~\ref{fig:sigmaDM}, we further
emphasize that the relative scatter in \dmcosmic\
will be largest for low-$z$ FRBs where the total
value is small.

\begin{figure}
	\includegraphics[width=\columnwidth]{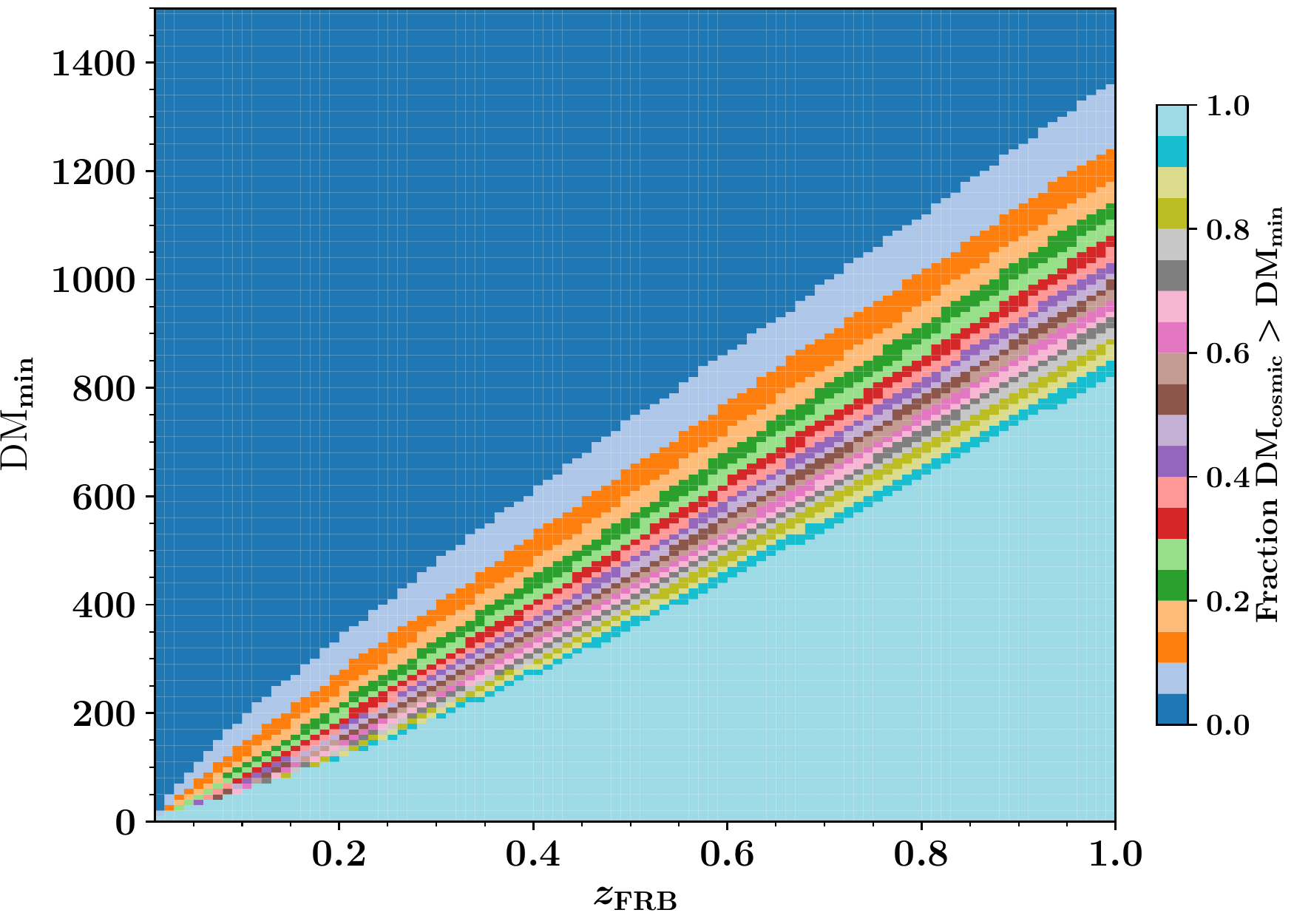}
    \caption{Fraction of 10,000 random sightlines with
    $\mdmcosmic > \, \rm DM_{min}$.
    The colored bands illustrate the scatter in the 
    predicted values.  The large blue swaths, meanwhile,
    show the excluded region (dark blue) and where the relation
    is always satisfied (light blue).
    \label{fig:fDM}
    }
\end{figure}

Continuing this exercise, we now consider the fraction of
FRB events that yield \dmcosmic\ greater than a given
minimum value. 
This may be used to set an upper limit to the redshift
of an FRB based on its measured DM.
Results are presented in Figure~\ref{fig:fDM}
where we have combined the \dmhalos\ values from our simulated
universe with \dmigm\ estimated from \dmacosmic\ and
\fhalo.  
Specifically, 
\begin{equation}
\mdmigm = \int [1- \mfhb \, \mfhalo(z)] \, \frac{d \mdmacosmic}{dz} \, dz \;\;  .
\end{equation}
with $d\mdmacosmic/dz$ the differential contribution 
to \dmacosmic\ in a $dz$ interval.
The trends are as expected, i.e.\ an
increasing fraction of sightlines have \dmcosmic\ exceeding
progressively higher $\dm{min}$ values.
At $z=1$, all of the sightlines have
$\mdmcosmic > 1000 \mdmunits$ due to the \dmigm\ component
which has no scatter in our model (the relative scatter
is estimated at $\sim 10 \%$; \citetalias{mcquinn14}).
It is also evident that 95\%\ the scatter in \dmcosmic\ at a given redshift
is generally greater than several hundred \dmunits.
Similarly, a given \dmcosmic\ value corresponds to a
range of \zfrb\ values with width $\Delta z_{\rm FRB} \approx 0.2$.

\begin{figure}
	\includegraphics[width=\columnwidth]{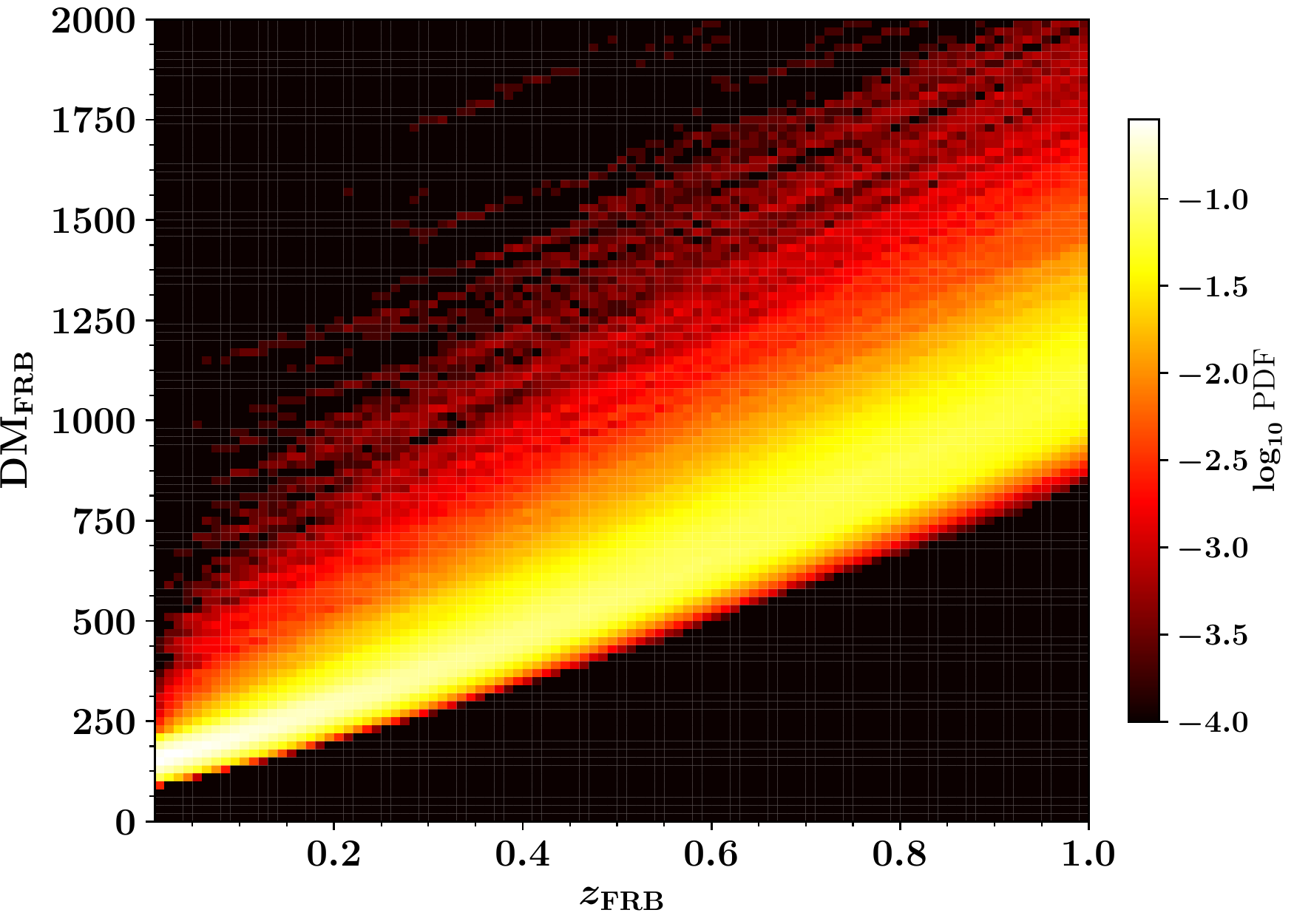}
    \caption{Estimations of the probability of a given 
    \dmfrb\ value ($\pm 10 \mdmunits$) at a given redshift
    ($\pm 0.005$).  The calculations attempt to capture
    all contributions to \dmfrb\ including simple estimations
    for the host and Local Group contributions (see the text
    for details).
    \label{fig:PDF}
    }
\end{figure}

Lastly, Figure~\ref{fig:PDF} 
show the probability distribution function (PDF) 
for recovering a specific \dmfrb\ value in boxes
of $\Delta z = 0.01, \Delta \rm DM = 20$.
For this calculation, we have included a 
nominal Local Group term 
${\rm DM}_{\rm LG} = 100 \pm 30 \mdmunits$ (uniform deviate)
and a host term
${\rm DM}_{\rm host} = 50 \pm 30 \mdmunits$ (uniform deviate).
One edge of the PDF smoothly follows the \dmigm\ track,
in part because we do not include any scatter in that term.
The other side shows the dispersion related to the Poisson
nature of massive halos.
As an example, for DM$_{\rm FRB} = 360 \mdmunits$,
we find the 95\%\ interval spans from 
$\mzfrb = [0.11, 0.35]$.
Clearly, \dmfrb\ measurements offer only a 
rough redshift estimate for any given FRB.

\begin{figure*}
	\includegraphics[width=2\columnwidth]{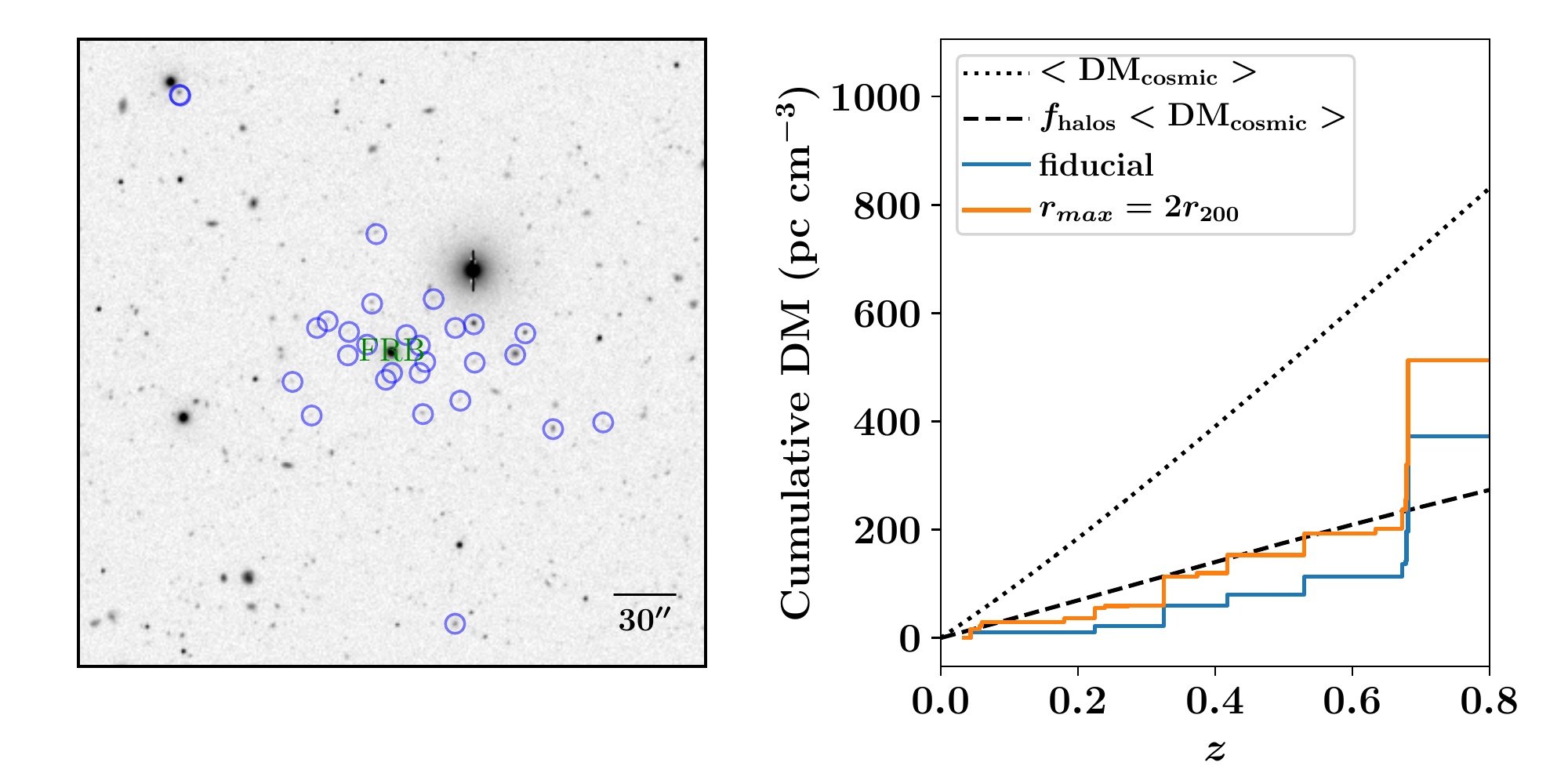}
    \caption{(Left: Image of the field surrounging PG1407+265 taken from the
    CASBaH survey \citep{x+18}.  In the following, we assume that an FRB
    event has occured at the redshift ($z=0.94$) and position of the quasar.
    Circles on the image identify galaxies foreground to PG1407+265 that also
    lie within 300\,pkpc of the sightline.
    Right: Colored solid lines show 
    integrated DM curves for the galactic halos identified in the field. 
    For these, we have adopted the halo mass estimated by 
    Prochaska et al.\ (2018)
    using abundance matching and then adopt the fiducial
    mNFW halo model with $\mfhb = 0.75$, $\mrmax = \mrvir$ (blue)
    or $\mrmax = 2 \mrvir$ (orange), 
    and concentration according to Equation~\ref{eqn:c}.
    For comparison, we show the \dmacosmic\ estimate and 
    $\mfhalo \mdmacosmic$ using the $\mrmax = \mrvir$ curve
    in Figure~\ref{fig:fhalo}.
    \label{fig:pg1407}
    }
\end{figure*}

\subsection{An Illustrative Example}
\label{sec:casbah}

As described above, 
future FRB analysis related to halo gas will synthesize
DM values with measurements of galaxies in the field surrounding
the event.  
To illustrate such analysis, we perform the following exercise
by leveraging the CASBaH galaxy database \citep{x+18}
which surveyed the galaxies surrounding 10~quasars at $z \sim 1$ observed
with HST/COS.
Designed to examine the relationship between galaxies and 
the IGM/CGM \citep{tripp+19,burchett18},  
the survey is relatively complete
to galaxies with stellar mass $M* \gtrsim 10^{9} \mmsun$ for 
redshifts $z<0.2$ with decreasing completeness and sensitivity at
higher redshifts.

Using these data we first assume 
that a FRB event occurred at the location and
redshift of a CASBaH quasar (e.g.\ PG1407+265; 
Figure~\ref{fig:pg1407}a).
Second, we consider all of the spectroscopically confirmed galaxies
with impact parameter $\mrperp < \mrmax$ from the FRB sightline
and consider two values of \rmax\ ($\mrvir, 2\mrvir)$.
Third, we estimate a DM value for each galaxy using the fiducial mNFW
halo model with $\mfhb = 0.75$ for all galaxies.
For the halos, we take the estimated halo mass from the CASBaH
project which applies the Moster relation 
to link stellar mass estimates to \mhalo.  
Lastly, we integrate to \rmax\ 
and adopt concentration values from Equation~\ref{eqn:c}.
We emphasize
that this process may underpredict $\dm{halos}$ because
we have not attempted to connect clustered galaxies to yet
more massive halos (i.e. groups and clusters;  see
section~\ref{sec:clusters}).

The cumulative \dmhalos\ curves contributed by foreground galaxies 
is shown in Figure~\ref{fig:pg1407}. 
For comparison, we also show \dmacosmic\
and $\mfhalo \mdmacosmic$ with \fhalo\ given in Figure~\ref{fig:fhalo}.
Comparing the fiducial \dmhalos\ curve
to the $\mfhalo \mdmacosmic$ evaluation (which both adopt
$\mrmax = \mrvir$), we note the former lies systematically
below expectation for $z < 0.7$.
This implies a lower incidence of massive halos than on
average along this sightline.
The impact of large-scale (i.e.\ groups, filaments) is also evident
as significant jumps in \dmhalos\ occur at redshifts where multiple
galaxies contribute.  
At $z \sim 0.7$, the sightline crosses within 300~kpc of
three luminous galaxies with $\mmhalo \sim 10^{12.5} \mmsun$
that raise \dmhalos\ substantially.
This illustrates the stochastic and discrete nature 
of cosmic structure and the resulting
Poisson behavior of \dmhalos.

\begin{figure}
	\includegraphics[width=\columnwidth]{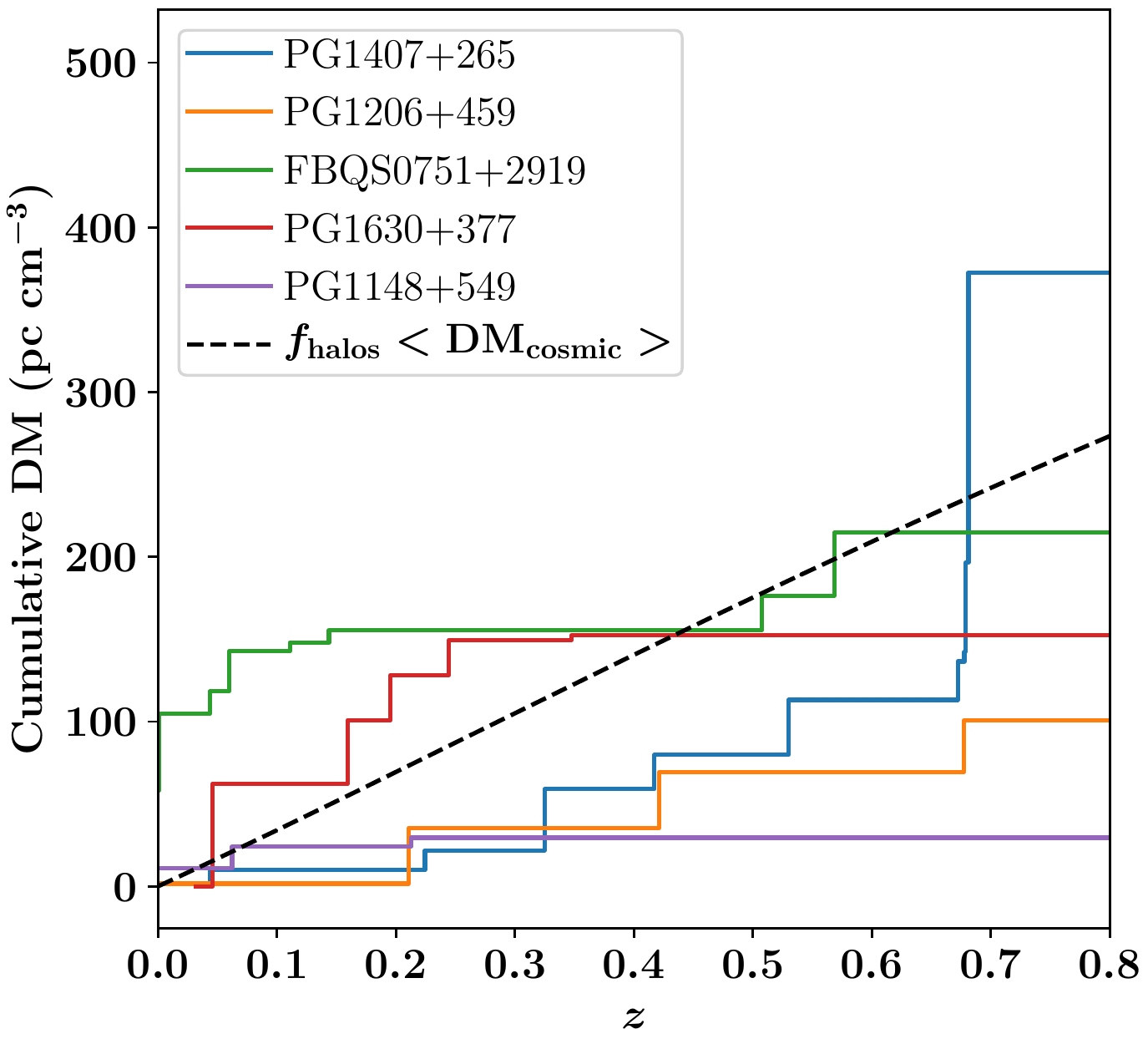}
    \caption{Similar to Figure~\ref{fig:pg1407} but for all of the
    fields with high survey completeness in the CASBaH project.
    \label{fig:allCASBaH}
    }
\end{figure}

Figure~\ref{fig:allCASBaH} extends the exercise to include all
5~CASBaH fields with high galaxy-survey completeness to $z \sim 0.7$.
The figure reveals a significant stochasticity in \dmhalos\
for the various sightlines which reflects scatter
in the number and masses of the halos intersected.
The most dramatic outlier is the FBQS0751+2919 field
where the sightline penetrates an overdensity of galaxies
at $z<0.1$.  The true \dmhalos\ value would even
higher if we assumed the galaxies occur within a 
$\mmhalo > 10^{13.5} \mmsun$ group.
By combining models of \dmhalos\ with measurements
of \dmfrb\ and the galaxies foreground to the events,
we will resolve the distribution of baryons within 
galactic halos.

\begin{figure*}
	\includegraphics[width=2\columnwidth]{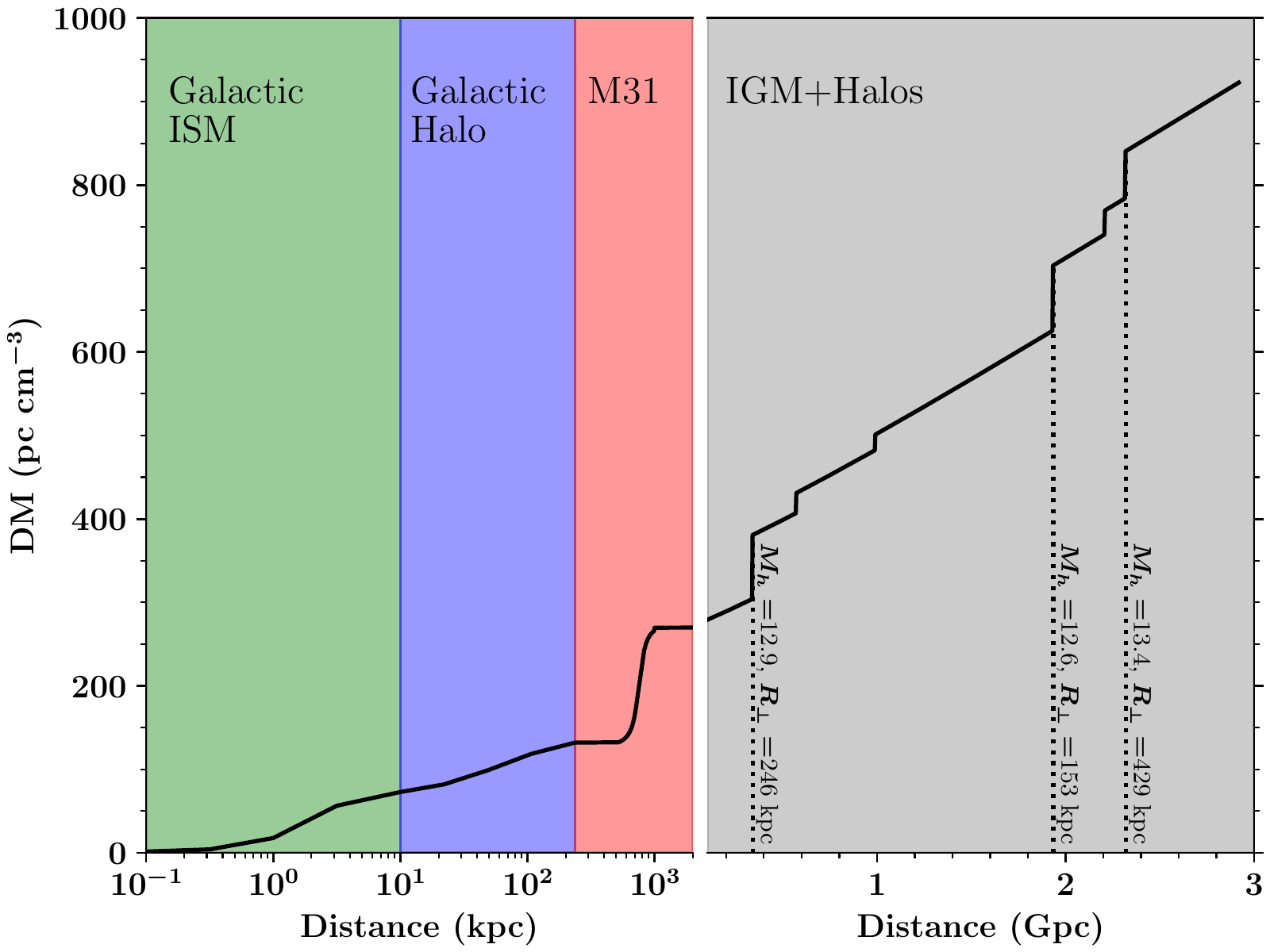}
    \caption{
    Illustrative estimation of the cumulative DM for a sightline
    traveling from Earth to the coordinate J004244.3+413009,
    chosen to intersect the halo of M31.
    The total distance depicted corresponds approximately 
    to an FRB at $z=0.5$.
    The figure details the contributions from the Galactic ISM,
    the Galactic halo, M31, and then the IGM and halos in the
    extragalactic universe.  The extragalactic halos with the 
    largest contribution are marked and their mass \mhalo\ and impact
    parameters \rperp\ are labeled.
    \label{fig:cartoon}
    }
\end{figure*}

\section{Summary and Concluding Remarks}
\label{sec:sum}
In this manuscript we have examined the contributions of
galactic halos to the integrated dispersion measures of
fast radio bursts.  Our treatment has spanned the scales
of our Galactic halo, the halos of our Local Group,
and the extragalactic halos of the distant universe.
We have not discussed extensively the DM signatures from
the halos hosting FRBs, but the models presented should
translate.

The principal results from our work may be summarized as:

\begin{itemize}
\item A range of models introduced to describe the
gas profiles of $\sim L*$ galaxy halos predict a diverse set
of DM profiles for sightlines intersecting them.  The
DM distributions are further dependent on the radius used 
to define a halo's physical extent (Section \ref{sec:models}).
\item The cool CGM of the Galaxy traced by HVCs yields a
small but non-negligible contribution to FRB DM 
measurements ($\sim 10 - 20 \mdmunits$; Section~\ref{sec:hvc}).
The warm/hot plasma traced by \ovi\ and \ovii\ absorption,
imposes an additional, estimated $\mdmmwh = 50-80\mdmunits$ (Section~\ref{sec:hot_halo}).
Hydrostatic equilibrium models of the Galactic halo with
the cosmic baryonic fraction can satisfy this constraint and
reproduce current estimates of the integrated DM to the LMC.
\item Halos of the Local Group galaxies -- especially M31 -- will 
have a detectable signal in all-sky DM maps from a large ensemble
of FRB observations.  One may also search for signatures of 
a Local Group medium (Section~\ref{sec:LG}).
\item Absorption-line analysis of CGM gas in $\sim L*$ galaxies
yields an estimated DM$_{\rm CGM}^{\rm cool} \sim 20 \mdmunits$
at $\mrperp = 100$\,kpc from cool ($T\sim 10^4$K) gas alone (see Figure~\ref{fig:cgm_empirical}).  Observations of \ovi\ and other
highly ionized species suggest a comparable DM signature from
a warm/hot plasma.
\item The DM values of more massive halos -- groups and clusters --
are sufficiently large that these will dominate the scatter in 
extragalactic DM signals.  We assess this scatter from random
sightlines intersecting simulated halos in a mock universe
and through select fields previously surveyed for galactic halos (Sections~\ref{sec:exgal} \&\ \ref{sec:discuss}).
\end{itemize}

As a means of further summarizing these results, 
Figure~\ref{fig:cartoon} presents the cumulative DM
from Earth to a notional $z=0.5$ FRB selected to lie behind
M31's halo.  For this illustration, we have adopted our fiducial
models for the Galactic and M31 halos and have generated a
random realization of halos in the extragalactic universe.
In this direction, the DM from the our Galaxy and its
Local Group is substantial.
The Galactic ISM is estimated to contribute $\approx 75 \mdmunits$,
the Galactic halo adds $\approx 50 \mdmunits$, 
and the putative M31 halo imposes an additional $\approx 100\mdmunits$.
This large signal underscores the potential of FRB DM
measurements to resolve halo gas in our local environment.

Examining the cosmic contributions,
the figure shows a simple, monotonic rise to the cumulative DM
from the IGM component.  A more realistic treatment may
show variability from intersections with
filaments in the cosmic web, as 
revealed by the \lya\ forest \citep[e.g.][]{shull18}.
The analysis does show, however, the stochastic
impact of individual halos
with $\mmhalo > 10^{11} \mmsun$ with DM jumping discretely
at the halo locations.  In this example, the largest
DM increments in the extragalactic regime
are from massive halos with $\mmhalo > 10^{12.5} \mmsun$.
Lower mass halos are also included (and shown) yet have 
DM contributions that are small and therefore similar
to the integrated IGM component.  In short, one predicts 
scatter in the cosmic DM contribution to be driven predominantly
by galaxy groups and clusters.  Looking forward to galaxy
surveys of the fields surrounding FRB events, the emphasis
should be placed on more massive, foreground halos.

These results motivate a further question
on survey design, namely: what is the optimal FRB sample in terms
of \zfrb\ for exploring the \dmhalos\ contribution to \dmfrb.
Qualitatively, there are three main considerations:
 (1) require large enough \dmcosmic\ values to dominate over the
 intrinsic scatter in \dmfrb\ from the Local Group and the host galaxy;
 (2) maximize the fractional variation in \dmhalos\ by minimizing the number
 of halos intersected;
 (3) have sufficient sensitivity with follow-up observations to 
 map the foreground galaxy population.
On the first point, if we assume 
$\sigma^2(\dm{LG+host}) = \sigma^2(\dm{LG}) 
+ \sigma^2(\dm{host})$
and assert for sake of example that 
$\sigma(\dm{LG}) = 50 \mdmunits$ and
$\sigma(\dm{host}) = 40 \mdmunits$ we
recover $\sigma^2(\dm{LG+host}) \approx 65 \mdmunits$.
For a nominal halo model with $\mmmin = 10^{10} \mmsun$
and $\mrmax = 2$, we estimate 
$\mdmigm > 5 \sigma(\dm{LG+host})$ at $\mzfrb = 0.44$.
At this redshift, we estimate the average intersection 
with $\mmhalo \ge 10^{12} \mmsun$ halos to be $N=1.2$
for $\mrperp \le \mrvir$, i.e. we are well within the
Poisson regime and expect large fluctuations from field
to field. 
And, conveniently, this redshift is sufficiently low that
one can build a deep spectroscopic survey
in finite observing time with 4m to 10m-class telescopes
(e.g.\ CASBaH; Figure~\ref{fig:pg1407}).

This paper has focused on the impacts of halo gas on the DM values encoded in
FRB events.  There are, however, additional physical effects that halo gas
may impose on these signals.  This includes, for example, temporal broadening
of the signal due to turbulence in the gas \citep{macquart13,mcquinn14,pn18,vedantham19}.
The inferences from photonionization modeling that the cool CGM may be
clumped on scales of $\sim 1$\,pc is quite favorable for this effect
\citep[e.g.][]{cantalupo14}.
One may also search for signatures of magnetic fields in galactic halos
on the rotation measure RM, as suggested by previous studies using radio
loud quasars \citep{bernet+08}.
We may address each of these in greater detail in future works.

Last, we remind the reader that all of the codes 
for the analysis presented here are public at
https://github.com/frbs.  This includes links to
the all-sky DM maps generated for the Local Group
(e.g.\ Figure~\ref{fig:hvc}).
These software will be updated as the field progresses
and new observations offer constraints on the gas within halos.
We encourage community development.


\section*{Acknowledgements}

We greatly appreciate the comments provided by
M. McQuinn, J. Werk, and T.-W. Lan on a earlier
version of the manuscript.
We thank A. Fox for helpful discussions and input
on HVCs and the Magellanic Clouds. We thank D. Lenz for generously sharing the HI4PI high-velocity N$_{\rm HI}$ datasets and for helpful discussions on interpreting the maps. 
Y. Zheng acknowledges support from the Miller Institute for Basic Research in Science. 





\begin{thebibliography}{}
\makeatletter
\relax
\def\mn@urlcharsother{\let\do\@makeother \do\$\do\&\do\#\do\^\do\_\do\%\do\~}
\def\mn@doi{\begingroup\mn@urlcharsother \@ifnextchar [ {\mn@doi@}
  {\mn@doi@[]}}
\def\mn@doi@[#1]#2{\def\@tempa{#1}\ifx\@tempa\@empty \href
  {http://dx.doi.org/#2} {doi:#2}\else \href {http://dx.doi.org/#2} {#1}\fi
  \endgroup}
\def\mn@eprint#1#2{\mn@eprint@#1:#2::\@nil}
\def\mn@eprint@arXiv#1{\href {http://arxiv.org/abs/#1} {{\tt arXiv:#1}}}
\def\mn@eprint@dblp#1{\href {http://dblp.uni-trier.de/rec/bibtex/#1.xml}
  {dblp:#1}}
\def\mn@eprint@#1:#2:#3:#4\@nil{\def\@tempa {#1}\def\@tempb {#2}\def\@tempc
  {#3}\ifx \@tempc \@empty \let \@tempc \@tempb \let \@tempb \@tempa \fi \ifx
  \@tempb \@empty \def\@tempb {arXiv}\fi \@ifundefined
  {mn@eprint@\@tempb}{\@tempb:\@tempc}{\expandafter \expandafter \csname
  mn@eprint@\@tempb\endcsname \expandafter{\@tempc}}}

\bibitem[\protect\citeauthoryear{{Albert} \& {Danly}}{{Albert} \&
  {Danly}}{2004}]{albert04}
{Albert} C.~E.,  {Danly} L.,  2004, in {van Woerden} H.,  {Wakker} B.~P.,
  {Schwarz} U.~J.,   {de Boer} K.~S.,  eds,  Astrophysics and Space Science
  Library Vol. 312, High Velocity Clouds. p.~73

\bibitem[\protect\citeauthoryear{{Allen}, {Schmidt}  \& {Fabian}}{{Allen}
  et~al.}{2002}]{allen02}
{Allen} S.~W.,  {Schmidt} R.~W.,   {Fabian} A.~C.,  2002, \mn@doi [\mnras]
  {10.1046/j.1365-8711.2002.05601.x}, \href
  {https://ui.adsabs.harvard.edu/#abs/2002MNRAS.334L..11A} {334, L11}

\bibitem[\protect\citeauthoryear{{Anderson} \& {Bregman}}{{Anderson} \&
  {Bregman}}{2010}]{anderson10}
{Anderson} M.~E.,  {Bregman} J.~N.,  2010, \mn@doi [\apj]
  {10.1088/0004-637X/714/1/320}, \href
  {http://adsabs.harvard.edu/abs/2010ApJ...714..320A} {714, 320}

\bibitem[\protect\citeauthoryear{{Anderson}, {Bregman}  \& {Dai}}{{Anderson}
  et~al.}{2013}]{anderson13}
{Anderson} M.~E.,  {Bregman} J.~N.,   {Dai} X.,  2013, \mn@doi [\apj]
  {10.1088/0004-637X/762/2/106}, \href
  {http://adsabs.harvard.edu/abs/2013ApJ...762..106A} {762, 106}

\bibitem[\protect\citeauthoryear{{Asplund}, {Grevesse}, {Sauval}  \&
  {Scott}}{{Asplund} et~al.}{2009}]{asplund09}
{Asplund} M.,  {Grevesse} N.,  {Sauval} A.~J.,   {Scott} P.,  2009, \mn@doi
  [\araa] {10.1146/annurev.astro.46.060407.145222}, \href
  {http://adsabs.harvard.edu/abs/2009ARA%26A..47..481A} {47, 481}

\bibitem[\protect\citeauthoryear{{Bahcall} \& {Peebles}}{{Bahcall} \&
  {Peebles}}{1969}]{bp69}
{Bahcall} J.~N.,  {Peebles} P.~J.~E.,  1969, \apjl, \href
  {http://adsabs.harvard.edu/cgi-bin/nph-bib_query?bibcode=1969ApJ...156L...7B&db_key=AST}
  {156, L7+}

\bibitem[\protect\citeauthoryear{{Bandura} et~al.,}{{Bandura}
  et~al.}{2014}]{CHIME}
{Bandura} K.,  et~al., 2014, in Ground-based and Airborne Telescopes V. p.
  914522 (\mn@eprint {arXiv} {1406.2288}), \mn@doi{10.1117/12.2054950}

\bibitem[\protect\citeauthoryear{{Bannister} \& {et al.}}{{Bannister} \& {et
  al.}}{2019}]{kb19}
{Bannister} K.,  {et al.} 2019, \mn@doi [Science] {10.1088/0067-0049/217/2/21},
  \href {http://adsabs.harvard.edu/abs/2015ApJS..217...21F} {in prep}

\bibitem[\protect\citeauthoryear{{Barger}, {Lehner}  \& {Howk}}{{Barger}
  et~al.}{2016}]{barger16}
{Barger} K.~A.,  {Lehner} N.,   {Howk} J.~C.,  2016, \mn@doi [\apj]
  {10.3847/0004-637X/817/2/91}, \href
  {http://adsabs.harvard.edu/abs/2016ApJ...817...91B} {817, 91}

\bibitem[\protect\citeauthoryear{{Behroozi}, {Conroy}  \&
  {Wechsler}}{{Behroozi} et~al.}{2010}]{behroozi10}
{Behroozi} P.~S.,  {Conroy} C.,   {Wechsler} R.~H.,  2010, \mn@doi [\apj]
  {10.1088/0004-637X/717/1/379}, \href
  {https://ui.adsabs.harvard.edu/#abs/2010ApJ...717..379B} {717, 379}

\bibitem[\protect\citeauthoryear{{Bernet}, {Miniati}, {Lilly}, {Kronberg}  \&
  {Dessauges-Zavadsky}}{{Bernet} et~al.}{2008}]{bernet+08}
{Bernet} M.~L.,  {Miniati} F.,  {Lilly} S.~J.,  {Kronberg} P.~P.,
  {Dessauges-Zavadsky} M.,  2008, \mn@doi [\nat] {10.1038/nature07105}, \href
  {http://adsabs.harvard.edu/abs/2008Natur.454..302B} {454, 302}

\bibitem[\protect\citeauthoryear{{Blitz}, {Spergel}, {Teuben}, {Hartmann}  \&
  {Burton}}{{Blitz} et~al.}{1999}]{blitz99}
{Blitz} L.,  {Spergel} D.~N.,  {Teuben} P.~J.,  {Hartmann} D.,   {Burton}
  W.~B.,  1999, \mn@doi [\apj] {10.1086/306963}, \href
  {http://adsabs.harvard.edu/abs/1999ApJ...514..818B} {514, 818}

\bibitem[\protect\citeauthoryear{{Bolton}, {Treu}, {Koopmans}, {Gavazzi},
  {Moustakas}, {Burles}, {Schlegel}  \& {Wayth}}{{Bolton}
  et~al.}{2008}]{slacsVII}
{Bolton} A.~S.,  {Treu} T.,  {Koopmans} L.~V.~E.,  {Gavazzi} R.,  {Moustakas}
  L.~A.,  {Burles} S.,  {Schlegel} D.~J.,   {Wayth} R.,  2008, \mn@doi [\apj]
  {10.1086/589989}, \href {http://adsabs.harvard.edu/abs/2008ApJ...684..248B}
  {684, 248}

\bibitem[\protect\citeauthoryear{{Booth}, {Schaye}, {Delgado}  \& {Dalla
  Vecchia}}{{Booth} et~al.}{2012}]{booth+12}
{Booth} C.~M.,  {Schaye} J.,  {Delgado} J.~D.,   {Dalla Vecchia} C.,  2012,
  \mn@doi [\mnras] {10.1111/j.1365-2966.2011.20047.x}, \href
  {http://adsabs.harvard.edu/abs/2012MNRAS.420.1053B} {420, 1053}

\bibitem[\protect\citeauthoryear{{Bordoloi} et~al.,}{{Bordoloi}
  et~al.}{2014}]{bordoloi14}
{Bordoloi} R.,  et~al., 2014, \mn@doi [\apj] {10.1088/0004-637X/796/2/136},
  \href {http://adsabs.harvard.edu/abs/2014ApJ...796..136B} {796, 136}

\bibitem[\protect\citeauthoryear{{Bowen} et~al.,}{{Bowen}
  et~al.}{2008}]{bjt+08}
{Bowen} D.~V.,  et~al., 2008, \mn@doi [\apjs] {10.1086/524773}, \href
  {http://adsabs.harvard.edu/abs/2008ApJS..176...59B} {176, 59}

\bibitem[\protect\citeauthoryear{{Burchett} et~al.,}{{Burchett}
  et~al.}{2018}]{burchett18}
{Burchett} J.~N.,  et~al., 2018, preprint, \href
  {http://adsabs.harvard.edu/abs/2018arXiv181006560B} {} (\mn@eprint {arXiv}
  {1810.06560})

\bibitem[\protect\citeauthoryear{{Burles} \& {Tytler}}{{Burles} \&
  {Tytler}}{1996}]{bt96}
{Burles} S.,  {Tytler} D.,  1996, \mn@doi [\apj] {10.1086/176994}, \href
  {http://adsabs.harvard.edu/abs/1996ApJ...460..584B} {460, 584}

\bibitem[\protect\citeauthoryear{{Cantalupo}, {Arrigoni-Battaia}, {Prochaska},
  {Hennawi}  \& {Madau}}{{Cantalupo} et~al.}{2014}]{cantalupo14}
{Cantalupo} S.,  {Arrigoni-Battaia} F.,  {Prochaska} J.~X.,  {Hennawi} J.~F.,
  {Madau} P.,  2014, \mn@doi [\nat] {10.1038/nature12898}, \href
  {http://adsabs.harvard.edu/abs/2014Natur.506...63C} {506, 63}

\bibitem[\protect\citeauthoryear{{Chatterjee} et~al.,}{{Chatterjee}
  et~al.}{2017}]{chatterjee17}
{Chatterjee} S.,  et~al., 2017, \mn@doi [\nat] {10.1038/nature20797}, \href
  {http://adsabs.harvard.edu/abs/2017Natur.541...58C} {541, 58}

\bibitem[\protect\citeauthoryear{{Chen}, {Wild}, {Tinker}, {Gauthier},
  {Helsby}, {Shectman}  \& {Thompson}}{{Chen} et~al.}{2010}]{cwt+10}
{Chen} H.-W.,  {Wild} V.,  {Tinker} J.~L.,  {Gauthier} J.-R.,  {Helsby} J.~E.,
  {Shectman} S.~A.,   {Thompson} I.~B.,  2010, \mn@doi [\apjl]
  {10.1088/2041-8205/724/2/L176}, \href
  {http://adsabs.harvard.edu/abs/2010ApJ...724L.176C} {724, L176}

\bibitem[\protect\citeauthoryear{{Christensen}, {Dave}, {Brooks}, {Quinn}  \&
  {Shen}}{{Christensen} et~al.}{2018}]{christensen18}
{Christensen} C.~R.,  {Dave} R.,  {Brooks} A.,  {Quinn} T.,   {Shen} S.,  2018,
  preprint, \href {http://adsabs.harvard.edu/abs/2018arXiv180807872C} {}
  (\mn@eprint {arXiv} {1808.07872})

\bibitem[\protect\citeauthoryear{{Collins}, {Shull}  \& {Giroux}}{{Collins}
  et~al.}{2009}]{collins09}
{Collins} J.~A.,  {Shull} J.~M.,   {Giroux} M.~L.,  2009, \mn@doi [\apj]
  {10.1088/0004-637X/705/1/962}, \href
  {http://adsabs.harvard.edu/abs/2009ApJ...705..962C} {705, 962}

\bibitem[\protect\citeauthoryear{{Cooke}, {Pettini}  \& {Steidel}}{{Cooke}
  et~al.}{2018}]{cooke18}
{Cooke} R.~J.,  {Pettini} M.,   {Steidel} C.~C.,  2018, \mn@doi [\apj]
  {10.3847/1538-4357/aaab53}, \href
  {https://ui.adsabs.harvard.edu/#abs/2018ApJ...855..102C} {855, 102}

\bibitem[\protect\citeauthoryear{{Cordes} \& {Lazio}}{{Cordes} \&
  {Lazio}}{2002}]{ne2001a}
{Cordes} J.~M.,  {Lazio} T.~J.~W.,  2002, arXiv Astrophysics e-prints, \href
  {http://adsabs.harvard.edu/abs/2002astro.ph..7156C} {}

\bibitem[\protect\citeauthoryear{{Cordes} \& {Lazio}}{{Cordes} \&
  {Lazio}}{2003}]{ne2001b}
{Cordes} J.~M.,  {Lazio} T.~J.~W.,  2003, ArXiv Astrophysics e-prints, \href
  {http://adsabs.harvard.edu/abs/2003astro.ph..1598C} {}

\bibitem[\protect\citeauthoryear{{D'Onghia} \& {Fox}}{{D'Onghia} \&
  {Fox}}{2016}]{df16}
{D'Onghia} E.,  {Fox} A.~J.,  2016, \mn@doi [\araa]
  {10.1146/annurev-astro-081915-023251}, \href
  {http://adsabs.harvard.edu/abs/2016ARA%26A..54..363D} {54, 363}

\bibitem[\protect\citeauthoryear{{Dai}, {Bregman}, {Kochanek}  \&
  {Rasia}}{{Dai} et~al.}{2010}]{dai10}
{Dai} X.,  {Bregman} J.~N.,  {Kochanek} C.~S.,   {Rasia} E.,  2010, \mn@doi
  [\apj] {10.1088/0004-637X/719/1/119}, \href
  {http://adsabs.harvard.edu/abs/2010ApJ...719..119D} {719, 119}

\bibitem[\protect\citeauthoryear{{Diemer} \& {Kravtsov}}{{Diemer} \&
  {Kravtsov}}{2014}]{dimer14}
{Diemer} B.,  {Kravtsov} A.~V.,  2014, \mn@doi [\apj]
  {10.1088/0004-637X/789/1/1}, \href
  {http://adsabs.harvard.edu/abs/2014ApJ...789....1D} {789, 1}

\bibitem[\protect\citeauthoryear{{Dolag}, {Gaensler}, {Beck}  \&
  {Beck}}{{Dolag} et~al.}{2015}]{dolag15}
{Dolag} K.,  {Gaensler} B.~M.,  {Beck} A.~M.,   {Beck} M.~C.,  2015, \mn@doi
  [\mnras] {10.1093/mnras/stv1190}, \href
  {http://adsabs.harvard.edu/abs/2015MNRAS.451.4277D} {451, 4277}

\bibitem[\protect\citeauthoryear{{Faerman}, {Sternberg}  \& {McKee}}{{Faerman}
  et~al.}{2017}]{faerman17}
{Faerman} Y.,  {Sternberg} A.,   {McKee} C.~F.,  2017, \mn@doi [\apj]
  {10.3847/1538-4357/835/1/52}, \href
  {http://adsabs.harvard.edu/abs/2017ApJ...835...52F} {835, 52}

\bibitem[\protect\citeauthoryear{{Fang}, {Bullock}  \& {Boylan-Kolchin}}{{Fang}
  et~al.}{2013}]{fang+13}
{Fang} T.,  {Bullock} J.,   {Boylan-Kolchin} M.,  2013, \mn@doi [\apj]
  {10.1088/0004-637X/762/1/20}, \href
  {http://adsabs.harvard.edu/abs/2013ApJ...762...20F} {762, 20}

\bibitem[\protect\citeauthoryear{{Fang}, {Buote}, {Bullock}  \& {Ma}}{{Fang}
  et~al.}{2015}]{fang+15}
{Fang} T.,  {Buote} D.,  {Bullock} J.,   {Ma} R.,  2015, \mn@doi [\apjs]
  {10.1088/0067-0049/217/2/21}, \href
  {http://adsabs.harvard.edu/abs/2015ApJS..217...21F} {217, 21}

\bibitem[\protect\citeauthoryear{{Faucher-Gigu{\`e}re}, {Prochaska}, {Lidz},
  {Hernquist}  \& {Zaldarriaga}}{{Faucher-Gigu{\`e}re} et~al.}{2008a}]{fpl+08}
{Faucher-Gigu{\`e}re} C.-A.,  {Prochaska} J.~X.,  {Lidz} A.,  {Hernquist} L.,
  {Zaldarriaga} M.,  2008a, \mn@doi [\apj] {10.1086/588648}, \href
  {http://adsabs.harvard.edu/abs/2008ApJ...681..831F} {681, 831}

\bibitem[\protect\citeauthoryear{{Faucher-Gigu{\`e}re}, {Lidz}, {Hernquist}  \&
  {Zaldarriaga}}{{Faucher-Gigu{\`e}re} et~al.}{2008b}]{flh+08}
{Faucher-Gigu{\`e}re} C.-A.,  {Lidz} A.,  {Hernquist} L.,   {Zaldarriaga} M.,
  2008b, \mn@doi [\apj] {10.1086/592289}, \href
  {http://adsabs.harvard.edu/abs/2008ApJ...688...85F} {688, 85}

\bibitem[\protect\citeauthoryear{{Fielding}, {Quataert}, {McCourt}  \&
  {Thompson}}{{Fielding} et~al.}{2017}]{fielding+17}
{Fielding} D.,  {Quataert} E.,  {McCourt} M.,   {Thompson} T.~A.,  2017,
  \mn@doi [\mnras] {10.1093/mnras/stw3326}, \href
  {http://adsabs.harvard.edu/abs/2017MNRAS.466.3810F} {466, 3810}

\bibitem[\protect\citeauthoryear{{Fox}, {Wakker}, {Savage}, {Tripp}, {Sembach}
  \& {Bland-Hawthorn}}{{Fox} et~al.}{2005}]{fox+05}
{Fox} A.~J.,  {Wakker} B.~P.,  {Savage} B.~D.,  {Tripp} T.~M.,  {Sembach}
  K.~R.,   {Bland-Hawthorn} J.,  2005, \mn@doi [\apj] {10.1086/431915}, \href
  {http://adsabs.harvard.edu/abs/2005ApJ...630..332F} {630, 332}

\bibitem[\protect\citeauthoryear{{Fox}, {Savage}  \& {Wakker}}{{Fox}
  et~al.}{2006}]{foxetal06}
{Fox} A.~J.,  {Savage} B.~D.,   {Wakker} B.~P.,  2006, \mn@doi [\apjs]
  {10.1086/504800}, \href
  {http://adsabs.harvard.edu/cgi-bin/nph-bib_query?bibcode=2006ApJS..165..229F&db_key=AST}
  {165, 229}

\bibitem[\protect\citeauthoryear{{Fox} et~al.,}{{Fox} et~al.}{2014}]{fox+14}
{Fox} A.~J.,  et~al., 2014, \mn@doi [\apj] {10.1088/0004-637X/787/2/147}, \href
  {http://adsabs.harvard.edu/abs/2014ApJ...787..147F} {787, 147}

\bibitem[\protect\citeauthoryear{{Frenk} et~al.,}{{Frenk}
  et~al.}{1999}]{frenk99}
{Frenk} C.~S.,  et~al., 1999, \mn@doi [\apj] {10.1086/307908}, \href
  {https://ui.adsabs.harvard.edu/#abs/1999ApJ...525..554F} {525, 554}

\bibitem[\protect\citeauthoryear{{Fukugita}, {Hogan}  \& {Peebles}}{{Fukugita}
  et~al.}{1998}]{fhp98}
{Fukugita} M.,  {Hogan} C.~J.,   {Peebles} P.~J.~E.,  1998, \apj, \href
  {http://adsabs.harvard.edu/cgi-bin/nph-bib_query?bibcode=1998ApJ...503..518F&amp;db_key=AST}
  {503, 518}

\bibitem[\protect\citeauthoryear{{Gaensler}, {Madsen}, {Chatterjee}  \&
  {Mao}}{{Gaensler} et~al.}{2008}]{gaensler08}
{Gaensler} B.~M.,  {Madsen} G.~J.,  {Chatterjee} S.,   {Mao} S.~A.,  2008,
  \mn@doi [\pasa] {10.1071/AS08004}, \href
  {http://adsabs.harvard.edu/abs/2008PASA...25..184G} {25, 184}

\bibitem[\protect\citeauthoryear{{Gibson}, {Giroux}, {Penton}, {Stocke},
  {Shull}  \& {Tumlinson}}{{Gibson} et~al.}{2001}]{gibson01}
{Gibson} B.~K.,  {Giroux} M.~L.,  {Penton} S.~V.,  {Stocke} J.~T.,  {Shull}
  J.~M.,   {Tumlinson} J.,  2001, \mn@doi [\aj] {10.1086/324227}, \href
  {http://adsabs.harvard.edu/abs/2001AJ....122.3280G} {122, 3280}

\bibitem[\protect\citeauthoryear{{HI4PI Collaboration} et~al.,}{{HI4PI
  Collaboration} et~al.}{2016}]{hi4pi}
{HI4PI Collaboration} et~al., 2016, \mn@doi [\aap]
  {10.1051/0004-6361/201629178}, \href
  {http://adsabs.harvard.edu/abs/2016A%26A...594A.116H} {594, A116}

\bibitem[\protect\citeauthoryear{{Hafen} et~al.,}{{Hafen}
  et~al.}{2018}]{hafen+18}
{Hafen} Z.,  et~al., 2018, arXiv e-prints, \href
  {http://adsabs.harvard.edu/abs/2018arXiv181111753H} {}

\bibitem[\protect\citeauthoryear{{Hill}, {Baxter}, {Lidz}, {Greco}  \&
  {Jain}}{{Hill} et~al.}{2018}]{hill18}
{Hill} J.~C.,  {Baxter} E.~J.,  {Lidz} A.,  {Greco} J.~P.,   {Jain} B.,  2018,
  \mn@doi [\prd] {10.1103/PhysRevD.97.083501}, \href
  {http://adsabs.harvard.edu/abs/2018PhRvD..97h3501H} {97, 083501}

\bibitem[\protect\citeauthoryear{{Hirata} \& {McQuinn}}{{Hirata} \&
  {McQuinn}}{2014}]{hirata14}
{Hirata} C.~M.,  {McQuinn} M.,  2014, \mn@doi [\mnras] {10.1093/mnras/stu509},
  \href {http://adsabs.harvard.edu/abs/2014MNRAS.440.3613H} {440, 3613}

\bibitem[\protect\citeauthoryear{{Hodges-Kluck}, {Miller}  \&
  {Bregman}}{{Hodges-Kluck} et~al.}{2016}]{hk2016}
{Hodges-Kluck} E.~J.,  {Miller} M.~J.,   {Bregman} J.~N.,  2016, \mn@doi [\apj]
  {10.3847/0004-637X/822/1/21}, \href
  {http://adsabs.harvard.edu/abs/2016ApJ...822...21H} {822, 21}

\bibitem[\protect\citeauthoryear{{Hoopes}, {Sembach}, {Howk}, {Savage}  \&
  {Fullerton}}{{Hoopes} et~al.}{2002}]{hoopes+02}
{Hoopes} C.~G.,  {Sembach} K.~R.,  {Howk} J.~C.,  {Savage} B.~D.,   {Fullerton}
  A.~W.,  2002, \mn@doi [\apj] {10.1086/339323}, \href
  {http://adsabs.harvard.edu/abs/2002ApJ...569..233H} {569, 233}

\bibitem[\protect\citeauthoryear{{Howk}, {Sembach}  \& {Savage}}{{Howk}
  et~al.}{2006}]{howk06}
{Howk} J.~C.,  {Sembach} K.~R.,   {Savage} B.~D.,  2006, \mn@doi [\apj]
  {10.1086/497352}, \href
  {https://ui.adsabs.harvard.edu/#abs/2006ApJ...637..333H} {637, 333}

\bibitem[\protect\citeauthoryear{{Howk} et~al.,}{{Howk} et~al.}{2017}]{howk+17}
{Howk} J.~C.,  et~al., 2017, \mn@doi [\apj] {10.3847/1538-4357/aa87b4}, \href
  {https://ui.adsabs.harvard.edu/#abs/2017ApJ...846..141H} {846, 141}

\bibitem[\protect\citeauthoryear{{Inoue}}{{Inoue}}{2004}]{inoue04}
{Inoue} S.,  2004, \mn@doi [\mnras] {10.1111/j.1365-2966.2004.07359.x}, \href
  {http://adsabs.harvard.edu/abs/2004MNRAS.348..999I} {348, 999}

\bibitem[\protect\citeauthoryear{{Kalberla} \& {Haud}}{{Kalberla} \&
  {Haud}}{2015}]{gass}
{Kalberla} P.~M.~W.,  {Haud} U.,  2015, \mn@doi [\aap]
  {10.1051/0004-6361/201525859}, \href
  {https://ui.adsabs.harvard.edu/#abs/2015A&A...578A..78K} {578, A78}

\bibitem[\protect\citeauthoryear{{Kalberla}, {Burton}, {Hartmann}, {Arnal},
  {Bajaja}, {Morras}  \& {P{\"o}ppel}}{{Kalberla} et~al.}{2005}]{LAB05}
{Kalberla} P.~M.~W.,  {Burton} W.~B.,  {Hartmann} D.,  {Arnal} E.~M.,  {Bajaja}
  E.,  {Morras} R.,   {P{\"o}ppel} W.~G.~L.,  2005, \mn@doi [\aap]
  {10.1051/0004-6361:20041864}, \href
  {http://adsabs.harvard.edu/abs/2005A%26A...440..775K} {440, 775}

\bibitem[\protect\citeauthoryear{{Keeney} et~al.,}{{Keeney}
  et~al.}{2017}]{keeney17}
{Keeney} B.~A.,  et~al., 2017, \mn@doi [The Astrophysical Journal Supplement
  Series] {10.3847/1538-4365/aa6b59}, \href
  {https://ui.adsabs.harvard.edu/#abs/2017ApJS..230....6K} {230, 6}

\bibitem[\protect\citeauthoryear{{Kunth}, {Lequeux}, {Sargent}  \&
  {Viallefond}}{{Kunth} et~al.}{1994}]{kunth94}
{Kunth} D.,  {Lequeux} J.,  {Sargent} W.~L.~W.,   {Viallefond} F.,  1994, \aap,
  \href {http://adsabs.harvard.edu/abs/1994A%26A...282..709K} {282, 709}

\bibitem[\protect\citeauthoryear{{Lanzetta}, {Bowen}, {Tytler}  \&
  {Webb}}{{Lanzetta} et~al.}{1995}]{lbt+95}
{Lanzetta} K.~M.,  {Bowen} D.~V.,  {Tytler} D.,   {Webb} J.~K.,  1995, \mn@doi
  [\apj] {10.1086/175459}, \href
  {http://adsabs.harvard.edu/abs/1995ApJ...442..538L} {442, 538}

\bibitem[\protect\citeauthoryear{{Lehner} \& {Howk}}{{Lehner} \&
  {Howk}}{2007}]{lh07}
{Lehner} N.,  {Howk} J.~C.,  2007, \mn@doi [\mnras]
  {10.1111/j.1365-2966.2007.11631.x}, \href
  {http://adsabs.harvard.edu/abs/2007MNRAS.377..687L} {377, 687}

\bibitem[\protect\citeauthoryear{{Lehner} \& {Howk}}{{Lehner} \&
  {Howk}}{2010}]{lh10}
{Lehner} N.,  {Howk} J.~C.,  2010, \mn@doi [\apjl]
  {10.1088/2041-8205/709/2/L138}, \href
  {http://adsabs.harvard.edu/abs/2010ApJ...709L.138L} {709, L138}

\bibitem[\protect\citeauthoryear{{Lehner} \& {Howk}}{{Lehner} \&
  {Howk}}{2011}]{lh11}
{Lehner} N.,  {Howk} J.~C.,  2011, \mn@doi [Science] {10.1126/science.1209069},
  \href {http://adsabs.harvard.edu/abs/2011Sci...334..955L} {334, 955}

\bibitem[\protect\citeauthoryear{{Lehner}, {Howk}, {Thom}, {Fox}, {Tumlinson},
  {Tripp}  \& {Meiring}}{{Lehner} et~al.}{2012}]{lehner+12}
{Lehner} N.,  {Howk} J.~C.,  {Thom} C.,  {Fox} A.~J.,  {Tumlinson} J.,  {Tripp}
  T.~M.,   {Meiring} J.~D.,  2012, \mn@doi [\mnras]
  {10.1111/j.1365-2966.2012.21428.x}, \href
  {http://adsabs.harvard.edu/abs/2012MNRAS.424.2896L} {424, 2896}

\bibitem[\protect\citeauthoryear{{Lehner}, {Howk}  \& {Wakker}}{{Lehner}
  et~al.}{2015}]{lehner+15}
{Lehner} N.,  {Howk} J.~C.,   {Wakker} B.~P.,  2015, \mn@doi [\apj]
  {10.1088/0004-637X/804/2/79}, \href
  {http://adsabs.harvard.edu/abs/2015ApJ...804...79L} {804, 79}

\bibitem[\protect\citeauthoryear{{Li}, {Bregman}, {Wang}, {Crain}  \&
  {Anderson}}{{Li} et~al.}{2018}]{li+18}
{Li} J.-T.,  {Bregman} J.~N.,  {Wang} Q.~D.,  {Crain} R.~A.,   {Anderson}
  M.~E.,  2018, \mn@doi [\apjl] {10.3847/2041-8213/aab2af}, \href
  {http://adsabs.harvard.edu/abs/2018ApJ...855L..24L} {855, L24}

\bibitem[\protect\citeauthoryear{{Macquart} \& {Koay}}{{Macquart} \&
  {Koay}}{2013}]{macquart13}
{Macquart} J.-P.,  {Koay} J.~Y.,  2013, \mn@doi [\apj]
  {10.1088/0004-637X/776/2/125}, \href
  {http://adsabs.harvard.edu/abs/2013ApJ...776..125M} {776, 125}

\bibitem[\protect\citeauthoryear{{Mahony} et~al.,}{{Mahony}
  et~al.}{2018}]{mahony18}
{Mahony} E.~K.,  et~al., 2018, preprint, \href
  {http://adsabs.harvard.edu/abs/2018arXiv181004354M} {} (\mn@eprint {arXiv}
  {1810.04354})

\bibitem[\protect\citeauthoryear{{Maller} \& {Bullock}}{{Maller} \&
  {Bullock}}{2004}]{mb04}
{Maller} A.~H.,  {Bullock} J.~S.,  2004, \mn@doi [\mnras]
  {10.1111/j.1365-2966.2004.08349.x}, \href
  {http://adsabs.harvard.edu/abs/2004MNRAS.355..694M} {355, 694}

\bibitem[\protect\citeauthoryear{{Manchester}, {Fan}, {Lyne}, {Kaspi}  \&
  {Crawford}}{{Manchester} et~al.}{2006}]{manchester06}
{Manchester} R.~N.,  {Fan} G.,  {Lyne} A.~G.,  {Kaspi} V.~M.,   {Crawford} F.,
  2006, \mn@doi [\apj] {10.1086/505461}, \href
  {http://adsabs.harvard.edu/abs/2006ApJ...649..235M} {649, 235}

\bibitem[\protect\citeauthoryear{{Martini} et~al.,}{{Martini}
  et~al.}{2018}]{DESI}
{Martini} P.,  et~al., 2018, in Society of Photo-Optical Instrumentation
  Engineers (SPIE) Conference Series. p. 107021F (\mn@eprint {arXiv}
  {1807.09287}), \mn@doi{10.1117/12.2313063}

\bibitem[\protect\citeauthoryear{{Mathews} \& {Prochaska}}{{Mathews} \&
  {Prochaska}}{2017}]{mp17}
{Mathews} W.~G.,  {Prochaska} J.~X.,  2017, \mn@doi [\apjl]
  {10.3847/2041-8213/aa8861}, \href
  {http://adsabs.harvard.edu/abs/2017ApJ...846L..24M} {846, L24}

\bibitem[\protect\citeauthoryear{{McConnachie}}{{McConnachie}}{2012}]{McConnachie12}
{McConnachie} A.~W.,  2012, \mn@doi [\aj] {10.1088/0004-6256/144/1/4}, \href
  {https://ui.adsabs.harvard.edu/#abs/2012AJ....144....4M} {144, 4}

\bibitem[\protect\citeauthoryear{{McQuinn}}{{McQuinn}}{2014}]{mcquinn14}
{McQuinn} M.,  2014, \mn@doi [\apjl] {10.1088/2041-8205/780/2/L33}, \href
  {http://adsabs.harvard.edu/abs/2014ApJ...780L..33M} {780, L33}

\bibitem[\protect\citeauthoryear{{Miralda-Escud{\'e}}, {Cen}, {Ostriker}  \&
  {Rauch}}{{Miralda-Escud{\'e}} et~al.}{1996}]{mco+96}
{Miralda-Escud{\'e}} J.,  {Cen} R.,  {Ostriker} J.~P.,   {Rauch} M.,  1996,
  \mn@doi [\apj] {10.1086/177992}, \href
  {http://adsabs.harvard.edu/abs/1996ApJ...471..582M} {471, 582}

\bibitem[\protect\citeauthoryear{{Mo} \& {Miralda-Escude}}{{Mo} \&
  {Miralda-Escude}}{1996}]{mm96}
{Mo} H.~J.,  {Miralda-Escude} J.,  1996, \mn@doi [\apj] {10.1086/177808}, \href
  {http://adsabs.harvard.edu/cgi-bin/nph-bib_query?bibcode=1996ApJ...469..589M&db_key=AST}
  {469, 589}

\bibitem[\protect\citeauthoryear{{Moster}, {Somerville}, {Maulbetsch}, {van den
  Bosch}, {Macci{\`o}}, {Naab}  \& {Oser}}{{Moster} et~al.}{2010}]{moster+10}
{Moster} B.~P.,  {Somerville} R.~S.,  {Maulbetsch} C.,  {van den Bosch} F.~C.,
  {Macci{\`o}} A.~V.,  {Naab} T.,   {Oser} L.,  2010, \mn@doi [\apj]
  {10.1088/0004-637X/710/2/903}, \href
  {http://adsabs.harvard.edu/abs/2010ApJ...710..903M} {710, 903}

\bibitem[\protect\citeauthoryear{{Muratov}, {Kere{\v s}},
  {Faucher-Gigu{\`e}re}, {Hopkins}, {Quataert}  \& {Murray}}{{Muratov}
  et~al.}{2015}]{muratov15}
{Muratov} A.~L.,  {Kere{\v s}} D.,  {Faucher-Gigu{\`e}re} C.-A.,  {Hopkins}
  P.~F.,  {Quataert} E.,   {Murray} N.,  2015, \mn@doi [\mnras]
  {10.1093/mnras/stv2126}, \href
  {http://adsabs.harvard.edu/abs/2015MNRAS.454.2691M} {454, 2691}

\bibitem[\protect\citeauthoryear{{Navarro}, {Frenk}  \& {White}}{{Navarro}
  et~al.}{1997}]{nfw97}
{Navarro} J.~F.,  {Frenk} C.~S.,   {White} S.~D.~M.,  1997, \mn@doi [\apj]
  {10.1086/304888}, \href {http://adsabs.harvard.edu/abs/1997ApJ...490..493N}
  {490, 493}

\bibitem[\protect\citeauthoryear{{Nicastro} et~al.,}{{Nicastro}
  et~al.}{2005}]{nme+05}
{Nicastro} F.,  et~al., 2005, \mn@doi [\nat] {10.1038/nature03245}, \href
  {http://adsabs.harvard.edu/abs/2005Natur.433..495N} {433, 495}

\bibitem[\protect\citeauthoryear{{Nicastro} et~al.,}{{Nicastro}
  et~al.}{2018}]{nicastro18}
{Nicastro} F.,  et~al., 2018, \mn@doi [\nat] {10.1038/s41586-018-0204-1}, \href
  {http://adsabs.harvard.edu/abs/2018Natur.558..406N} {558, 406}

\bibitem[\protect\citeauthoryear{{Nidever}, {Majewski}  \& {Butler
  Burton}}{{Nidever} et~al.}{2008}]{nidever08}
{Nidever} D.~L.,  {Majewski} S.~R.,   {Butler Burton} W.,  2008, \mn@doi [\apj]
  {10.1086/587042}, \href {http://adsabs.harvard.edu/abs/2008ApJ...679..432N}
  {679, 432}

\bibitem[\protect\citeauthoryear{{Nidever}, {Majewski}, {Butler Burton}  \&
  {Nigra}}{{Nidever} et~al.}{2010}]{nidever10}
{Nidever} D.~L.,  {Majewski} S.~R.,  {Butler Burton} W.,   {Nigra} L.,  2010,
  \mn@doi [\apj] {10.1088/0004-637X/723/2/1618}, \href
  {http://adsabs.harvard.edu/abs/2010ApJ...723.1618N} {723, 1618}

\bibitem[\protect\citeauthoryear{{O'Meara}, {Tytler}, {Kirkman}, {Suzuki},
  {Prochaska}, {Lubin}  \& {Wolfe}}{{O'Meara} et~al.}{2001}]{omeara+01}
{O'Meara} J.~M.,  {Tytler} D.,  {Kirkman} D.,  {Suzuki} N.,  {Prochaska} J.~X.,
   {Lubin} D.,   {Wolfe} A.~M.,  2001, \mn@doi [\apj] {10.1086/320579}, \href
  {http://adsabs.harvard.edu/abs/2001ApJ...552..718O} {552, 718}

\bibitem[\protect\citeauthoryear{{Oppenheimer} et~al.,}{{Oppenheimer}
  et~al.}{2016}]{beno+16}
{Oppenheimer} B.~D.,  et~al., 2016, \mn@doi [\mnras] {10.1093/mnras/stw1066},
  \href {http://adsabs.harvard.edu/abs/2016MNRAS.460.2157O} {460, 2157}

\bibitem[\protect\citeauthoryear{{Palanque-Delabrouille}
  et~al.,}{{Palanque-Delabrouille} et~al.}{2013}]{palanque+13}
{Palanque-Delabrouille} N.,  et~al., 2013, \mn@doi [\aap]
  {10.1051/0004-6361/201322130}, \href
  {http://adsabs.harvard.edu/abs/2013A%26A...559A..85P} {559, A85}

\bibitem[\protect\citeauthoryear{{Planck Collaboration} et~al.,}{{Planck
  Collaboration} et~al.}{2013}]{planckXI}
{Planck Collaboration} et~al., 2013, \mn@doi [\aap]
  {10.1051/0004-6361/201220941}, \href
  {http://adsabs.harvard.edu/abs/2013A%26A...557A..52P} {557, A52}

\bibitem[\protect\citeauthoryear{{Planck Collaboration} et~al.,}{{Planck
  Collaboration} et~al.}{2016}]{Planck2015}
{Planck Collaboration} et~al., 2016, \mn@doi [\aap]
  {10.1051/0004-6361/201525830}, \href
  {http://adsabs.harvard.edu/abs/2016A%26A...594A..13P} {594, A13}

\bibitem[\protect\citeauthoryear{{Prochaska}}{{Prochaska}}{2006}]{pro06}
{Prochaska} J.~X.,  2006, \mn@doi [\apj] {10.1086/507126}, \href
  {http://adsabs.harvard.edu/cgi-bin/nph-bib_query?bibcode=2006ApJ...650..272P&db_key=AST}
  {650, 272}

\bibitem[\protect\citeauthoryear{{Prochaska} \& {Neeleman}}{{Prochaska} \&
  {Neeleman}}{2018}]{pn18}
{Prochaska} J.~X.,  {Neeleman} M.,  2018, \mn@doi [\mnras]
  {10.1093/mnras/stx2824}, \href
  {http://adsabs.harvard.edu/abs/2018MNRAS.474..318P} {474, 318}

\bibitem[\protect\citeauthoryear{{Prochaska} \& {Tumlinson}}{{Prochaska} \&
  {Tumlinson}}{2009}]{pt09}
{Prochaska} J.~X.,  {Tumlinson} J.,  2009, {Baryons: What,When and Where?}.
p.~419, \mn@doi{10.1007/978-1-4020-9457-6_16}

\bibitem[\protect\citeauthoryear{{Prochaska} \& {et al.}}{{Prochaska} \& {et
  al.}}{2019}]{x+18}
{Prochaska} J.,  {et al.} 2019, \mn@doi [ApJ] {10.1088/0067-0049/217/2/21},
  \href {http://adsabs.harvard.edu/abs/2015ApJS..217...21F} {in prep}

\bibitem[\protect\citeauthoryear{{Prochaska}, {Weiner}, {Chen}, {Mulchaey}  \&
  {Cooksey}}{{Prochaska} et~al.}{2011}]{pwc+11}
{Prochaska} J.~X.,  {Weiner} B.,  {Chen} H.-W.,  {Mulchaey} J.,   {Cooksey} K.,
   2011, \mn@doi [\apj] {10.1088/0004-637X/740/2/91}, \href
  {http://adsabs.harvard.edu/abs/2011ApJ...740...91P} {740, 91}

\bibitem[\protect\citeauthoryear{{Prochaska} et~al.,}{{Prochaska}
  et~al.}{2017}]{prochaska17}
{Prochaska} J.~X.,  et~al., 2017, \mn@doi [\apj] {10.3847/1538-4357/aa6007},
  \href {http://adsabs.harvard.edu/abs/2017ApJ...837..169P} {837, 169}

\bibitem[\protect\citeauthoryear{{Putman}, {Staveley-Smith}, {Freeman},
  {Gibson}  \& {Barnes}}{{Putman} et~al.}{2003}]{putman03}
{Putman} M.~E.,  {Staveley-Smith} L.,  {Freeman} K.~C.,  {Gibson} B.~K.,
  {Barnes} D.~G.,  2003, \mn@doi [\apj] {10.1086/344477}, \href
  {http://adsabs.harvard.edu/abs/2003ApJ...586..170P} {586, 170}

\bibitem[\protect\citeauthoryear{{Putman}, {Peek}  \& {Joung}}{{Putman}
  et~al.}{2012}]{putman12}
{Putman} M.~E.,  {Peek} J.~E.~G.,   {Joung} M.~R.,  2012, \mn@doi [\araa]
  {10.1146/annurev-astro-081811-125612}, \href
  {http://adsabs.harvard.edu/abs/2012ARA%26A..50..491P} {50, 491}

\bibitem[\protect\citeauthoryear{{Rauch}}{{Rauch}}{1998}]{rau98}
{Rauch} M.,  1998, \mn@doi [\araa] {10.1146/annurev.astro.36.1.267}, \href
  {http://adsabs.harvard.edu/abs/1998ARA%26A..36..267R} {36, 267}

\bibitem[\protect\citeauthoryear{{Reynolds}}{{Reynolds}}{1991}]{reynolds91}
{Reynolds} R.~J.,  1991, in {Bloemen} H.,  ed.,  IAU Symposium Vol. 144, The
  Interstellar Disk-Halo Connection in Galaxies. pp 67--76

\bibitem[\protect\citeauthoryear{{Richter} et~al.,}{{Richter}
  et~al.}{2017}]{richter+17}
{Richter} P.,  et~al., 2017, \mn@doi [\aap] {10.1051/0004-6361/201630081},
  \href {http://adsabs.harvard.edu/abs/2017A%26A...607A..48R} {607, A48}

\bibitem[\protect\citeauthoryear{{Riess}, {Fliri}  \& {Valls-Gabaud}}{{Riess}
  et~al.}{2012}]{riess12}
{Riess} A.~G.,  {Fliri} J.,   {Valls-Gabaud} D.,  2012, \mn@doi [\apj]
  {10.1088/0004-637X/745/2/156}, \href
  {http://adsabs.harvard.edu/abs/2012ApJ...745..156R} {745, 156}

\bibitem[\protect\citeauthoryear{{Rykoff} et~al.,}{{Rykoff}
  et~al.}{2016}]{rykoff16}
{Rykoff} E.~S.,  et~al., 2016, \mn@doi [\apjs] {10.3847/0067-0049/224/1/1},
  \href {http://adsabs.harvard.edu/abs/2016ApJS..224....1R} {224, 1}

\bibitem[\protect\citeauthoryear{{Salem}, {Besla}, {Bryan}, {Putman}, {van der
  Marel}  \& {Tonnesen}}{{Salem} et~al.}{2015}]{salem+15}
{Salem} M.,  {Besla} G.,  {Bryan} G.,  {Putman} M.,  {van der Marel} R.~P.,
  {Tonnesen} S.,  2015, \mn@doi [\apj] {10.1088/0004-637X/815/1/77}, \href
  {http://adsabs.harvard.edu/abs/2015ApJ...815...77S} {815, 77}

\bibitem[\protect\citeauthoryear{{Sargent}, {Young}, {Boksenberg}  \&
  {Tytler}}{{Sargent} et~al.}{1980}]{sargent80}
{Sargent} W.~L.~W.,  {Young} P.~J.,  {Boksenberg} A.,   {Tytler} D.,  1980,
  \mn@doi [\apjs] {10.1086/190644}, \href
  {http://adsabs.harvard.edu/abs/1980ApJS...42...41S} {42, 41}

\bibitem[\protect\citeauthoryear{{Savage} et~al.,}{{Savage}
  et~al.}{2003}]{ssw+03}
{Savage} B.~D.,  et~al., 2003, \mn@doi [\apjs] {10.1086/346229}, \href
  {http://adsabs.harvard.edu/abs/2003ApJS..146..125S} {146, 125}

\bibitem[\protect\citeauthoryear{{Savage}, {Narayanan}, {Lehner}  \&
  {Wakker}}{{Savage} et~al.}{2011}]{savage+11b}
{Savage} B.~D.,  {Narayanan} A.,  {Lehner} N.,   {Wakker} B.~P.,  2011, \mn@doi
  [\apj] {10.1088/0004-637X/731/1/14}, \href
  {http://adsabs.harvard.edu/abs/2011ApJ...731...14S} {731, 14}

\bibitem[\protect\citeauthoryear{{Sawala} et~al.,}{{Sawala}
  et~al.}{2016}]{APOSTLE}
{Sawala} T.,  et~al., 2016, \mn@doi [\mnras] {10.1093/mnras/stw145}, \href
  {http://adsabs.harvard.edu/abs/2016MNRAS.457.1931S} {457, 1931}

\bibitem[\protect\citeauthoryear{{Sembach}, {Howk}, {Ryans}  \&
  {Keenan}}{{Sembach} et~al.}{2000}]{sembach+00}
{Sembach} K.~R.,  {Howk} J.~C.,  {Ryans} R.~S.~I.,   {Keenan} F.~P.,  2000,
  \mn@doi [\apj] {10.1086/308173}, \href
  {http://adsabs.harvard.edu/abs/2000ApJ...528..310S} {528, 310}

\bibitem[\protect\citeauthoryear{{Sembach} et~al.,}{{Sembach}
  et~al.}{2003}]{sws+03}
{Sembach} K.~R.,  et~al., 2003, \mn@doi [\apjs] {10.1086/346231}, \href
  {http://adsabs.harvard.edu/cgi-bin/nph-bib_query?bibcode=2003ApJS..146..165S&db_key=AST}
  {146, 165}

\bibitem[\protect\citeauthoryear{{Sharma}, {McCourt}, {Parrish}  \&
  {Quataert}}{{Sharma} et~al.}{2012}]{sharma12}
{Sharma} P.,  {McCourt} M.,  {Parrish} I.~J.,   {Quataert} E.,  2012, \mn@doi
  [\mnras] {10.1111/j.1365-2966.2012.22050.x}, \href
  {https://ui.adsabs.harvard.edu/#abs/2012MNRAS.427.1219S} {427, 1219}

\bibitem[\protect\citeauthoryear{{Shen}, {Madau}, {Conroy}, {Governato}  \&
  {Mayer}}{{Shen} et~al.}{2014}]{shen+14}
{Shen} S.,  {Madau} P.,  {Conroy} C.,  {Governato} F.,   {Mayer} L.,  2014,
  \mn@doi [\apj] {10.1088/0004-637X/792/2/99}, \href
  {http://adsabs.harvard.edu/abs/2014ApJ...792...99S} {792, 99}

\bibitem[\protect\citeauthoryear{{Shull} \& {Danforth}}{{Shull} \&
  {Danforth}}{2018}]{shull18}
{Shull} J.~M.,  {Danforth} C.~W.,  2018, \mn@doi [\apjl]
  {10.3847/2041-8213/aaa2fa}, \href
  {http://adsabs.harvard.edu/abs/2018ApJ...852L..11S} {852, L11}

\bibitem[\protect\citeauthoryear{{Shull}, {Jones}, {Danforth}  \&
  {Collins}}{{Shull} et~al.}{2009}]{Shull09}
{Shull} J.~M.,  {Jones} J.~R.,  {Danforth} C.~W.,   {Collins} J.~A.,  2009,
  \mn@doi [\apj] {10.1088/0004-637X/699/1/754}, \href
  {http://adsabs.harvard.edu/abs/2009ApJ...699..754S} {699, 754}

\bibitem[\protect\citeauthoryear{{Somerville} \& {Dav{\'e}}}{{Somerville} \&
  {Dav{\'e}}}{2015}]{somerville+15}
{Somerville} R.~S.,  {Dav{\'e}} R.,  2015, \mn@doi [\araa]
  {10.1146/annurev-astro-082812-140951}, \href
  {http://adsabs.harvard.edu/abs/2015ARA%26A..53...51S} {53, 51}

\bibitem[\protect\citeauthoryear{{Sonnenfeld}, {Leauthaud}, {Auger}, {Gavazzi},
  {Treu}, {More}  \& {Komiyama}}{{Sonnenfeld} et~al.}{2018}]{sonnenfeld18}
{Sonnenfeld} A.,  {Leauthaud} A.,  {Auger} M.~W.,  {Gavazzi} R.,  {Treu} T.,
  {More} S.,   {Komiyama} Y.,  2018, \mn@doi [\mnras] {10.1093/mnras/sty2262},
  \href {http://adsabs.harvard.edu/abs/2018MNRAS.481..164S} {481, 164}

\bibitem[\protect\citeauthoryear{{Steigman}}{{Steigman}}{2010}]{steigman10}
{Steigman} G.,  2010, preprint, \href
  {https://ui.adsabs.harvard.edu/#abs/2010arXiv1008.4765S} {p. arXiv:1008.4765}
  (\mn@eprint {arXiv} {1008.4765})

\bibitem[\protect\citeauthoryear{{Stern}, {Hennawi}, {Prochaska}  \&
  {Werk}}{{Stern} et~al.}{2016}]{stern+16}
{Stern} J.,  {Hennawi} J.~F.,  {Prochaska} J.~X.,   {Werk} J.~K.,  2016,
  \mn@doi [\apj] {10.3847/0004-637X/830/2/87}, \href
  {http://adsabs.harvard.edu/abs/2016ApJ...830...87S} {830, 87}

\bibitem[\protect\citeauthoryear{{Stern}, {Faucher-Gigu{\`e}re}, {Hennawi},
  {Hafen}, {Johnson}  \& {Fielding}}{{Stern} et~al.}{2018}]{stern+18}
{Stern} J.,  {Faucher-Gigu{\`e}re} C.-A.,  {Hennawi} J.~F.,  {Hafen} Z.,
  {Johnson} S.~D.,   {Fielding} D.,  2018, \mn@doi [\apj]
  {10.3847/1538-4357/aac884}, \href
  {https://ui.adsabs.harvard.edu/#abs/2018ApJ...865...91S} {865, 91}

\bibitem[\protect\citeauthoryear{{Stocke}, {Keeney}, {Danforth}, {Shull},
  {Froning}, {Green}, {Penton}  \& {Savage}}{{Stocke} et~al.}{2013}]{stocke13}
{Stocke} J.~T.,  {Keeney} B.~A.,  {Danforth} C.~W.,  {Shull} J.~M.,  {Froning}
  C.~S.,  {Green} J.~C.,  {Penton} S.~V.,   {Savage} B.~D.,  2013, \mn@doi
  [\apj] {10.1088/0004-637X/763/2/148}, \href
  {http://adsabs.harvard.edu/abs/2013ApJ...763..148S} {763, 148}

\bibitem[\protect\citeauthoryear{{Tendulkar} et~al.,}{{Tendulkar}
  et~al.}{2017}]{tendulkar17}
{Tendulkar} S.~P.,  et~al., 2017, \mn@doi [\apjl] {10.3847/2041-8213/834/2/L7},
  \href {http://adsabs.harvard.edu/abs/2017ApJ...834L...7T} {834, L7}

\bibitem[\protect\citeauthoryear{{Thom}, {Peek}, {Putman}, {Heiles}, {Peek}  \&
  {Wilhelm}}{{Thom} et~al.}{2008}]{thom08}
{Thom} C.,  {Peek} J.~E.~G.,  {Putman} M.~E.,  {Heiles} C.,  {Peek} K.~M.~G.,
  {Wilhelm} R.,  2008, \mn@doi [\apj] {10.1086/589960}, \href
  {http://adsabs.harvard.edu/abs/2008ApJ...684..364T} {684, 364}

\bibitem[\protect\citeauthoryear{{Thom} et~al.,}{{Thom} et~al.}{2012}]{thom12}
{Thom} C.,  et~al., 2012, \mn@doi [\apjl] {10.1088/2041-8205/758/2/L41}, \href
  {http://adsabs.harvard.edu/abs/2012ApJ...758L..41T} {758, L41}

\bibitem[\protect\citeauthoryear{{Tripp} \& {et al.}}{{Tripp} \& {et
  al.}}{2019}]{tripp+19}
{Tripp} T.,  {et al.} 2019, \mn@doi [ApJ] {10.1088/0067-0049/217/2/21}, \href
  {http://adsabs.harvard.edu/abs/2015ApJS..217...21F} {in prep}

\bibitem[\protect\citeauthoryear{{Tumlinson} et~al.,}{{Tumlinson}
  et~al.}{2011}]{ttw+11}
{Tumlinson} J.,  et~al., 2011, \mn@doi [Science] {10.1126/science.1209840},
  \href {http://adsabs.harvard.edu/abs/2011Sci...334..948T} {334, 948}

\bibitem[\protect\citeauthoryear{{Tumlinson} et~al.,}{{Tumlinson}
  et~al.}{2013}]{tumlinson+13}
{Tumlinson} J.,  et~al., 2013, \mn@doi [\apj] {10.1088/0004-637X/777/1/59},
  \href {http://adsabs.harvard.edu/abs/2013ApJ...777...59T} {777, 59}

\bibitem[\protect\citeauthoryear{{Vedantham} \& {Phinney}}{{Vedantham} \&
  {Phinney}}{2019}]{vedantham19}
{Vedantham} H.~K.,  {Phinney} E.~S.,  2019, \mn@doi [\mnras]
  {10.1093/mnras/sty2948}, \href
  {http://adsabs.harvard.edu/abs/2019MNRAS.483..971V} {483, 971}

\bibitem[\protect\citeauthoryear{{Verschuur}}{{Verschuur}}{1975}]{verschuur75}
{Verschuur} G.~L.,  1975, \mn@doi [Annual Review of Astronomy and Astrophysics]
  {10.1146/annurev.aa.13.090175.001353}, \href
  {https://ui.adsabs.harvard.edu/\#abs/1975ARA&A..13..257V} {13, 257}

\bibitem[\protect\citeauthoryear{{Vikhlinin}, {Kravtsov}, {Forman}, {Jones},
  {Markevitch}, {Murray}  \& {Van Speybroeck}}{{Vikhlinin}
  et~al.}{2006}]{vik06}
{Vikhlinin} A.,  {Kravtsov} A.,  {Forman} W.,  {Jones} C.,  {Markevitch} M.,
  {Murray} S.~S.,   {Van Speybroeck} L.,  2006, \mn@doi [\apj]
  {10.1086/500288}, \href {http://adsabs.harvard.edu/abs/2006ApJ...640..691V}
  {640, 691}

\bibitem[\protect\citeauthoryear{{Wakker}}{{Wakker}}{1991}]{Wakker91}
{Wakker} B.~P.,  1991, \aap, \href
  {http://adsabs.harvard.edu/abs/1991A%26A...250..499W} {250, 499}

\bibitem[\protect\citeauthoryear{{Wakker}}{{Wakker}}{2001}]{wakker01}
{Wakker} B.~P.,  2001, \mn@doi [\apjs] {10.1086/321783}, \href
  {http://adsabs.harvard.edu/abs/2001ApJS..136..463W} {136, 463}

\bibitem[\protect\citeauthoryear{{Wakker}}{{Wakker}}{2004}]{wakker04}
{Wakker} B.~P.,  2004, in {van Woerden} H.,  {Wakker} B.~P.,  {Schwarz} U.~J.,
   {de Boer} K.~S.,  eds,  Astrophysics and Space Science Library Vol. 312,
  High Velocity Clouds. p.~25, \mn@doi{10.1007/1-4020-2579-3_2}

\bibitem[\protect\citeauthoryear{{Wakker} et~al.,}{{Wakker}
  et~al.}{2003}]{wakkeretal03}
{Wakker} B.~P.,  et~al., 2003, \mn@doi [\apjs] {10.1086/346230}, \href
  {http://adsabs.harvard.edu/cgi-bin/nph-bib_query?bibcode=2003ApJS..146....1W&db_key=AST}
  {146, 1}

\bibitem[\protect\citeauthoryear{{Wakker}, {Savage}, {Fox}, {Benjamin}  \&
  {Shapiro}}{{Wakker} et~al.}{2012}]{wakker+12}
{Wakker} B.~P.,  {Savage} B.~D.,  {Fox} A.~J.,  {Benjamin} R.~A.,   {Shapiro}
  P.~R.,  2012, \mn@doi [\apj] {10.1088/0004-637X/749/2/157}, \href
  {http://adsabs.harvard.edu/abs/2012ApJ...749..157W} {749, 157}

\bibitem[\protect\citeauthoryear{{Wang} et~al.,}{{Wang}
  et~al.}{2005}]{wangetal05}
{Wang} Q.~D.,  et~al., 2005, \mn@doi [\apj] {10.1086/497584}, \href
  {http://adsabs.harvard.edu/cgi-bin/nph-bib_query?bibcode=2005ApJ...635..386W&db_key=AST}
  {635, 386}

\bibitem[\protect\citeauthoryear{{Werk}, {Prochaska}, {Thom}, {Tumlinson},
  {Tripp}, {O'Meara}  \& {Peeples}}{{Werk} et~al.}{2013}]{werk+13}
{Werk} J.~K.,  {Prochaska} J.~X.,  {Thom} C.,  {Tumlinson} J.,  {Tripp} T.~M.,
  {O'Meara} J.~M.,   {Peeples} M.~S.,  2013, \mn@doi [\apjs]
  {10.1088/0067-0049/204/2/17}, \href
  {http://adsabs.harvard.edu/abs/2013ApJS..204...17W} {204, 17}

\bibitem[\protect\citeauthoryear{{Werk} et~al.,}{{Werk} et~al.}{2014}]{werk+14}
{Werk} J.~K.,  et~al., 2014, \mn@doi [\apj] {10.1088/0004-637X/792/1/8}, \href
  {http://adsabs.harvard.edu/abs/2014ApJ...792....8W} {792, 8}

\bibitem[\protect\citeauthoryear{{Winkel}, {Kerp}, {Fl{\"o}er}, {Kalberla},
  {Ben Bekhti}, {Keller}  \& {Lenz}}{{Winkel} et~al.}{2016}]{ebhis}
{Winkel} B.,  {Kerp} J.,  {Fl{\"o}er} L.,  {Kalberla} P.~M.~W.,  {Ben Bekhti}
  N.,  {Keller} R.,   {Lenz} D.,  2016, \mn@doi [\aap]
  {10.1051/0004-6361/201527007}, \href
  {https://ui.adsabs.harvard.edu/#abs/2016A&A...585A..41W} {585, A41}

\bibitem[\protect\citeauthoryear{{Xu} \& {Han}}{{Xu} \& {Han}}{2015}]{xu15}
{Xu} J.,  {Han} J.~L.,  2015, \mn@doi [Research in Astronomy and Astrophysics]
  {10.1088/1674-4527/15/10/002}, \href
  {http://adsabs.harvard.edu/abs/2015RAA....15.1629X} {15, 1629}

\bibitem[\protect\citeauthoryear{{Yao}, {Manchester}  \& {Wang}}{{Yao}
  et~al.}{2017}]{YMW17}
{Yao} J.~M.,  {Manchester} R.~N.,   {Wang} N.,  2017, \mn@doi [\apj]
  {10.3847/1538-4357/835/1/29}, \href
  {http://adsabs.harvard.edu/abs/2017ApJ...835...29Y} {835, 29}

\bibitem[\protect\citeauthoryear{{Zhai} et~al.,}{{Zhai} et~al.}{2017}]{zhai17}
{Zhai} Z.,  et~al., 2017, \mn@doi [\apj] {10.3847/1538-4357/aa8eee}, \href
  {http://adsabs.harvard.edu/abs/2017ApJ...848...76Z} {848, 76}

\bibitem[\protect\citeauthoryear{{Zheng}, {Ofek}, {Kulkarni}, {Neill}  \&
  {Juric}}{{Zheng} et~al.}{2014}]{zheng14}
{Zheng} Z.,  {Ofek} E.~O.,  {Kulkarni} S.~R.,  {Neill} J.~D.,   {Juric} M.,
  2014, \mn@doi [\apj] {10.1088/0004-637X/797/1/71}, \href
  {http://adsabs.harvard.edu/abs/2014ApJ...797...71Z} {797, 71}

\bibitem[\protect\citeauthoryear{{Zheng}, {Putman}, {Peek}  \& {Joung}}{{Zheng}
  et~al.}{2015}]{zheng+15}
{Zheng} Y.,  {Putman} M.~E.,  {Peek} J.~E.~G.,   {Joung} M.~R.,  2015, \mn@doi
  [\apj] {10.1088/0004-637X/807/1/103}, \href
  {http://adsabs.harvard.edu/abs/2015ApJ...807..103Z} {807, 103}

\bibitem[\protect\citeauthoryear{{van Woerden}, {Wakker}, {Schwarz}  \& {de
  Boer}}{{van Woerden} et~al.}{2004}]{vanwoerden04}
{van Woerden} H.,  {Wakker} B.~P.,  {Schwarz} U.~J.,   {de Boer} K.~S.,  eds,
  2004, {High Velocity Clouds}  Astrophysics and Space Science Library Vol.
  312, \mn@doi{10.1007/1-4020-2579-3.
}

\bibitem[\protect\citeauthoryear{{van der Marel}, {Fardal}, {Besla}, {Beaton},
  {Sohn}, {Anderson}, {Brown}  \& {Guhathakurta}}{{van der Marel}
  et~al.}{2012a}]{vdm2012a}
{van der Marel} R.~P.,  {Fardal} M.,  {Besla} G.,  {Beaton} R.~L.,  {Sohn}
  S.~T.,  {Anderson} J.,  {Brown} T.,   {Guhathakurta} P.,  2012a, \mn@doi
  [\apj] {10.1088/0004-637X/753/1/8}, \href
  {http://adsabs.harvard.edu/abs/2012ApJ...753....8V} {753, 8}

\bibitem[\protect\citeauthoryear{{van der Marel}, {Besla}, {Cox}, {Sohn}  \&
  {Anderson}}{{van der Marel} et~al.}{2012b}]{vandermar12b}
{van der Marel} R.~P.,  {Besla} G.,  {Cox} T.~J.,  {Sohn} S.~T.,   {Anderson}
  J.,  2012b, \mn@doi [\apj] {10.1088/0004-637X/753/1/9}, \href
  {http://adsabs.harvard.edu/abs/2012ApJ...753....9V} {753, 9}

\makeatother
\end{thebibliography}


\bsp	
\label{lastpage}
\end{document}